\documentclass[preprint,12pt]{elsarticle}
\usepackage{subfigure}
\usepackage{lineno,hyperref}
\usepackage{indentfirst}
\usepackage{bm}
\usepackage{threeparttable}
\usepackage{multirow}
\usepackage{mathrsfs}
\usepackage{amsmath}
\usepackage{array}
\usepackage{amssymb}
\usepackage{graphicx}
\usepackage{url}
\usepackage{float}
\usepackage{epstopdf}
\usepackage{booktabs}
\usepackage{color}
\usepackage{lscape}
\usepackage{fancyhdr}
\usepackage{dsfont}
\usepackage[top=2.5cm, bottom=2.5cm, left=2cm, right=2cm]{geometry}
\modulolinenumbers[5]

\journal{Elsevier}

\biboptions{numbers,sort&compress}

\begin{document}

\begin{frontmatter}

\title{A second-order unified gas-kinetic wave-particle method with enhanced mesh independence for hypersonic flows}

\author{Junzhe Cao$^a$}
\ead{jcaobb@connect.ust.hk}

\author{Rui Zhang$^b$}
\ead{zhangruinwpu@mail.nwpu.edu.cn}

\author{Wenpei Long$^a$}
\ead{wlongab@connect.ust.hk}

\author{Chengwen Zhong$^b$}
\ead{zhongcw@nwpu.edu.cn}

\author[]{Kun Xu$^a$$^c$$^d$$^*$\corref{mycorrespondingauthor}}
\ead{makxu@ust.hk}

\address{$^a$Department of Mathematics, The Hong Kong University of Science and Technology, Hong Kong, China\\
$^b$School of Aeronautics, Northwestern Polytechnical University, Xi'an, Shaanxi 710072, China\\
$^c$Department of Mechanical and Aerospace Engineering, The Hong Kong University of Science and Technology, Hong Kong, China\\
$^d$HKUST Shenzhen Research Institute, Shenzhen, 518057, China}

\begin{abstract}
Benefiting from the direct modeling of physical laws in a discretized space and the automatic decomposition of the gas distribution function into hydrodynamic waves and particles, the unified gas-kinetic wave-particle (UGKWP) method offers significant advantages for multiscale flows such as hypersonic flows, plasma transport, and radiation transport. In this study, the particle sampling accuracy in the UGKWP method is improved from first order to second order, so that the second-order spatial and temporal accuracy is preserved across the full scheme. Specifically, the modifications include second-order particle sampling based on local macroscopic gradients, a weighted least-squares gradient reconstruction that incorporates wall values, a revised Venkatakrishnan limiter for highly stretched cells, and conservation corrections after particle sampling. Moreover, the first-order Chapman--Enskog term is considered in the free-transport part of the hydrodynamic wave flux, enabling better recovery of the gas-kinetic scheme (GKS) in the near-continuum regime. Based on these improvements, the mesh-independence behavior of the UGKWP method is notably enhanced, which is more consistent with the performance of the unified gas-kinetic scheme (UGKS), validated by a detailed hypersonic cylinder flow test case. Furthermore, systematic comparisons with the single-scale direct simulation Monte Carlo (DSMC) method are performed for two-dimensional hypersonic flow over a cylinder and three-dimensional flow over a blunt cone. Wall pressure, shear stress, and heat flux coefficients ($C_P$, $C_F$, and $C_Q$) are examined in the cylinder case, while the overall aerodynamic coefficients ($C_L$, $C_D$, and $L/D$) are assessed in the cone case. The multiscale UGKWP method exhibits significantly better mesh-independence performance than DSMC for mesh-sensitive quantities such as $C_F$, $C_Q$, $C_D$, and $L/D$, which are critical for aerodynamic and thermal protection design of near-space hypersonic vehicles.
\end{abstract}

\begin{keyword}
Unified gas-kinetic wave-particle method; multiscale flow; hypersonic flow; mesh independence; second-order spatial accuracy
\end{keyword}

\end{frontmatter}

\section{Introduction}\label{Sec:introduction}
Multiscale phenomena play a prominent role in a wide range of physical processes. By introducing an observation scale into the numerical scheme, physical laws can be modeled directly in a discretized space~\cite{dm1,dm2}. Built on the integral solution of the kinetic equation, the unified gas-kinetic scheme (UGKS)~\cite{ugks} and the unified gas-kinetic wave-particle (UGKWP) method~\cite{ugkwp1,ugkwp2} have demonstrated substantial advantages in diverse multiscale problems, including hypersonic flows~\cite{ugks-hyper,wp-hyper,wp-hyper2}, gas-solid two-phase flows~\cite{ugks-phase,wp-phase}, radiation transport~\cite{ugks-rad,wp-rad}, plasma transport~\cite{ugks-plasma,wp-plasma}, and turbulent flows~\cite{wp-turb}. In recent years, both methods have been successfully extended to increasingly complex physical settings~\cite{wp2025a,wp2025b,ugks2026a,ugks2026b,wp2026}.

For continuum-rarefied multiscale flows, the UGKS and UGKWP methods can recover the gas-kinetic scheme (GKS)~\cite{gks2001} as a Navier--Stokes solver in the continuum regime, and serve as a Boltzmann solver in the rarefied regime. In the transitional regime, high accuracy can be achieved because the flux is constructed from the characteristic integral solution. For hypersonic flows, the continuum-rarefied multiscale transition extends from the intensely compressed region at the vehicle head to the strong expansion at the leeward side, accompanied by an increasingly thickened Knudsen ($\rm{Kn}$) layer. As shown in Ref.~\cite{jiangaia}, near the surface of an X-38-like vehicle, the local Knudsen number $\rm{Kn_{GLL}}$ can span four to five orders of magnitude. When simulating such flows, the computational efficiency depends strongly on the height of the first mesh layer. On one hand, the local time step is limited by the cell size due to the Courant--Friedrichs--Lewy (CFL) condition, and the total number of mesh cells increases with smaller cell sizes within the boundary layer. On the other hand, to simplify the mesh topology for complex vehicle geometries, the first-layer heights are often kept uniform, determined by the most continuum-like region, such as the stagnation point on the windward side. A key advantage of multiscale methods is that they enable physically resolved meshes in the near-continuum flow regime~\cite{principle}, without being restricted by the molecular mean free path and mean collision time. As a result, the mesh-independent performance of multiscale methods is essential for improving computational efficiency. A detailed study on mesh independence can also provide valuable guidance for engineering applications. Currently, relatively few studies have addressed this issue~\cite{pfeiffer,fp}, and previous UGKS studies have been limited to low-speed and one-dimensional cases~\cite{ugks,ugks-chen}. In this paper, we first conduct a detailed study on the mesh independence of the UGKWP method for hypersonic flows, incorporating results from the UGKS and the direct simulation Monte Carlo (DSMC) method~\cite{dsmc} for comparison.

From another perspective, for hypersonic flows, deterministic methods such as the UGKS are constrained by the high computational cost of velocity space discretization. This is because when constructing the velocity space mesh, both high-temperature and low-temperature regions must be considered simultaneously. High-temperature regions require a wide velocity space window, whereas low-temperature regions require local refinements. Using unstructured meshes~\cite{unstructured} or adaptive Cartesian meshes~\cite{avs}, recent extensions~\cite{wyf,zr,cjf} have reduced the velocity space mesh to about $5000$ points for three-dimensional hypersonic problems. Although the number of iterations can also be decreased by $1$ to $3$ orders of magnitude through implicit methods~\cite{iugks1,iugks2} and multigrid methods~\cite{multigrid}, the achievable simulation size remains limited by an excessively large velocity space due to memory constraints, which grow further for multispecies mixtures and thermal nonequilibrium flows. In contrast, this problem is naturally mitigated in stochastic particle methods. Only $100$ to $400$ particles are required per cell, which automatically provide the necessary velocity-space resolution. Moreover, particles offer additional flexibility in modifying microscopic velocities and free-transport times~\cite{xu-tau}. However, the implicit treatment of the UGKWP method is considerably more complicated because of randomness~\cite{iugkwp}, and particle noise makes it difficult to maintain spatial reconstruction accuracy. This paper focuses on the latter problem. By enhancing particle sampling, second-order accuracy is preserved across the full scheme, leading to substantially better mesh-independent performance in hypersonic tests. Additionally, the limiter and wave flux calculations are improved.

The remainder of this paper is organized as follows. The basis of gas kinetic theory is introduced in Sec.~\ref{sec:gkt}. Section~\ref{sec:wp} introduces the modified UGKWP method. Numerical test cases are presented in Sec.~\ref{sec:cases}, and concluding remarks are provided in Sec.~\ref{sec:conclusion}.

\section{Gas kinetic theory}\label{sec:gkt}
In gas-kinetic theory, the microscopic gas distribution function $f\left(\rho,\boldsymbol{x},\boldsymbol{u},\boldsymbol{\xi},t\right)$ is used to describe the physical system, where $\boldsymbol{x}$ denotes the position, $t$ denotes the time, $\rho$ is the macroscopic density, $\boldsymbol{u}$ is the molecular velocity vector, and the internal energy is represented by the equivalent velocity $\boldsymbol{\xi}$. The distribution function at the equilibrium state is entirely determined by macroscopic variables, known as the Maxwellian distribution function:
\begin{equation}\label{eq:maxwellian}
g = \rho{\left( {\frac{1}{{2\pi RT}}} \right)^{\frac{{D + 3}}{2}}}\exp \left( { - \frac{{\left|\boldsymbol{c}\right|^2 + \left|\boldsymbol{\xi}\right|^2}}{2RT}} \right),
\end{equation}
where $R$ is the specific gas constant, $T$ is the macroscopic temperature, $D$ denotes the number of internal degrees of freedom associated with rotational and vibrational modes, $\boldsymbol{c}=\boldsymbol{u}-\boldsymbol{U}$ is the peculiar velocity, and $\boldsymbol{U}$ is the macroscopic velocity vector. The macroscopic conservative variables $\boldsymbol{W}=\left(\rho, \rho \boldsymbol{U}, \rho E\right)^T$ are obtained from the distribution function $f$ as follows:
\begin{equation}\label{eq:gkt_moment0}
\begin{aligned}
&\boldsymbol{W} = \int_{\mathbb{R}^D}\int_{\mathbb{R}^3} \boldsymbol{\Psi}f {\rm d}\boldsymbol{u} {\rm d}\boldsymbol{\xi},\\
&\boldsymbol{\Psi}=\left({ 1,\boldsymbol{u}, {1\over 2}\left|\boldsymbol{u}\right|^2+{1\over 2}\left|\boldsymbol{\xi}\right|^2 }\right)^T.
\end{aligned}
\nonumber
\end{equation}
These integrals are referred to as ``moments'' in gas-kinetic theory. The heat flux vector $\boldsymbol{Q}$ can also be obtained by calculating moments:
\begin{equation}\label{eq:gkt_moment1}
{\boldsymbol{Q}} = \frac{1}{2}\int_{\mathbb{R}^D}\int_{\mathbb{R}^3} \boldsymbol{c}\left[{ {1\over 2}\left({ \left|\boldsymbol{c}\right|^2 + \left|\boldsymbol{\xi}\right|^2 }\right)f }\right] {\rm d}\boldsymbol{u} {\rm d}\boldsymbol{\xi}.
\nonumber
\end{equation}
In this paper, the Shakhov model~\cite{shakhov} is used, which can recover both the viscosity coefficient $\mu$ and Prandtl number $\rm{Pr}$ in the continuum flow regime, as follows:
\begin{equation}\label{eq:shakhov}
\begin{aligned}
& \frac{\partial f}{\partial t}+\boldsymbol{u}\cdot\frac{\partial f}{\partial \boldsymbol{x}}=\frac{g^S-f}{\tau},\\
& g^S=g\left[ {1+(1-{\rm{Pr}})\frac{\boldsymbol{c}\cdot\boldsymbol{Q}}{5pRT}\left( {\frac{\left|\boldsymbol{c}\right|^2}{RT}-5} \right)} \right],
\end{aligned}
\end{equation}
where $p=\rho RT$ is the macroscopic pressure and $\tau=\mu/p$ is the relaxation time. The Shakhov model reduces to the Bhatnagar-Gross-Krook (BGK) model~\cite{bgk} at ${\rm{Pr}}=1$.

Additionally, in this paper, the molecular mean free path $\lambda$ is defined by the variable soft sphere (VSS) model:
\begin{equation}\label{eq:lambda}
\begin{aligned}
\lambda &= \frac{1}{\beta}\frac{\mu}{p}\sqrt{\frac{RT}{2\pi}},\\
\beta  &= \frac{{5(\alpha  + 1)(\alpha  + 2)}}{{4\alpha (5 - 2\omega )(7 - 2\omega )}},
\end{aligned}
\end{equation}
where $\alpha$ is the molecular scattering factor and $\omega$ is the viscosity index. The variable hard sphere (VHS) model is used in this study to calculate viscosity as $\mu\sim T^{\omega}$, and $\alpha$ is taken as unity to calculate $\lambda$. The farfield Knudsen number is defined as ${\rm{Kn}}_{\infty}=\lambda_{\infty}/L$, where $L$ is the reference length.

\section{Unified gas-kinetic wave-particle method}\label{sec:wp}
Within the general FVM framework, the conservation of both macroscopic variables and microscopic gas distribution functions is considered:
\begin{equation}\label{eq:conservation}
\begin{aligned}
\boldsymbol{W}^{n+1}_i &= \boldsymbol{W}^{n}_i - \frac{1}{\Omega_i}\sum\limits_{j\in \mathcal{M}\left(i\right)}\boldsymbol{F}_js_j,\\
f^{n+1}_i &= f^{n}_i - \frac{1}{\Omega_i}\sum\limits_{j\in \mathcal{M}\left(i\right)}\int^{\Delta t}_0\boldsymbol{u}\cdot\boldsymbol{n}_jf_js_j{\rm d}t + \int^{\Delta t}_0\mathcal{J}\left(f,f\right){\rm d}t,
\nonumber
\end{aligned}
\end{equation}
where ``$i$'' denotes the cell index, $\mathcal{M}\left(i\right)$ is the collection of interfaces surrounding cell ``$i$'', $\Omega_i$ is the area or volume of cell ``$i$'', $s_j$ is the length or area of interface ``$j$'', and $\boldsymbol{n}_j$ is the outer unit normal vector. In the Shakhov model equation, Eq.~\eqref{eq:shakhov}, the collision term is modeled as $\mathcal{J}\left(f,f\right)=\frac{g^S-f}{\tau}$. The macroscopic flux $\boldsymbol{F}_j$ is calculated as:
\begin{equation}\label{eq:macrof}
\boldsymbol{F}_j = \int_{\mathbb{R}^D} \int_{\mathbb{R}^3}\int^{\Delta t}_0\boldsymbol{u}\cdot\boldsymbol{n}_jf_j\boldsymbol{\Psi} {\rm d}t{\rm d}\boldsymbol{u}{\rm d}\boldsymbol{\xi},
\end{equation}
where $\Delta t$ is the time step. Setting the coordinate origin at the center of an interface, the integral solution along the characteristic line can be derived as:
\begin{equation}\label{eq:integral}
f\left( {\boldsymbol{0},t} \right) = \frac{1}{\tau}\int^t_0 g^S\left[{ -\boldsymbol{u}\left({ t-\tilde{t} }\right),\tilde{t} }\right]e^{\frac{\tilde{t}-t}{\tau}} d\tilde{t} + e^{-t/\tau}f\left( {-\boldsymbol{u}t,0} \right).
\end{equation}
In this equation, both collision and free transport are considered. The first term denotes the accumulation of the equilibrium state, while the second term denotes the transport of the non-equilibrium initial state. Expanding $g^S$ and $f$ as:
\begin{equation}\label{eq:taylor}
\begin{aligned}
g^S\left({ \boldsymbol{x},t }\right)&=g^S_0+\frac{\partial g^S}{\partial \boldsymbol{x}}\cdot\boldsymbol{x}+\frac{\partial g^S}{\partial t}t,\\
f\left({ \boldsymbol{x},0 }\right)&=f_0+\frac{\partial f}{\partial \boldsymbol{x}}\cdot\boldsymbol{x},
\nonumber
\end{aligned}
\end{equation}
it can be derived from Eq.~\eqref{eq:integral} that:
\begin{equation}\label{eq:integrala1}
f\left( {\boldsymbol{0},t} \right) = \varepsilon_ag^S_0 + \varepsilon_b\frac{\partial g^S}{\partial \boldsymbol{x}}\cdot\boldsymbol{u}+\varepsilon_c\frac{\partial g^S}{\partial t}+\varepsilon_df_0 + \varepsilon_e\frac{\partial f}{\partial \boldsymbol{x}}\cdot\boldsymbol{u},
\end{equation}
where,
\begin{equation}\label{eq:integrala2}
\begin{aligned}
\varepsilon_a &= 1-e^{-t/\tau},\\
\varepsilon_b &= te^{-t/\tau}-\tau\left(1-e^{-t/\tau}\right),\\
\varepsilon_c &= t-\tau\left(1-e^{-t/\tau}\right),\\
\varepsilon_d &= e^{-t/\tau},\\
\varepsilon_e &= -te^{-t/\tau}.
\nonumber
\end{aligned}
\end{equation}
The flux can then be obtained by substituting Eq.~\eqref{eq:integrala1} into Eq.~\eqref{eq:macrof}:
\begin{equation}\label{eq:integralb1}
\boldsymbol{F} = \boldsymbol{F}^{eq}+\boldsymbol{F}^{fr},
\end{equation}
where,
\begin{equation}\label{eq:integralb2}
\begin{aligned}
\boldsymbol{F}^{eq} =& \int_{\mathbb{R}^D} \int_{\mathbb{R}^3}{\boldsymbol{\Psi}\left(\delta_ag^S_0+\delta_b\frac{\partial g^S}{\partial \boldsymbol{x}}\cdot\boldsymbol{u}+ \delta_c\frac{\partial g^S}{\partial t} \right)\left(\boldsymbol{u}\cdot\boldsymbol{n}\right) {\rm d}\boldsymbol{u}} {\rm d}\boldsymbol{\xi},\\
\boldsymbol{F}^{fr} =& \int_{\mathbb{R}^D} \int_{\mathbb{R}^3}{ \boldsymbol{\Psi}\left(\delta_df_0+\delta_e\frac{\partial f}{\partial \boldsymbol{x}}\cdot\boldsymbol{u} \right)} \left(\boldsymbol{u}\cdot\boldsymbol{n}\right) {\rm d}\boldsymbol{u} {\rm d}\boldsymbol{\xi},
\end{aligned}
\end{equation}
and,
\begin{equation}\label{eq:integralb3}
\begin{aligned}
\delta_a &= \Delta t-\tau\left( {1-e^{-\Delta t/\tau}} \right),\\
\delta_b &= 2\tau^2\left( {1-e^{-\Delta t/\tau}} \right)-\tau\Delta t-\tau\Delta te^{-\Delta t/\tau},\\
\delta_c &= \frac{\Delta t^2}{2}-\tau\Delta t+\tau^2\left( {1-e^{-\Delta t/\tau}} \right),\\
\delta_d &= \tau\left( {1-e^{-\Delta t/\tau}} \right),\\
\delta_e &= \tau\Delta te^{-\Delta t/\tau}-\tau^2\left( {1-e^{-\Delta t/\tau}} \right).
\nonumber
\end{aligned}
\end{equation}
The $\boldsymbol{F}^{eq}$ term in Eq.~\eqref{eq:integralb1} plays a dominant role in the continuum flow regime, where the gas evolution is governed by deterministic hydrodynamic waves. Conversely, the $\boldsymbol{F}^{fr}$ term gradually dominates the gas evolution as the flow becomes more rarefied. Numerical particles are employed to simulate this latter part.

\subsection{Wave evolution}\label{sec:macro}
As the first term in Eq.~\eqref{eq:integralb1}, $\boldsymbol{F}^{eq}$ can be calculated using the GKS framework~\cite{gks2001}, where the first step is to calculate $g^S_0$ and its derivatives $\frac{\partial g^S}{\partial \boldsymbol{x}}$ and $\frac{\partial g^S}{\partial t}$. It has been shown in Ref.~\cite{wp-hyper} that $\frac{\partial g^S}{\partial \boldsymbol{x}}$ and $\frac{\partial g^S}{\partial t}$ can be simplified using $\frac{\partial g}{\partial \boldsymbol{x}}$ and $\frac{\partial g}{\partial t}$. For simplicity, $g_0$ is used instead of $g^S_0$, in conjunction with the Prandtl correction method in the GKS~\cite{may}. Here, $g_0$ denotes the equilibrium state at the interface, which is calculated assuming sufficient collisions as follows:
\begin{equation}\label{eq:mac1}
\int_{\mathbb{R}^D} \int_{\mathbb{R}^3} \boldsymbol{\Psi} g_0 {\rm d}\boldsymbol{u} {\rm d}\boldsymbol{\xi} = \boldsymbol{W}_0 = \int_{\mathbb{R}^D} \int_{\boldsymbol{u}\cdot\boldsymbol{n}>0} \boldsymbol{\Psi} g_L {\rm d}\boldsymbol{u} {\rm d}\boldsymbol{\xi} + \int_{\mathbb{R}^D} \int_{\boldsymbol{u}\cdot\boldsymbol{n}<0} \boldsymbol{\Psi} g_R {\rm d}\boldsymbol{u} {\rm d}\boldsymbol{\xi},
\nonumber
\end{equation}
where $g_L$, $g_R$, and $g_0$ are calculated from the macroscopic variables $\boldsymbol{W}_L$, $\boldsymbol{W}_R$, and $\boldsymbol{W}_0$ through Eq.~\eqref{eq:maxwellian}. $\boldsymbol{W}_L$ and $\boldsymbol{W}_R$ are the left-hand and right-hand limits at the interface after reconstruction. Meanwhile, the spatial derivative $\frac{\partial g}{\partial \boldsymbol{x}}$ and temporal derivative $\frac{\partial g}{\partial t}$ are derived as:
\begin{equation}\label{eq:mac2}
\begin{aligned}
\frac{\partial g}{\partial \boldsymbol{x}} &= \boldsymbol{a}g,\\
\frac{\partial g}{\partial t} &= Ag,
\nonumber
\end{aligned}
\end{equation}
where,
\begin{equation}
\begin{aligned}
\boldsymbol{a}&=\frac{1}{g}\frac{\partial g}{\partial \boldsymbol{x}}=\frac{\partial \left[ {\rm{ln}}(g) \right]}{\partial \boldsymbol{x}},\\
A&=\frac{1}{g}\frac{\partial g}{\partial t}=\frac{\partial \left[ {\rm{ln}}(g) \right]}{\partial t}.
\nonumber
\end{aligned}
\end{equation}
For the spatial derivative, $\boldsymbol{a}$ is written in the form:
\begin{equation}
\boldsymbol{a}=\boldsymbol{a}_0+\boldsymbol{a}_1u_1+\boldsymbol{a}_2u_2+\boldsymbol{a}_3u_3+\boldsymbol{a}_4\frac{|\boldsymbol{u}|^2+|\boldsymbol{\xi}|^2}{2},
\nonumber
\end{equation}
and,
\begin{equation}
\begin{aligned}
\boldsymbol{a}_0 &= \frac{ 2\frac{\partial\rho}{\partial\boldsymbol{x}}\Lambda + \rho\left[{ -4\Lambda^2\boldsymbol{U}\cdot\frac{\partial\boldsymbol{U}}{\partial\boldsymbol{x}} + \left({3+D-2|\boldsymbol{U}|^2\Lambda}\right)\frac{\partial\Lambda}{\partial\boldsymbol{x}} }\right] }{2\rho\Lambda},\\
\boldsymbol{a}_1 &= 2\left( {\frac{\partial U_1}{\partial\boldsymbol{x}}\Lambda + U_1\frac{\partial\Lambda}{\partial\boldsymbol{x}}} \right),\\
\boldsymbol{a}_2 &= 2\left( {\frac{\partial U_2}{\partial\boldsymbol{x}}\Lambda + U_2\frac{\partial\Lambda}{\partial\boldsymbol{x}}} \right),\\
\boldsymbol{a}_3 &= 2\left( {\frac{\partial U_3}{\partial\boldsymbol{x}}\Lambda + U_3\frac{\partial\Lambda}{\partial\boldsymbol{x}}} \right),\\
\boldsymbol{a}_4 &= -2\frac{\partial\Lambda}{\partial\boldsymbol{x}},
\nonumber
\end{aligned}
\end{equation}
where $\Lambda=\frac{1}{2RT}$. When using the Prandtl correction method, the derivative of $\Lambda$ is modified as follows:
\begin{equation}\label{eq:mac3}
\begin{aligned}
&\frac{\partial\Lambda}{\partial\boldsymbol{x}}\Rightarrow\frac{\partial\Lambda}{\partial\boldsymbol{x}}/{\rm{Pr}}.
\nonumber
\end{aligned}
\end{equation}

On the other hand, the time derivative is determined by enforcing the conservation relation on the right-hand side of the BGK-type model:
\begin{equation}
\int_{\mathbb{R}^D} \int_{\mathbb{R}^3} gA \boldsymbol{\Psi} {\rm d}\boldsymbol{u} {\rm d}\boldsymbol{\xi} = -\int_{\mathbb{R}^D} \int_{\mathbb{R}^3} g \boldsymbol{a}\cdot\boldsymbol{u} \boldsymbol{\Psi} {\rm d}\boldsymbol{u} {\rm d}\boldsymbol{\xi} = \boldsymbol{b}.
\nonumber
\end{equation}
Expanding $A$ as $A=A_0+A_1u_1+A_2u_2+A_3u_3+A_4\frac{|\boldsymbol{u}|^2+|\boldsymbol{\xi}|^2}{2}$, it is rewritten as:
\begin{equation}
A_{\beta} M_{\alpha\beta} = b_{\alpha},
\nonumber
\end{equation}
where subscripts $\alpha$ and $\beta$ denote components, and the Einstein summation convention is used. For example, in the two-dimensional case:
\begin{equation}
\boldsymbol{M}=
\left(
\begin{array}{cccc}
1    &U_1                       &U_2                      &B_1 \\
U_1  &U_1^2+\frac{1}{2\Lambda}  &U_1U_2                   &B_2 \\
U_2  &U_1U_2                    &U_2^2+\frac{1}{2\Lambda} &B_3 \\
B_1  &B_2                       &B_3                      &B_4
\end{array} \right),
\nonumber
\end{equation}
where
\begin{equation}
\begin{aligned}
&B_1 = \frac{1}{2}\left( {U_1^2+U_2^2+\frac{D+3}{2\Lambda}} \right),\\
&B_2 = \frac{1}{2}\left( {U_1^3+U_1U_2^2+\frac{D+5}{2\Lambda}U_1} \right),\\
&B_3 = \frac{1}{2}\left( {U_2^3+U_1^2U_2+\frac{D+5}{2\Lambda}U_2} \right),\\
&B_4 = \frac{1}{4}\left[ {\left({U_1^2+U_2^2}\right)^2 + \frac{D+5}{\Lambda}\left({U_1^2+U_2^2}\right) + \frac{D^2+8D+15}{4\Lambda^2}} \right].
\nonumber
\end{aligned}
\end{equation}
The resulting coefficients are:
\begin{equation}
\begin{aligned}
&A_4 = \frac{8\Lambda^2}{D+3}\left[ {b_4-U_1b_2-U_2b_3-(B_1+U_1^2+U_2^2)b_1} \right],\\
&A_3 = 2\Lambda(b_3-U_2b_1)-U_2A_4,\\
&A_2 = 2\Lambda(b_2-U_1b_1)-U_1A_4,\\
&A_1 = b_1-U_1A_2-U_2A_3-B_1A_4.
\nonumber
\end{aligned}
\end{equation}

Further details on calculating $g_0$, $\frac{\partial g}{\partial \boldsymbol{x}}$, and $\frac{\partial g}{\partial t}$ are provided in Refs.~\cite{gks2001,gks1998}. By substituting these into Eq.~\eqref{eq:integralb2}, the wave flux $\boldsymbol{F}^{eq}$ is obtained. To improve accuracy and robustness in flow regions with drastic scale variations, the scale-related coefficient $\delta$ is evaluated separately on the left and right sides of the interface ``$j$''. Consequently, $\delta_{a,L/R}$, $\delta_{b,L/R}$, and $\delta_{c,L/R}$ are calculated within the cell rather than directly at the interface. When sampling particles, the scale-related coefficient $e^{-\Delta t/\tau}$ is computed at the same location. This treatment aligns the wave flux with the particle evolution, preventing discrepancies in regions where scales change rapidly. Additional details can be found in Ref.~\cite{awp2}.
\begin{equation}\label{eq:feq}
\begin{aligned}
\boldsymbol{F}^{eq}_{j} =& \int_{\mathbb{R}^D}\int_{\boldsymbol{u}\cdot\boldsymbol{n}_j>0}{\boldsymbol{\Psi}\left(\delta_{a,L}g_0+\delta_{b,L}\frac{\partial g}{\partial \boldsymbol{x}}\cdot\boldsymbol{u}+ \delta_{c,L}\frac{\partial g}{\partial t} \right)\left(\boldsymbol{u}\cdot\boldsymbol{n}_j\right) {\rm d}\boldsymbol{u}}{\rm d}\boldsymbol{\xi}\\
+& \int_{\mathbb{R}^D}\int_{\boldsymbol{u}\cdot\boldsymbol{n}_j<0}{\boldsymbol{\Psi}\left(\delta_{a,R}g_0+\delta_{b,R}\frac{\partial g}{\partial \boldsymbol{x}}\cdot\boldsymbol{u}+ \delta_{c,R}\frac{\partial g}{\partial t} \right)\left(\boldsymbol{u}\cdot\boldsymbol{n}_j\right) {\rm d}\boldsymbol{u}}{\rm d}\boldsymbol{\xi},
\end{aligned}
\end{equation}
where the integration components are evaluated as:
\begin{equation}
\begin{aligned}
&\int_{\mathbb{R}} {\left( {\frac{1}{{2\pi RT}}} \right)^{\frac{{1}}{2}}}\exp \left( { - \frac{|u-U|^2}{{2RT}}} \right) du = 1,\\
&\int_{\mathbb{R}} u {\left( {\frac{1}{{2\pi RT}}} \right)^{\frac{{1}}{2}}}\exp \left( { - \frac{|u-U|^2}{{2RT}}} \right) du = U,\\
&\int_{u>0} {\left( {\frac{1}{{2\pi RT}}} \right)^{\frac{{1}}{2}}}\exp \left( { - \frac{|u-U|^2}{{2RT}}} \right) du = \frac{1}{2}\rm{erfc}\left( {-\sqrt{\Lambda}U} \right),\\
&\int_{u>0} u {\left( {\frac{1}{{2\pi RT}}} \right)^{\frac{{1}}{2}}}\exp \left( { - \frac{|u-U|^2}{{2RT}}} \right) du = \frac{U}{2}\rm{erfc}\left( {-\sqrt{\Lambda}U} \right) + \frac{1}{2}\frac{e^{-\Lambda U^2}}{\sqrt{\pi \Lambda}},\\
&\int_{u<0} {\left( {\frac{1}{{2\pi RT}}} \right)^{\frac{{1}}{2}}}\exp \left( { - \frac{|u-U|^2}{{2RT}}} \right) du = \frac{1}{2}\rm{erfc}\left( {\sqrt{\Lambda}U} \right),\\
&\int_{u<0} u {\left( {\frac{1}{{2\pi RT}}} \right)^{\frac{{1}}{2}}}\exp \left( { - \frac{|u-U|^2}{{2RT}}} \right) du = \frac{U}{2}\rm{erfc}\left( {\sqrt{\Lambda}U} \right) - \frac{1}{2}\frac{e^{-\Lambda U^2}}{\sqrt{\pi \Lambda}},
\nonumber
\end{aligned}
\end{equation}
and,
\begin{equation}
\begin{aligned}
&\int_{\mathbb{R}/u>0/u<0} u^{n+2} {\left( {\frac{1}{{2\pi RT}}} \right)^{\frac{{1}}{2}}}\exp \left( { - \frac{|u-U|^2}{{2RT}}} \right) du \\
= &U\int_{\mathbb{R}/u>0/u<0} u^{n+1} {\left( {\frac{1}{{2\pi RT}}} \right)^{\frac{{1}}{2}}}\exp \left( { - \frac{|u-U|^2}{{2RT}}} \right) du \\ + &\frac{n+1}{2\Lambda}\int_{\mathbb{R}/u>0/u<0} u^{n} {\left( {\frac{1}{{2\pi RT}}} \right)^{\frac{{1}}{2}}}\exp \left( { - \frac{|u-U|^2}{{2RT}}} \right) du,
\nonumber
\end{aligned}
\end{equation}
where $\rm{erfc}()=1-\rm{erf}()$ is the complementary error function.

\subsection{Particle evolution}\label{sec:micro}
In this method, numerical particles are employed to describe the nonequilibrium part of the distribution function. The particle parameters are mass $m$, position $\boldsymbol{x}$, velocity $\boldsymbol{u}$, and specific internal thermal energy $e=0.5\left|\boldsymbol{\xi}\right|^2$. As shown in Eq.~\eqref{eq:integral}, the distribution function after a time step $\Delta t$ is a combination of the initial state $f_0$ and the equilibrium state $g^S$. For a particle from the initial state, the probability of undergoing free transport without collision is $e^{-\frac{\Delta t}{\tau}}$; otherwise, it collides and relaxes into $g^S$. Thus, the cumulative distribution of a free-transport particle is $e^{-\frac{\Delta t}{\tau}}$. By drawing a random number $\epsilon$ uniformly distributed in $\left(0,1\right)$, the free-transport time $t_f$ of a particle is calculated as:
\begin{equation}\label{eq:mic1}
t_f = {\rm{min}}\left(-\tau{\rm{ln}}\left(\epsilon\right),\Delta t\right).
\end{equation}
The position of particle ``$k$'' is then updated by:
\begin{equation}\label{eq:mic2}
\boldsymbol{x}_k \Rightarrow \boldsymbol{x}_k + \boldsymbol{u}_kt_{f,k},
\end{equation}
and the contribution of all particles to the macroscopic conserved variables is given by:
\begin{equation}\label{eq:mic3}
\boldsymbol{W}_i^{fr,p} = \boldsymbol{W}_i^{p,+}-\boldsymbol{W}_i^{p,-},
\end{equation}
where,
\begin{equation}\label{eq:mic4}
\boldsymbol{W}_i^{p,+/-} = \frac{1}{\Omega_i}\sum\limits_{k\in \mathcal{N}^{+/-}\left(i\right)}m_{p,k}\boldsymbol{\Psi}_k,
\nonumber
\end{equation}
and $\mathcal{N}\left(i\right)$ is the collection of particles within cell ``$i$''. Superscripts ``$-$'' and ``$+$'' denote the states before and after free transport, respectively. Following the calculation of $\boldsymbol{W}_{i}^{fr,p}$, if $t_f<\Delta t$, the particle is relaxed into $g^S$ (collisional particles). If $t_f=\Delta t$, the particle is retained (collisionless particles).

According to Eq.~\eqref{eq:integral}, a fraction $e^{-\frac{\Delta t}{\tau}}$ of particles must be sampled from the hydrodynamic wave part, $\boldsymbol{W}_i^{h}=\boldsymbol{W}_i-\boldsymbol{W}_i^{p}$, with $t_f$ set to $\Delta t$. The total mass density of the sampled collisionless particles is:
\begin{equation}\label{eq:mic5}
\rho_i^{hp} = e^{-\frac{\Delta t}{\tau_i}}\rho_i^{h}.
\nonumber
\end{equation}
Unless $\rho_i^{h}=0$ (where no particles are sampled), the number of sampled particles is:
\begin{equation}\label{eq:mic6}
N^{hp}_i = \lceil{\frac{\rho^{hp}_i}{\rho_i}N^{hp,{\rm{ref}}}}\rceil,
\nonumber
\end{equation}
where $N^{hp,{\rm{ref}}}$ is a specified reference number. The corresponding mass of each particle is:
\begin{equation}\label{eq:mic7}
m_{p,k} = \frac{\rho_i^{hp}\Omega_i}{N^{hp}_i}.
\end{equation}
The particle positions are uniformly distributed within cell ``$i$''. The velocities are obtained via acceptance-rejection sampling. First, a velocity sample is drawn from the Maxwellian distribution:
\begin{equation}\label{eq:maxw}
\begin{aligned}
u_{k,1} &= U_{i,1} + \sqrt{2RT_i}{\rm{cos}}(2\pi \epsilon_{a1})\sqrt{-{\rm{ln}}(\epsilon_{a2})},\\
u_{k,2} &= U_{i,2} + \sqrt{2RT_i}{\rm{cos}}(2\pi \epsilon_{b1})\sqrt{-{\rm{ln}}(\epsilon_{b2})},\\
u_{k,3} &= U_{i,3} + \sqrt{2RT_i}{\rm{cos}}(2\pi \epsilon_{c1})\sqrt{-{\rm{ln}}(\epsilon_{c2})},
\end{aligned}
\end{equation}
where $\epsilon$ is a random number uniformly distributed in $\left(0,1\right)$. For the internal thermal energy associated with molecular rotational and vibrational motions, two reduced functions of the Shakhov model are derived:
\begin{equation}\label{eq:thermal}
\begin{aligned}
\int_{\mathbb{R}^D} g^S {\rm d}\boldsymbol{\xi} &= \left[1+\left(1-{\rm{Pr}}\right)\frac{\boldsymbol{c}\cdot\boldsymbol{Q}}{5pRT}\left(\frac{\left|\boldsymbol{c}\right|^2}{RT}-5\right)\right]\rho\left(\frac{1}{2\pi RT}\right)^{\frac{3}{2}}e^{-\frac{\left|\boldsymbol{c}\right|^2}{2RT}},\\
\int_{\mathbb{R}^D} \left|\boldsymbol{\xi}\right|^2 g^S {\rm d}\boldsymbol{\xi} &= DRT\left[1+\left(1-{\rm{Pr}}\right)\frac{\boldsymbol{c}\cdot\boldsymbol{Q}}{5pRT}\left(\frac{\left|\boldsymbol{c}\right|^2}{RT}-5\right)\right]\rho\left(\frac{1}{2\pi RT}\right)^{\frac{3}{2}}e^{-\frac{\left|\boldsymbol{c}\right|^2}{2RT}}.
\nonumber
\end{aligned}
\end{equation}
Consequently, the acceptance-rejection criterion is defined as:
\begin{equation}\label{eq:ajshak}
\frac{1+(1-{\rm{Pr}})\frac{\boldsymbol{c}_k\cdot\boldsymbol{Q}_i}{5p_iRT_i}\left( {\frac{|\boldsymbol{c}_k|^2}{RT_i}-5} \right)}{1+(1-{\rm{Pr}})\frac{20|\boldsymbol{Q}_i|}{p_i\sqrt{RT_i}}},
\end{equation}
and the specific internal thermal energy is:
\begin{equation}\label{eq:aje}
e_k=DRT_i.
\end{equation}

These steps are summarized by the sampling function:
\begin{equation}\label{eq:sample1st}
\left(m_{p,k},\boldsymbol{u}_{k},e_k\right)=SgS_{{\rm{1st}}}\left(\rho_i^{hp},\Omega_i,N^{hp}_i,\boldsymbol{U}_{i},T_i,\boldsymbol{Q}_i\right),
\end{equation}
which samples a particle from the distribution $g^S$. Briefly, the workflow of $SgS_{{\rm{1st}}}\left(\cdots\right)$ is as follows: the particle mass $m_{p,k}$ is calculated from the target sampling density $\rho_i^{hp}$, cell volume $\Omega_i$, and particle number $N^{hp}_i$ using Eq.~\eqref{eq:mic7}. Then, the acceptance-rejection method is used to determine the particle velocity $\boldsymbol{u}_{k}$. A proposal sample is generated from $\boldsymbol{U}_{i}$ and $T_i$ via Eq.~\eqref{eq:maxw}, and the acceptance criterion is evaluated using Eq.~\eqref{eq:ajshak}, incorporating the heat flux $\boldsymbol{Q}_i$. Finally, the specific internal thermal energy $e_k$ is calculated using Eq.~\eqref{eq:aje}. Because the inputs $\rho_i^{hp}$, $\boldsymbol{U}_{i}$, and $T_i$ are cell-averaged values, the spatial accuracy of this sampling is only first order. To improve this, local gradients are incorporated:
\begin{equation}\label{eq:sample2nd}
\begin{aligned}
\left(m_{p,k},\boldsymbol{u}_{k},e_k\right)=SgS_{{\rm{2nd}}}\left(\rho_i^{hp},\Omega_i,N^{hp}_i,\boldsymbol{U}_{i},T_i,\boldsymbol{Q}_i,\rho_i,\nabla\rho_i,\nabla\boldsymbol{U}_{i},\nabla T_i,\boldsymbol{x}_{i},\boldsymbol{x}_{k}\right),
\end{aligned}
\end{equation}
where $\boldsymbol{x}_{i}$ is the cell center coordinate and $\boldsymbol{x}_{k}$ is the particle coordinate, which is sampled from a uniform distribution in cell ``$i$''. The local macroscopic variables with second-order spatial accuracy are calculated as:
\begin{equation}
\begin{aligned}
\rho_k &= \rho_i + \nabla\rho_i\cdot\left(\boldsymbol{x}_{k}-\boldsymbol{x}_{i}\right),\\
\boldsymbol{U}_k &= \boldsymbol{U}_i + \nabla\boldsymbol{U}_i\cdot\left(\boldsymbol{x}_{k}-\boldsymbol{x}_{i}\right),\\
T_k &= T_i + \nabla T_i\cdot\left(\boldsymbol{x}_{k}-\boldsymbol{x}_{i}\right).
\nonumber
\end{aligned}
\end{equation}
To achieve second-order consistency in density, we use:
\begin{equation}\label{eq:massP2}
m_{p,k} = \frac{\rho_k}{\rho_i}\frac{\rho_i^{hp}\Omega_i}{N^{hp}_i},
\end{equation}
instead of Eq.~\eqref{eq:mic7}. For second-order velocity $\boldsymbol{U}_{k}$ and temperature $T_k$, we use:
\begin{equation}\label{eq:maxw2}
\begin{aligned}
u_{k,1} &= U_{k,1} + \sqrt{2RT_k}{\rm{cos}}(2\pi \epsilon_{a1})\sqrt{-{\rm{ln}}(\epsilon_{a2})},\\
u_{k,2} &= U_{k,2} + \sqrt{2RT_k}{\rm{cos}}(2\pi \epsilon_{b1})\sqrt{-{\rm{ln}}(\epsilon_{b2})},\\
u_{k,3} &= U_{k,3} + \sqrt{2RT_k}{\rm{cos}}(2\pi \epsilon_{c1})\sqrt{-{\rm{ln}}(\epsilon_{c2})},
\end{aligned}
\end{equation}
instead of Eq.~\eqref{eq:maxw}. When calculating the acceptance criterion and specific internal thermal energy, $\boldsymbol{U}_{i}$ and $T_i$ are replaced by $\boldsymbol{U}_{k}$ and $T_k$:
\begin{equation}\label{eq:ajshak2}
\frac{1+(1-{\rm{Pr}})\frac{\boldsymbol{c}_k\cdot\boldsymbol{Q}_i}{5p_kRT_k}\left( {\frac{|\boldsymbol{c}_k|^2}{RT_k}-5} \right)}{1+(1-{\rm{Pr}})\frac{20|\boldsymbol{Q}_i|}{p_k\sqrt{RT_k}}},
\end{equation}
and the specific internal thermal energy becomes:
\begin{equation}\label{eq:aje2}
e_k=DRT_k.
\nonumber
\end{equation}
After sampling second-order particles, conservation must be enforced due to stochastic variations. We examine all particles within cell ``$i$'':
\begin{equation}\label{eq:pForceConservation1}
\begin{aligned}
\tilde{\rho}_i^p &= \frac{1}{\Omega_i}\sum\limits_{k\in \mathcal{N}\left(i\right)}m_{p,k}\neq\rho_i^{hp}+\rho_i^{p},\\
\tilde{\boldsymbol{U}}_i &= \frac{1}{\tilde{\rho}_i^p\Omega_i}\sum\limits_{k\in \mathcal{N}\left(i\right)}m_{p,k}\boldsymbol{u}_{k}\neq\boldsymbol{U}_i,\\
\tilde{E}_{{\rm{in-tra}},i} &= \frac{1}{\tilde{\rho}_i^p\Omega_i}\sum\limits_{k\in \mathcal{N}\left(i\right)}\frac{1}{2}m_{p,k}\left|\boldsymbol{c}_{k}\right|^2\neq\frac{3}{2}RT_i,\\
\tilde{E}_{{\rm{in-rot+vib}},i} &= \frac{1}{\tilde{\rho}_i^p\Omega_i}\sum\limits_{k\in \mathcal{N}\left(i\right)}m_{p,k}e_k\neq\frac{D}{2}RT_i.
\end{aligned}
\end{equation}
These inequalities arise from three factors. First, when using the acceptance-rejection method in Eq.~\eqref{eq:ajshak2}, the accepted samples do not fully represent the Shakhov distribution because the negative tail is omitted. Thus, conservation is not strictly achieved when the $\boldsymbol{Q}$-related term is large enough to produce negative regions in $g^S$. Second, randomness in sampling $\boldsymbol{u}_{k}$ (Eqs.~\eqref{eq:maxw2} to \eqref{eq:ajshak2}) prevents exact momentum and energy conservation. Third, while first-order sampling (Eq.~\eqref{eq:mic7}) ensures exact mass conservation, second-order sampling (Eq.~\eqref{eq:massP2}) introduces stochastic mass fluctuations. Because conservation is critical, the mass and velocity of each particle are adjusted as follows:
\begin{equation}\label{eq:pForceConservation2}
\begin{aligned}
m_{p,k}&\Rightarrow m_{p,k}\frac{\rho_i^{hp}+\rho_i^{p}}{\tilde{\rho}_i^p},\\
\boldsymbol{u}_{k}&\Rightarrow \boldsymbol{U}_i+\left(\boldsymbol{u}_{k}-\tilde{\boldsymbol{U}}_i\right)\sqrt{\frac{3RT_i/2}{\tilde{E}_{{\rm{in-tra}},i}}},\\
e_k&\Rightarrow e_k\frac{DRT_i/2}{\tilde{E}_{{\rm{in-rot+vib}},i}}.
\end{aligned}
\end{equation}
Similar conservation corrections are widely used in stochastic particle methods (e.g., Ref.~\cite{pfeiffer-conservation}).

To calculate macroscopic gradients, the weighted least-squares method~\cite{wls} and the Venkatakrishnan limiter~\cite{venkata,jiri} are formulated as:
\begin{equation}\label{eq:gradient}
\nabla\varphi_i=\sigma_i\widetilde{\nabla\varphi_i},
\end{equation}
where $\varphi_i$ is an arbitrary macroscopic variable in cell ``$i$'', and $\widetilde{\nabla\varphi_i}$ is computed using the weighted least-squares method:
\begin{equation}\label{eq:wls}
\mathds{A}^T\mathds{A}\widetilde{\nabla\varphi_i}=\mathds{A}^T\boldsymbol{b},
\end{equation}
where $\mathds{A}_{j\ell}=w_j\left(x_{j,\ell}-x_{i,\ell}\right)$, $b_j=w_j\left(\varphi_{j}-\varphi_{i}\right)$, and the weight is $w_j=1/\left|\boldsymbol{x}_j-\boldsymbol{x}_i\right|$. Here, ``$j$'' is the index of the adjacent cell to target cell ``$i$'', and $\ell$ is the spatial dimension. Results in Fig.~\ref{cylinder-kn0.01_wallGrad} of Sec.~\ref{sec:cases} show that coarse-mesh accuracy is improved if the wall value is also considered:
\begin{equation}\label{eq:wls-wall}
\int_{\mathbb{R}^D} \int_{\mathbb{R}^3} \boldsymbol{\Psi} g_w {\rm d}\boldsymbol{u} {\rm d}\boldsymbol{\xi} = \boldsymbol{W}_w = \int_{\mathbb{R}^D} \int_{\boldsymbol{u}\cdot\boldsymbol{n}>0} \boldsymbol{\Psi} g_L {\rm d}\boldsymbol{u} {\rm d}\boldsymbol{\xi} + \int_{\mathbb{R}^D} \int_{\boldsymbol{u}\cdot\boldsymbol{n}<0} \boldsymbol{\Psi} g_R {\rm d}\boldsymbol{u} {\rm d}\boldsymbol{\xi},
\end{equation}
where $g_L$ is calculated from the reconstruction in the previous step and $g_R$ is the corresponding Maxwellian distribution function at the wall.

In Eq.~\eqref{eq:gradient}, $\sigma_i$ is calculated using the Venkatakrishnan limiter:
\begin{equation}\label{eq:venkata}
\begin{aligned}
&\sigma_i=\min\limits_{j}\left\{\begin{array}{cl}
\mathcal{L}\left(\varphi^M_i-\varphi_i,\Delta_{ij}\right),&\Delta_{ij}>0,\\
\mathcal{L}\left(\varphi^m_i-\varphi_i,\Delta_{ij}\right),&\Delta_{ij}<0,\\
1,&\Delta_{ij}=0.
\end{array}\right.
\end{aligned}
\end{equation}
Here, ``$j$'' is the index of the adjacent cell to target cell ``$i$'', and:
\begin{equation}\label{eq:venkata2}
\begin{aligned}
&\mathcal{L}\left(a,b\right)=\frac{a^2+2ab+\varepsilon^2}{a^2+ab+2b^2+\varepsilon^2},\\
&\Delta_{ij}=\widetilde{\nabla\varphi_i}\cdot\left(\boldsymbol{x}_{ij}-\boldsymbol{x}_{i}\right).
\end{aligned}
\end{equation}
The terms $\varphi^M_i=\max\limits_{j}\left(\varphi_i,\varphi_j\right)$ and $\varphi^m_i=\min\limits_{j}\left(\varphi_i,\varphi_j\right)$ are the maximum and minimum values, respectively, in the stencil of cell ``$i$'' and its neighbors. Here, $\boldsymbol{x}_{ij}$ denotes the coordinate of the interface center between cell ``$i$'' and cell ``$j$''. $\varepsilon^2$ is set to $\left(\zeta\sqrt{\Omega_i}\right)^3$ for two-dimensional cases and $\left(\zeta\sqrt[3]{\Omega_i}\right)^3$ for three-dimensional cases, with the global parameter $\zeta = 0.01$.

The limiter is modified to reduce the coupling effect of particle noise in highly stretched boundary-layer cells. Consider the condition where $\varphi^M_i=\varphi_i$. When calculating $\sigma_i$ using Eq.~\eqref{eq:venkata}, it may still select the branch $\Delta_{ij}>0$ due to a deviation in $\widetilde{\nabla\varphi_i}$. The formulation then becomes:
\begin{equation}
\mathcal{L}\left(\varphi^M_i-\varphi_i,\Delta_{ij}\right)=\mathcal{L}\left(0,\Delta_{ij}\right)=\frac{\varepsilon^2}{2\left[\widetilde{\nabla\varphi_i}\cdot\left(\boldsymbol{x}_{ij}-\boldsymbol{x}_{i}\right)\right]^2+\varepsilon^2}.
\nonumber
\end{equation}
For deterministic methods, the deviation in $\widetilde{\nabla\varphi_i}$ is small, meaning:
\begin{equation}
\widetilde{\nabla\varphi_i}\cdot\frac{\boldsymbol{x}_{ij}-\boldsymbol{x}_{i}}{\left|\boldsymbol{x}_{ij}-\boldsymbol{x}_{i}\right|}\rightarrow0,
\nonumber
\end{equation}
so that,
\begin{equation}
\mathcal{L}\left(\varphi^M_i-\varphi_i,\Delta_{ij}\right)\rightarrow\frac{\varepsilon^2}{\varepsilon^2}=1.
\nonumber
\end{equation}
However, with particle noise, this deviation is no longer negligible. When $\boldsymbol{x}_{ij}-\boldsymbol{x}_{i}$ is aligned with the long edge of a highly stretched cell, $\widetilde{\nabla\varphi_i}\cdot\left(\boldsymbol{x}_{ij}-\boldsymbol{x}_{i}\right)$ can be much larger than $\varepsilon$, leading to:
\begin{equation}
\mathcal{L}\left(\varphi^M_i-\varphi_i,\Delta_{ij}\right)=\frac{\varepsilon^2}{2\left[\widetilde{\nabla\varphi_i}\cdot\left(\boldsymbol{x}_{ij}-\boldsymbol{x}_{i}\right)\right]^2+\varepsilon^2}\rightarrow0.
\nonumber
\end{equation}
This causes a severe loss of spatial accuracy. To prevent this, the amplification effect of $\boldsymbol{x}_{ij}-\boldsymbol{x}_{i}$ is restricted by modifying the definition of $\Delta_{ij}$ in Eq.~\eqref{eq:venkata2} to:
\begin{equation}\label{eq:venkata3}
\Delta_{ij}=\widetilde{\nabla\varphi_i}\cdot\left(\boldsymbol{x}_{ij}-\boldsymbol{x}_{i}\right)\frac{\min\limits_{\jmath}\left|\boldsymbol{x}_{i\jmath}-\boldsymbol{x}_{i}\right|}{\left|\boldsymbol{x}_{ij}-\boldsymbol{x}_{i}\right|},
\end{equation}
where ``$\jmath$'' is an index considering all cells adjacent to cell ``$i$''. This modification reduces order reduction along the long edge without affecting the limiter's performance along the short edge. Without modifying the limiter, second-order particle sampling yields almost no benefit. Results in Sec.~\ref{sec:cases} demonstrate the effectiveness of second-order particle sampling with these modifications.

Additionally, in the original UGKWP method, because the analytical macroscopic flux of newly sampled particles corresponds to the second-order discrete velocity method (DVM), the free transport fluxes contributed by the collisional particles of $\left(\boldsymbol{W}^h-\boldsymbol{W}^{hp}\right)$ can be calculated as:
\begin{equation}\label{eq:ffrwave0}
\begin{aligned}
&\boldsymbol{F}^{fr,wave}=\boldsymbol{F}^{fr}_{{\rm{UGKS}}}\left(\boldsymbol{W}^h\right)-\boldsymbol{F}^{fr}_{{\rm{DVM}}}\left(\boldsymbol{W}^{hp}\right)\\
=&\int_{\mathbb{R}^D}\int_{\mathbb{R}^3}\boldsymbol{\Psi}\left(\delta_dg_0^h+\delta_e\frac{\partial g^h}{\partial \boldsymbol{x}}\cdot\boldsymbol{u}\right) \left(\boldsymbol{u}\cdot\boldsymbol{n}\right) {\rm d}\boldsymbol{u} {\rm d}\boldsymbol{\xi}\\
-&e^{-\frac{\Delta t}{\tau}}\int_{0}^{\Delta t}\int_{\mathbb{R}^D}\int_{\mathbb{R}^3}\boldsymbol{\Psi}\left(g_0^h-t\frac{\partial g^h}{\partial \boldsymbol{x}}\cdot\boldsymbol{u}\right) \left(\boldsymbol{u}\cdot\boldsymbol{n}\right) {\rm d}\boldsymbol{u} {\rm d}\boldsymbol{\xi}{\rm d}t\\
=&\int_{\mathbb{R}^D}\int_{\mathbb{R}^3}\boldsymbol{\Psi}\left[\left(\delta_d-\Delta te^{-\frac{\Delta t}{\tau}}\right)g_0^h+\left(\delta_e+\frac{\Delta t^2}{2}e^{-\frac{\Delta t}{\tau}}\right)\frac{\partial g^h}{\partial \boldsymbol{x}}\cdot\boldsymbol{u}\right] \left(\boldsymbol{u}\cdot\boldsymbol{n}\right) {\rm d}\boldsymbol{u} {\rm d}\boldsymbol{\xi}.
\end{aligned}
\end{equation}
Following the treatment in Eq.~\eqref{eq:feq}, the scale-related coefficient $\delta$ is evaluated separately on the left and right sides of interface ``$j$'' to ensure consistency with the sampled particles. In the original UGKWP method, the free-transport contribution of the hydrodynamic wave is approximated using the local Maxwellian distribution function $g$. As discussed in Ref.~\cite{wp-6}, this is a highly accurate approximation because molecules in the wave undergo frequent collisions and closely follow local equilibrium. While in the near-continuum regime, this approximation can be further improved by incorporating the first-order Chapman--Enskog expansion term. In the GKS~\cite{gks2001}, this term is implemented in the reconstruction of the initial conditions as $f_L=g_L\left[1+\boldsymbol{a}_L\cdot\boldsymbol{x}-\tau\left(\boldsymbol{a}_L\cdot\boldsymbol{u}+A_L\right)\right]$ and $f_R=g_R\left[1+\boldsymbol{a}_R\cdot\boldsymbol{x}-\tau\left(\boldsymbol{a}_R\cdot\boldsymbol{u}+A_R\right)\right]$ to recover viscous effects. This does not adversely affect rarefied-flow simulations because $\boldsymbol{W}^h\rightarrow\boldsymbol{0}$ when $\Delta t\ll\tau$. Consequently, the modification of Eq.~\eqref{eq:ffrwave0} is formulated as follows. The effect of including the $\delta_f$ term is examined in Figs.~\ref{cylinder-kn0.01_no6} and \ref{cylinder-kn0.01line_6} of Sec.~\ref{sec:cases}.
\begin{equation}\label{eq:ffrwave}
\begin{aligned}
\boldsymbol{F}^{fr,wave}_{j}=&\int_{\mathbb{R}^D}\int_{\boldsymbol{u}\cdot\boldsymbol{n}_j>0}\boldsymbol{\Psi}\left[\left(\delta_{d,L}-\Delta te^{-\frac{\Delta t}{\tau_{L}}}\right)g_0^h+\left(\delta_{e,L}+\frac{\Delta t^2}{2}e^{-\frac{\Delta t}{\tau_{L}}}\right)\frac{\partial g^h}{\partial \boldsymbol{x}}\cdot\boldsymbol{u}\right.\\
+&\left.\delta_{f,L}\left(\frac{\partial g^h}{\partial \boldsymbol{x}}\cdot\boldsymbol{u}+\frac{\partial g^h}{\partial t}\right)\right] \left(\boldsymbol{u}\cdot\boldsymbol{n}_j\right) {\rm d}\boldsymbol{u}{\rm d}\boldsymbol{\xi}\\
+& \int_{\mathbb{R}^D}\int_{\boldsymbol{u}\cdot\boldsymbol{n}_j<0}\boldsymbol{\Psi}\left[\left(\delta_{d,R}-\Delta te^{-\frac{\Delta t}{\tau_{R}}}\right)g_0^h+\left(\delta_{e,R}+\frac{\Delta t^2}{2}e^{-\frac{\Delta t}{\tau_{R}}}\right)\frac{\partial g^h}{\partial \boldsymbol{x}}\cdot\boldsymbol{u}\right.\\
+&\left.\delta_{f,R}\left(\frac{\partial g^h}{\partial \boldsymbol{x}}\cdot\boldsymbol{u}+\frac{\partial g^h}{\partial t}\right)\right] \left(\boldsymbol{u}\cdot\boldsymbol{n}_j\right) {\rm d}\boldsymbol{u}{\rm d}\boldsymbol{\xi},
\end{aligned}
\end{equation}
where,
\begin{equation}
\delta_{f}=-\tau^2\left(1-e^{-\frac{\Delta t}{\tau}}\right).
\nonumber
\end{equation}

Finally, the macroscopic update equation for the UGKWP method is:
\begin{equation}\label{eq:renew}
\boldsymbol{W}^{n+1}_i = \boldsymbol{W}^{n}_i - \frac{1}{\Omega_i}\sum\limits_{j\in \mathcal{M}\left(i\right)}\boldsymbol{F}_j^{eq}S_j - \frac{1}{\Omega_i}\sum\limits_{j\in \mathcal{M}\left(i\right)}\boldsymbol{F}_j^{fr,wave}S_j+\boldsymbol{W}_i^{fr,p}.
\end{equation}

\subsection{Algorithm of UGKWP}\label{sec:sum1}
\begin{figure}[H]
	\centering
	\subfigure[]{\label{fig1a}
			\includegraphics[width=0.22 \textwidth]{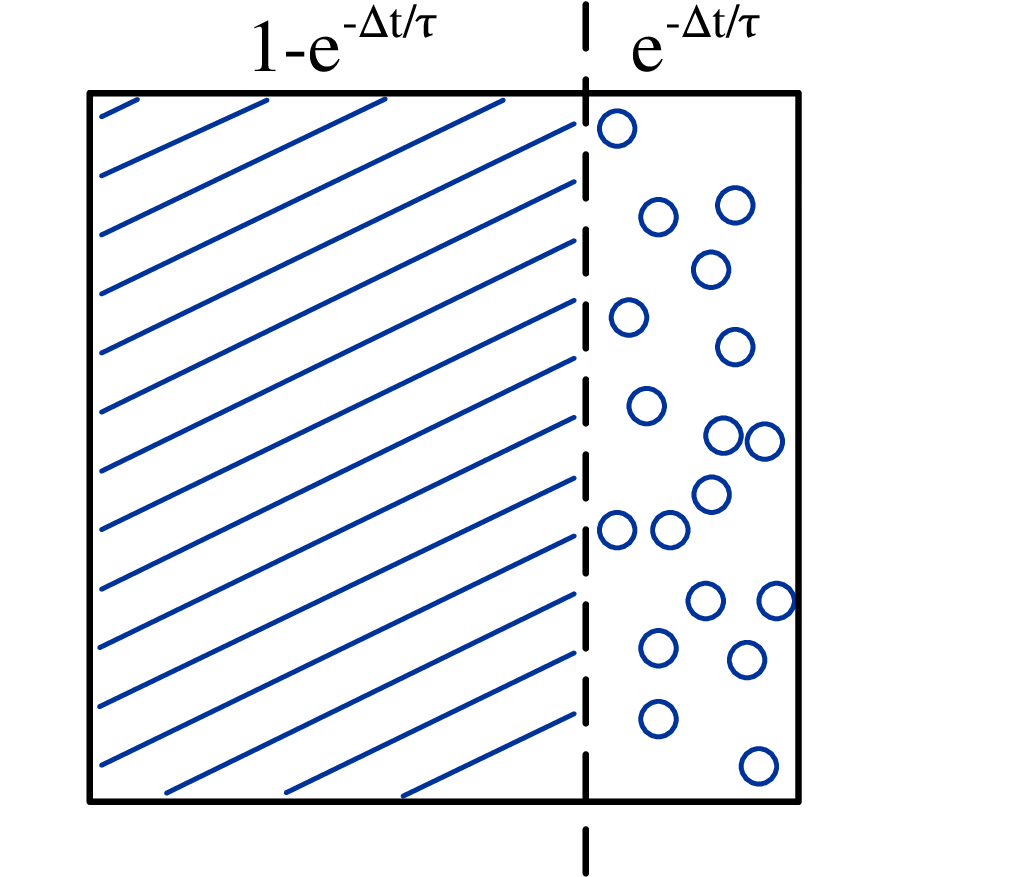}
		}
    \subfigure[]{\label{fig1b}
    		\includegraphics[width=0.22 \textwidth]{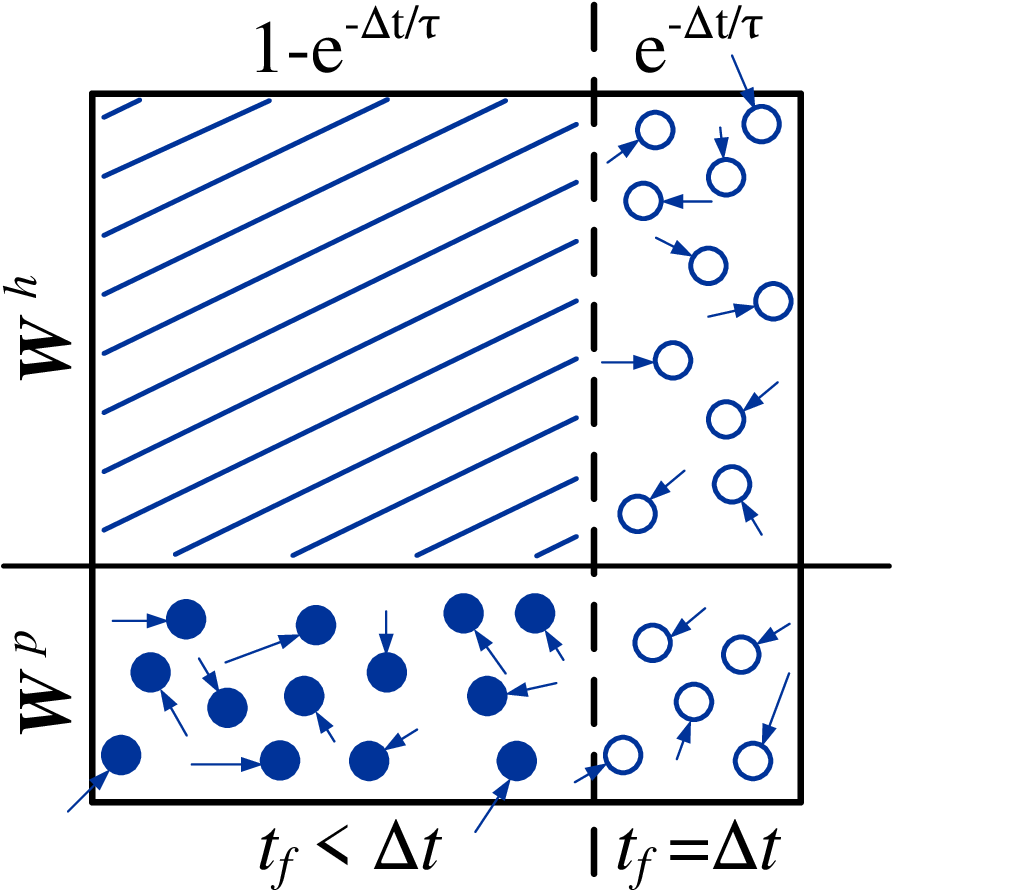}
    	}
    \subfigure[]{\label{fig1c}
    		\includegraphics[width=0.22 \textwidth]{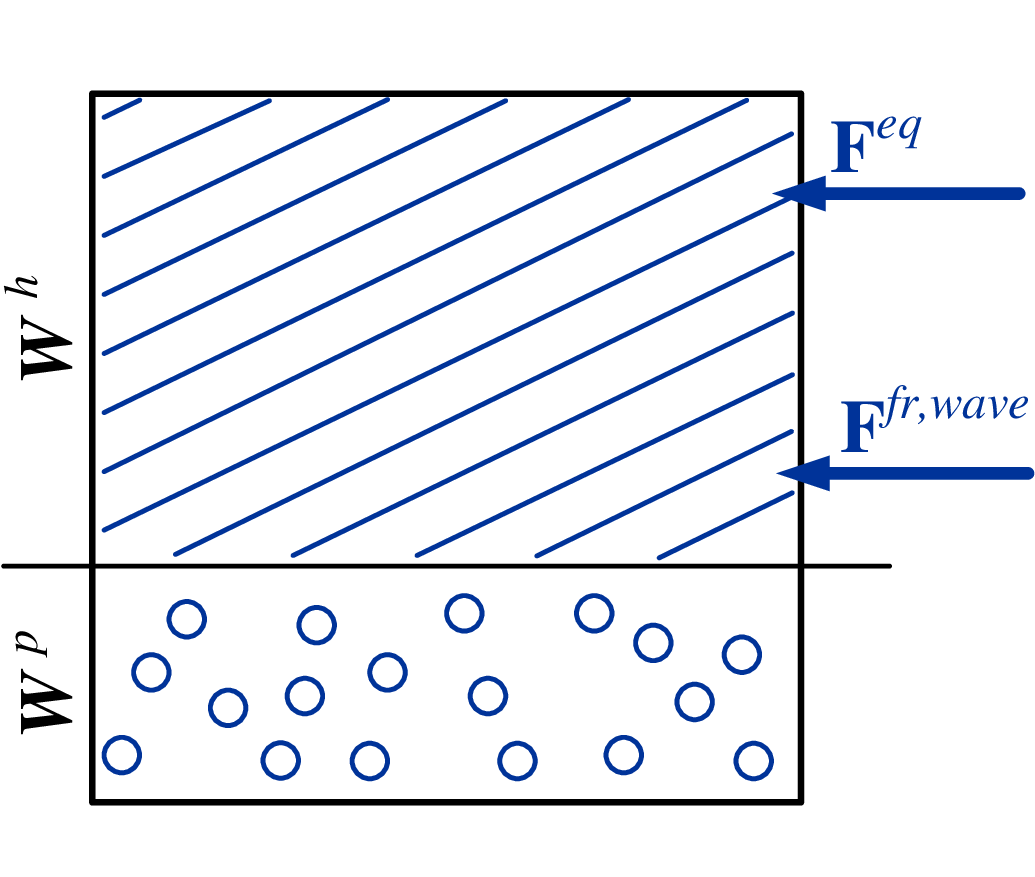}
    	}
    \subfigure[]{\label{fig1d}
    		\includegraphics[width=0.22 \textwidth]{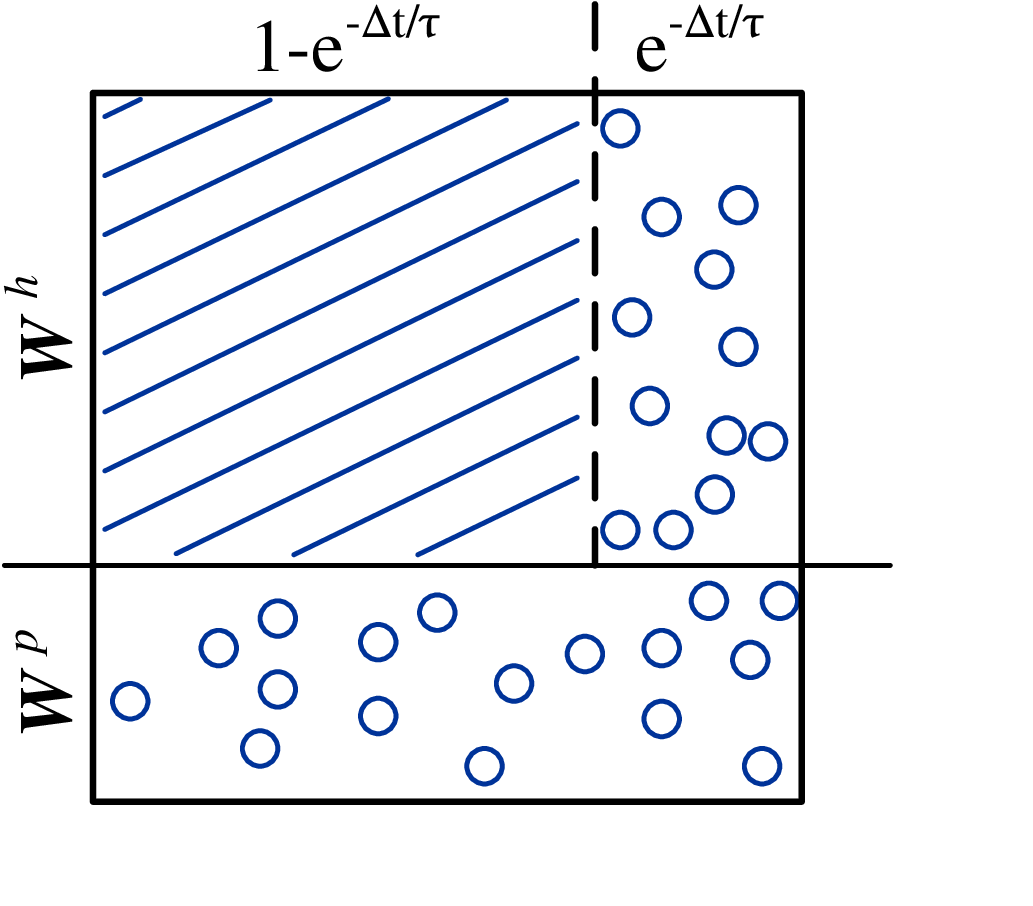}
    	}
	\caption{\label{fig1} Diagram illustrating the algorithm of the UGKWP method: (a) Initial state at $n=0$, (b) Step 2, (c) Step 3, and (d) initial state at subsequent steps.}
\end{figure}

In this paper, the primary methodological improvements include second-order particle sampling (Eq.~\eqref{eq:sample2nd}), the modified limiter (Eq.~\eqref{eq:venkata3}), and the implementation of the first-order Chapman--Enskog expansion term (Eq.~\eqref{eq:ffrwave}). Details regarding particle conservation (Eqs.~\eqref{eq:pForceConservation1} and \eqref{eq:pForceConservation2}) and gradient calculations (Eqs.~\eqref{eq:gradient} and \eqref{eq:wls}) are also provided. Figure~\ref{fig1} summarizes the UGKWP algorithm:
\begin{description}
    \item[Step (1)] Initial state. As a result of the previous time step $n-1$ (Fig.~\ref{fig1d}), numerical particles $\boldsymbol{W}^p$ coexist with the hydrodynamic wave $\boldsymbol{W}^h$. Gradients of macroscopic variables are calculated using Eqs.~\eqref{eq:gradient}, \eqref{eq:wls}, and \eqref{eq:venkata}. Collisionless particles $\boldsymbol{W}^{hp}$ are then sampled with second-order spatial accuracy using Eq.~\eqref{eq:sample2nd}. For the first step, $\boldsymbol{W}^p=\boldsymbol{0}$ (Fig.~\ref{fig1a}).
    \item[Step (2)] Free transport. First, particles $\boldsymbol{W}^p$ are divided into two parts according to Eq.~\eqref{eq:mic1}. In Fig.~\ref{fig1b}, collisionless particles are denoted by hollow circles and collisional particles by solid circles. All particles are transported using Eq.~\eqref{eq:mic2}, and their contributions to the macroscopic conserved variables are summed via Eq.~\eqref{eq:mic3}. Meanwhile, the free transport flux $\boldsymbol{F}_j^{fr,wave}$ contributed by the collisional particles of $\left(\boldsymbol{W}^h-\boldsymbol{W}^{hp}\right)$ is calculated using Eq.~\eqref{eq:ffrwave}. Finally, the collisional particles (solid circles) are deleted.
    \item[Step (3)] Collision. Compute the $\boldsymbol{F}^{eq}_j$ term in Eq.~\eqref{eq:feq} using the formulations in Sec.~\ref{sec:macro}. Then, update $\boldsymbol{W}^{n+1}$ using Eq.~\eqref{eq:renew}.
    \item[Step (4)] If the simulation continues, return to Step (1), where collisionless particles $\boldsymbol{W}^{hp}$ will be sampled.
\end{description}

\section{Test cases}\label{sec:cases}
Hypersonic flow around a cylinder is considered as a typical two-dimensional case, and hypersonic flow around a cone is conducted as a three-dimensional case. For the cylinder flow, the ${\rm{Ma}}_{\infty}=5$, ${\rm{Kn}}_{\infty}=0.01$ condition is selected as a classical multiscale case, and the mesh-independent performance of UGKWP, UGKS, and DSMC is examined. Modifications to the UGKWP method are validated, including second-order particle sampling (Eq.~\eqref{eq:sample2nd}) and the implementation of the first-order Chapman--Enskog expansion term (Eq.~\eqref{eq:ffrwave}). The effect of the reference particle number $N^{hp,{\rm{ref}}}$ is also tested. A more rarefied case at ${\rm{Kn}}_{\infty}=0.1$ is considered as well. For the cone flow, experimental conditions from the literature~\cite{cone-exp,cone-dsmc} are used for validation, and a condition at a $70{\rm{km}}$ altitude is used for mesh independence comparisons between the UGKWP and DSMC methods. Full accommodation is applied for the isothermal walls as the boundary condition. The CFL number is set to $N_{{\rm{CFL}}}=1.0$ in all cases, defined as follows~\cite{jiri}:
\begin{equation}\label{eq:cfl}
\Delta t=N_{{\rm{CFL}}}\min\limits_{i}\frac{\Omega_i}{\frac{1}{2}\sum\limits_{\ell=1,2,3}\left[{\left(\left|U_{i,\ell}\right|+3\sqrt{RT_i}\right)\sum\limits_{j\in\mathcal{M}\left(i\right)}s_j\left|n_{j,\ell}\right|}\right]},
\end{equation}
where $\ell$ is the dimension index and $\mathcal{M}\left(i\right)$ is the collection of interfaces surrounding cell ``$i$''.

\subsection{Hypersonic flow around a cylinder at ${\rm{Kn}}_{\infty}=0.01$ and ${\rm{Kn}}_{\infty}=0.1$}\label{sec:cylinder}
\begin{figure}[H]
	\centering
	\subfigure[]{
			\includegraphics[width=0.45 \textwidth]{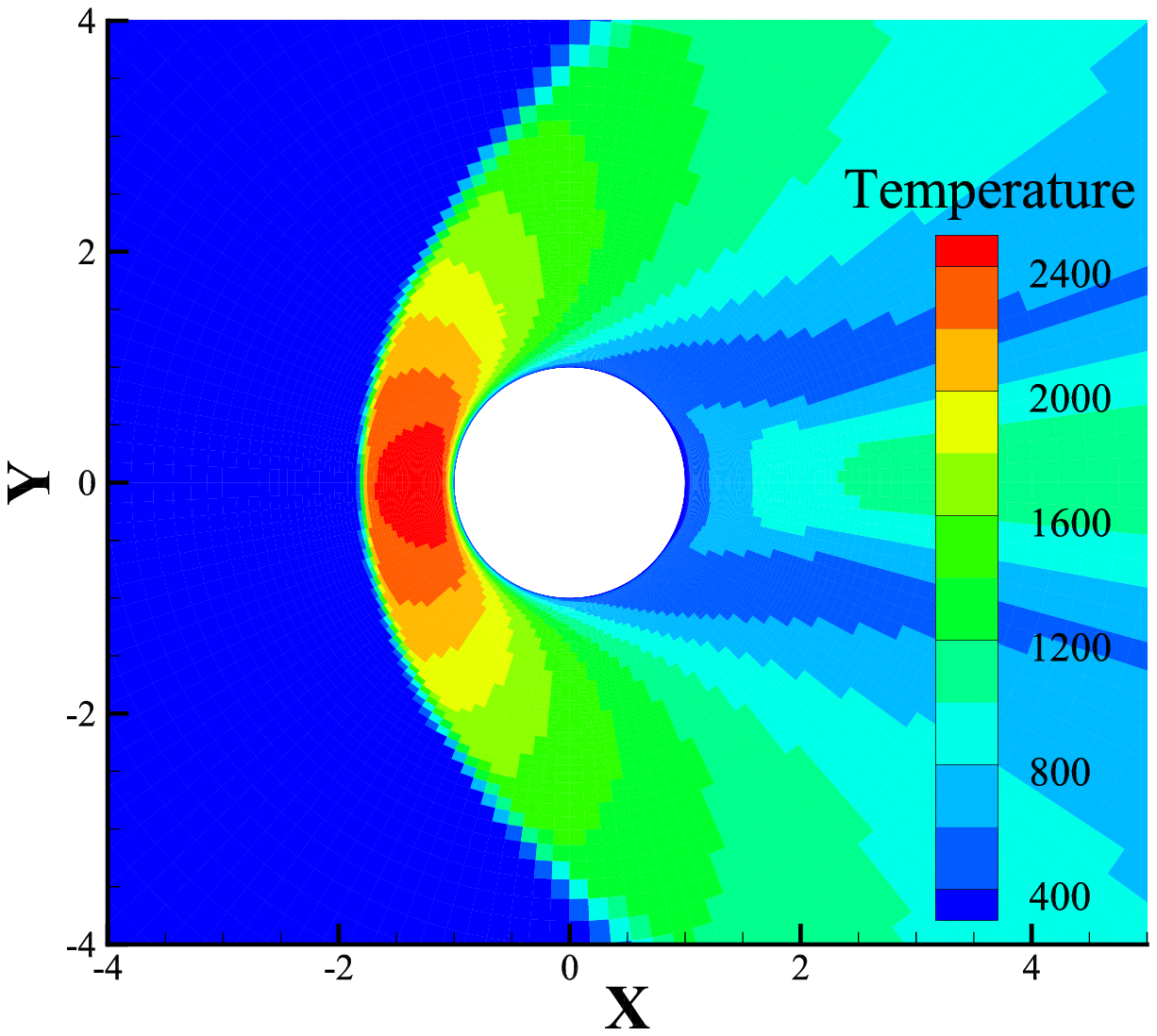}
		}
    \subfigure[]{
    		\includegraphics[width=0.45 \textwidth]{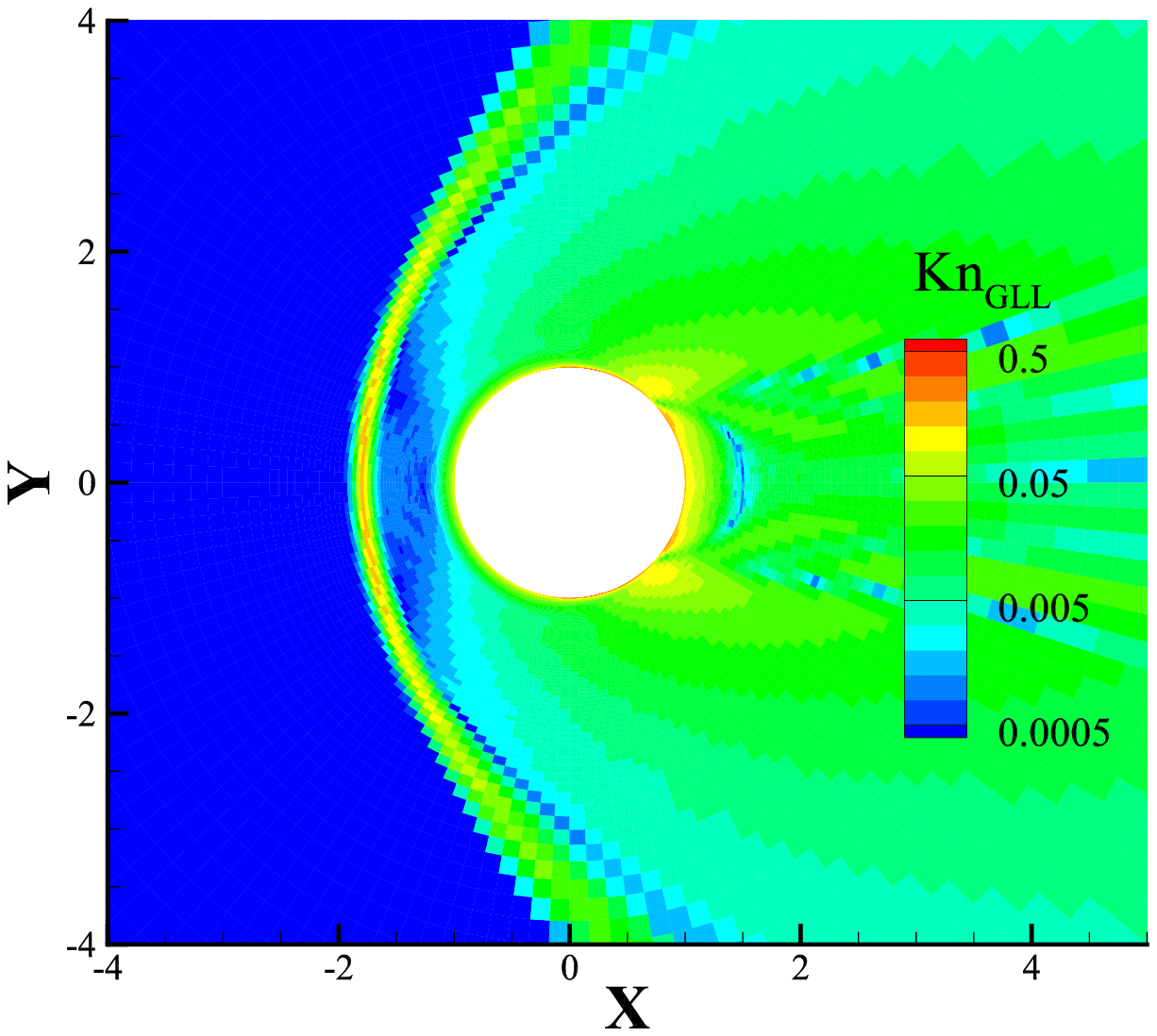}
    	}
	\caption{\label{cylinder-kn0.01-contour} Contours of hypersonic cylinder flow at ${\rm{Ma}}_{\infty}=5$, ${\rm{Kn}}_{\infty}=0.01$, ${\rm{Pr}}=1$: (a) Temperature, (b) $\rm{Kn_{GLL}}$.}
\end{figure}

Hypersonic flow around a cylinder at ${\rm{Ma}}_{\infty}=5$, ${\rm{Kn}}_{\infty}=0.01$ represents a typical multiscale case spanning the continuum to rarefied regimes. Here, the cylinder radius is set as the reference length. As shown in Fig.~\ref{cylinder-kn0.01-contour}, the bow shock strongly compresses the flow near the stagnation region, driving the local flow toward the continuum regime, while the flow in the wake region is more rarefied due to expansion. These features are representative of hypersonic flows around near-space vehicles. Consequently, a mesh independence study on this case can directly evaluate the advantages of the numerical method and provide practical guidance for engineering applications. In this case, the inflow temperature $T$ is set equal to the wall temperature, $N^{hp,{\rm{ref}}}$ is set to $200$, $\omega$ is set to $0.81$, and the gas is monatomic. Except for the comparison with the DSMC method, ${\rm{Pr}}=1$ is used. Regarding the multiscale phenomenon, the contour of the local Knudsen number $\rm{Kn_{GLL}}$~\cite{kngll1,kngll2,kngll3} is used to characterize the local scale variation, defined as:
\begin{equation}\label{eq:kngll}
{\rm{Kn_{GLL}}}=\frac{\lambda}{\rho/\Delta\rho}.
\end{equation}
The value of ${\rm{Kn_{GLL}}}$ increases significantly from the post-shock region to the leeward region by over three orders of magnitude, and there is a large region on the leeward side where ${\rm{Kn_{GLL}}}>0.05$, which is widely used as a criterion for the breakdown of the linear constitutive relations of the Navier-Stokes equations.

\begin{figure}[H]
	\centering
	\subfigure[]{
			\includegraphics[width=0.45 \textwidth]{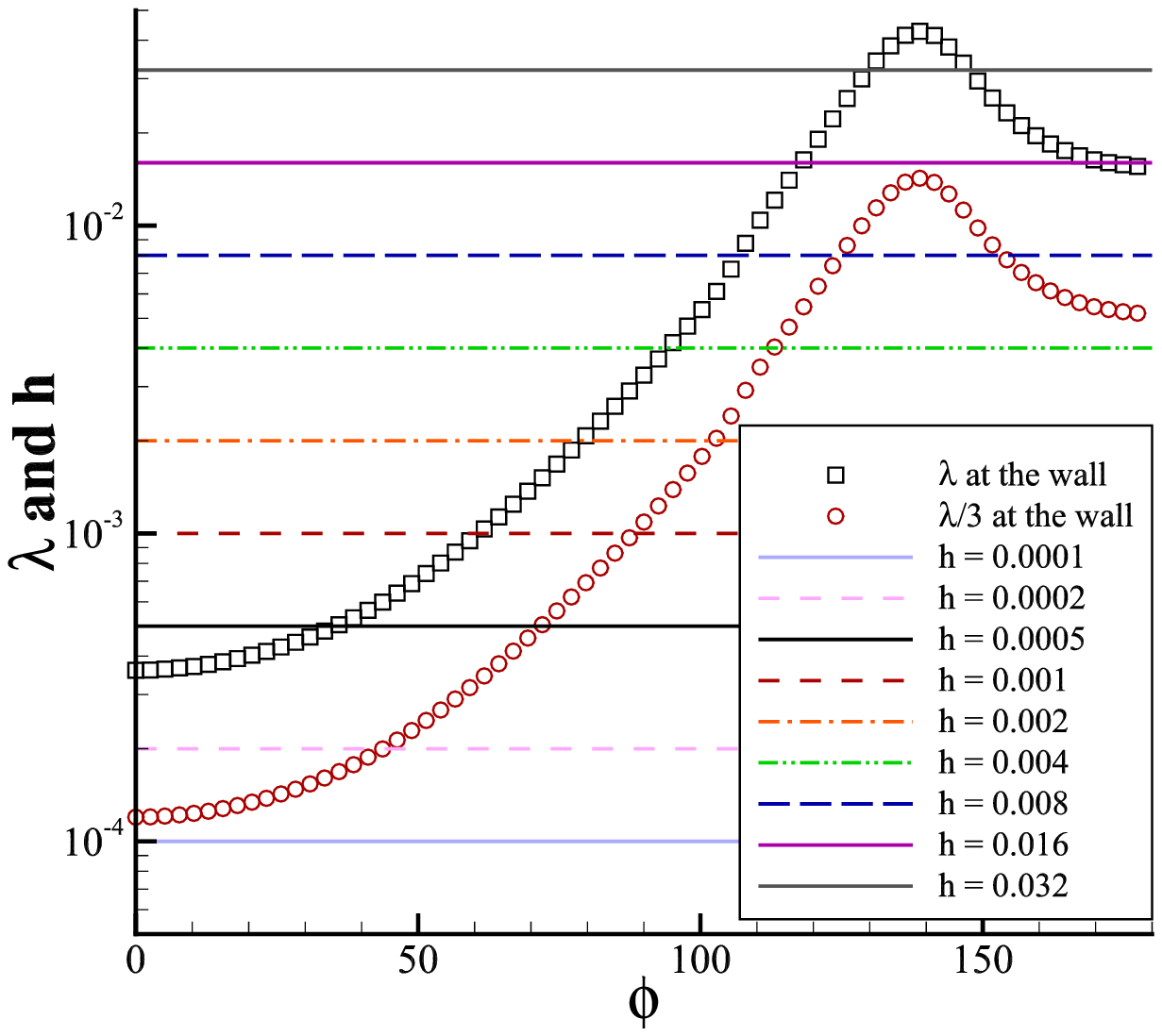}
		}
    \subfigure[]{
    		\includegraphics[width=0.45 \textwidth]{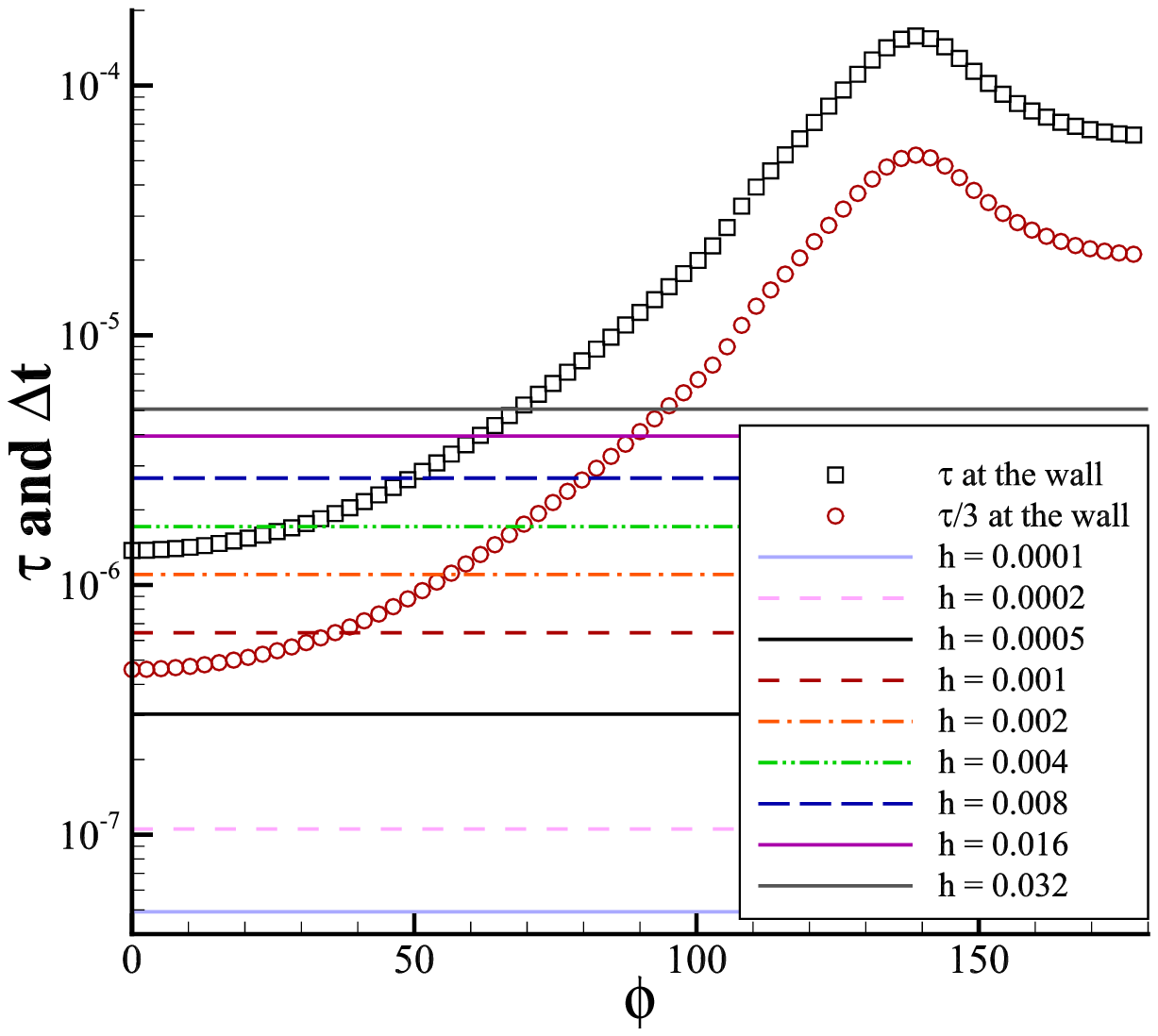}
    	}
	\caption{\label{cylinder-kn0.01-lambdatau} Scale-related variables at the wall of hypersonic cylinder flow at ${\rm{Ma}}_{\infty}=5$, ${\rm{Kn}}_{\infty}=0.01$, ${\rm{Pr}}=1$: (a) $\lambda$ and $h$, (b) $\tau$ and $\Delta t$.}
\end{figure}

For the mesh independence study, a series of meshes with increasing first-layer heights $h$ are implemented. The variation of $h$ compared with the local $\lambda$ and $\lambda/3$ along the wall is shown in Fig.~\ref{cylinder-kn0.01-lambdatau}. The value of $\lambda$ varies across over two orders of magnitude from the stagnation point to around $\theta=140^{\circ}$. The smallest $h$ value is lower than $\lambda/3$ at the stagnation point, and the largest is close to the maximum of $\lambda$. The relation between $\Delta t$ and $\tau$ is also shown because $\Delta t/\tau$ is always less than $h/\lambda$. Additionally, $140$ cells are set around the circumference.

\begin{figure}[H]
	\centering
	\subfigure[]{
			\includegraphics[width=0.45 \textwidth]{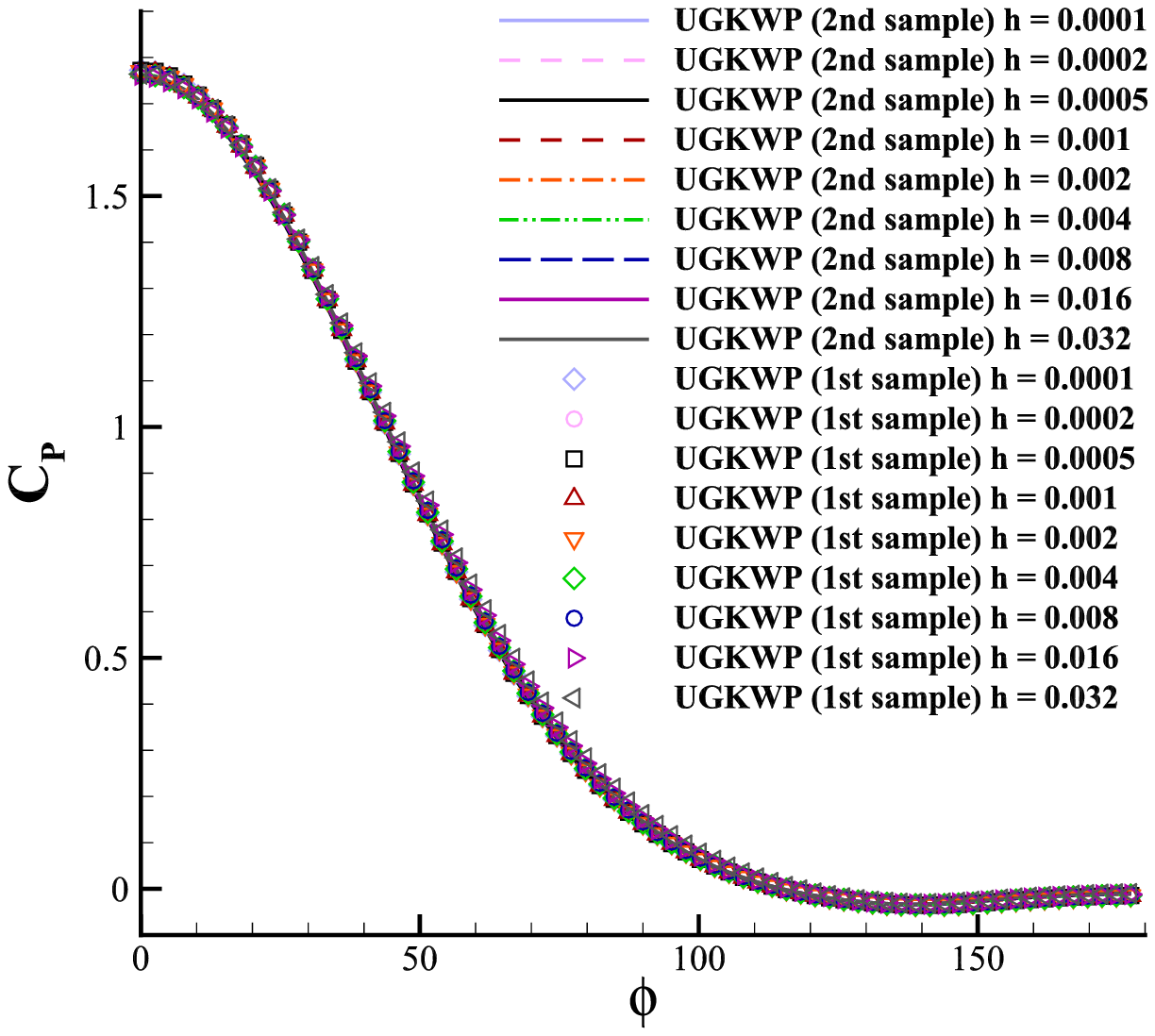}
		}
    \subfigure[]{
    		\includegraphics[width=0.45 \textwidth]{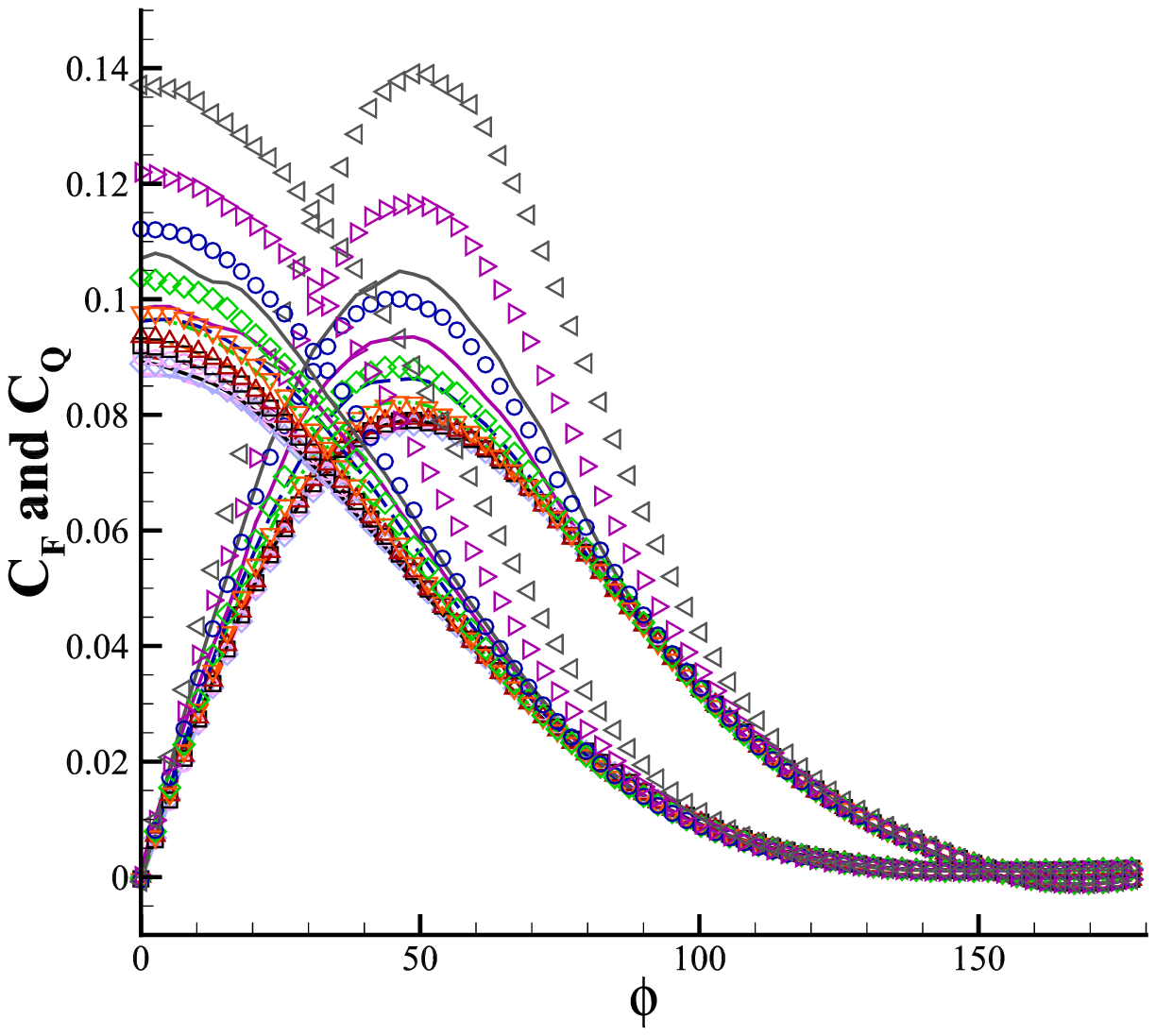}
    	}
    \subfigure[]{
			\includegraphics[width=0.45 \textwidth]{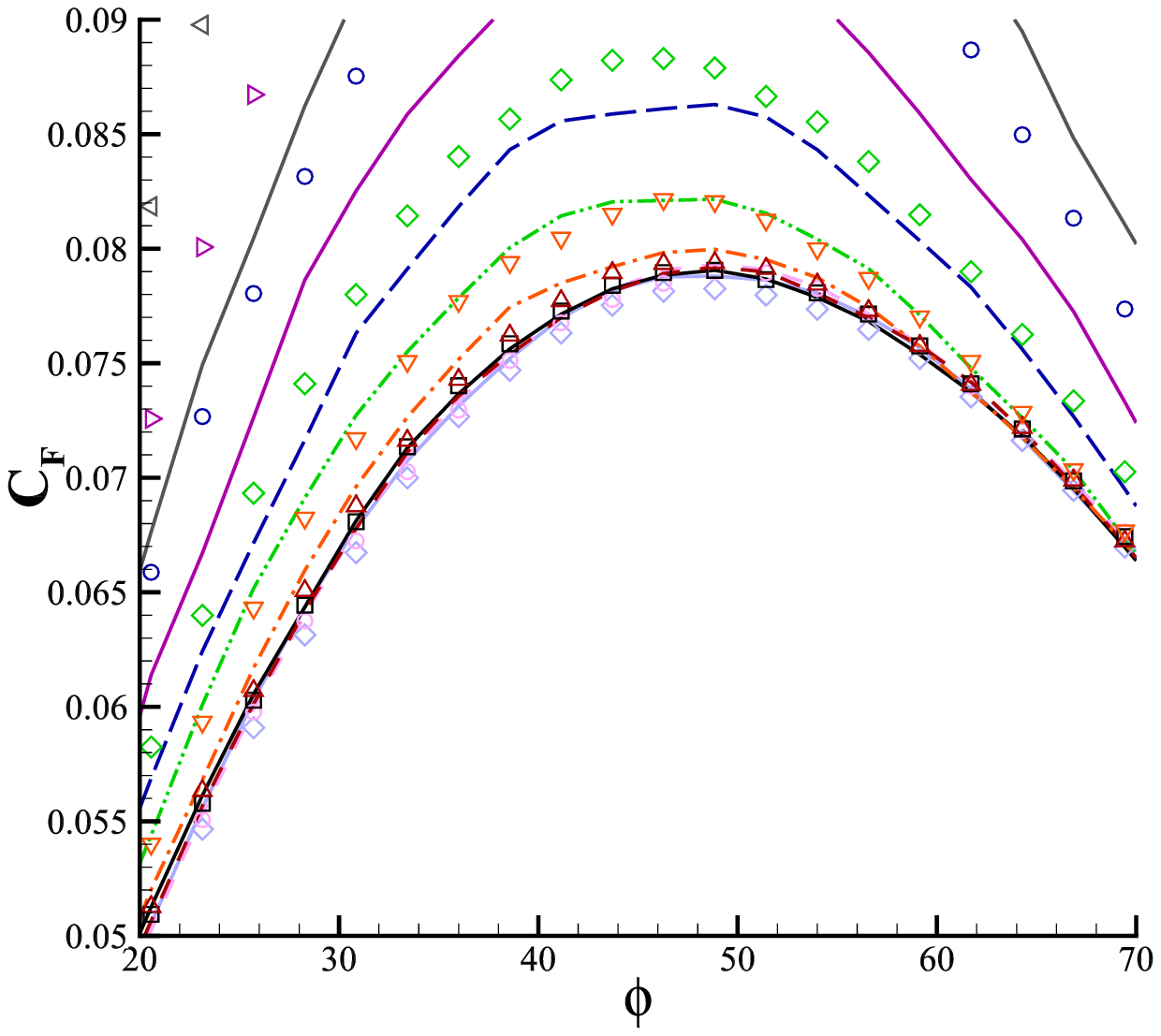}
		}
    \subfigure[]{
    		\includegraphics[width=0.45 \textwidth]{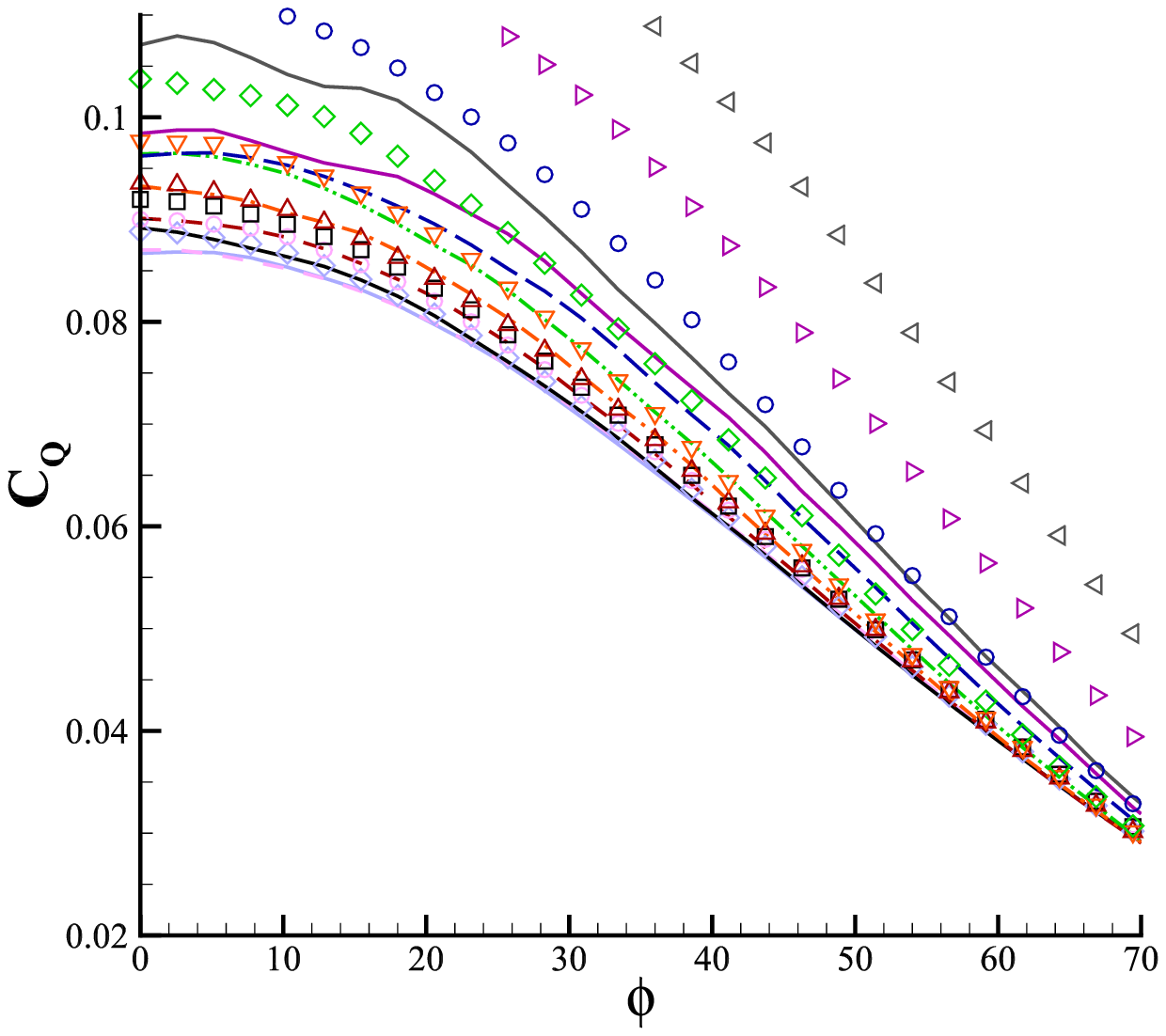}
    	}
	\caption{\label{cylinder-kn0.01_12} Comparisons of second-order particle sampling UGKWP with first-order particle sampling UGKWP on hypersonic cylinder flow at ${\rm{Ma}}_{\infty}=5$, ${\rm{Kn}}_{\infty}=0.01$, ${\rm{Pr}}=1$: (a) Pressure coefficient at the wall $C_P$, (b) shear stress coefficient $C_F$ and heat flux coefficient $C_Q$ at the wall, (c) enlarged view of $C_F$, (d) enlarged view of $C_Q$.}
\end{figure}

\begin{figure}[H]
	\centering
	\subfigure[]{
			\includegraphics[width=0.45 \textwidth]{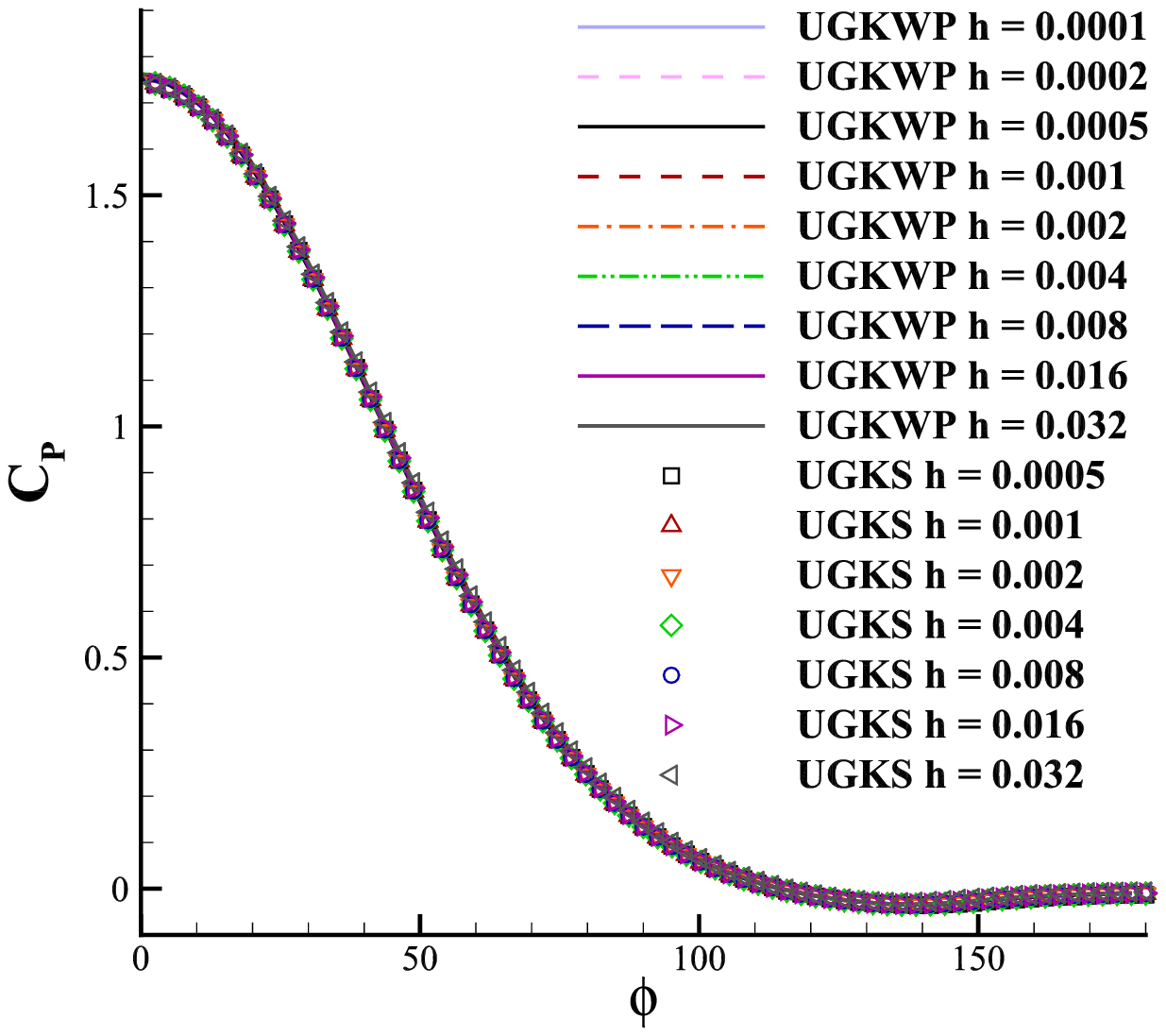}
		}
    \subfigure[]{
    		\includegraphics[width=0.45 \textwidth]{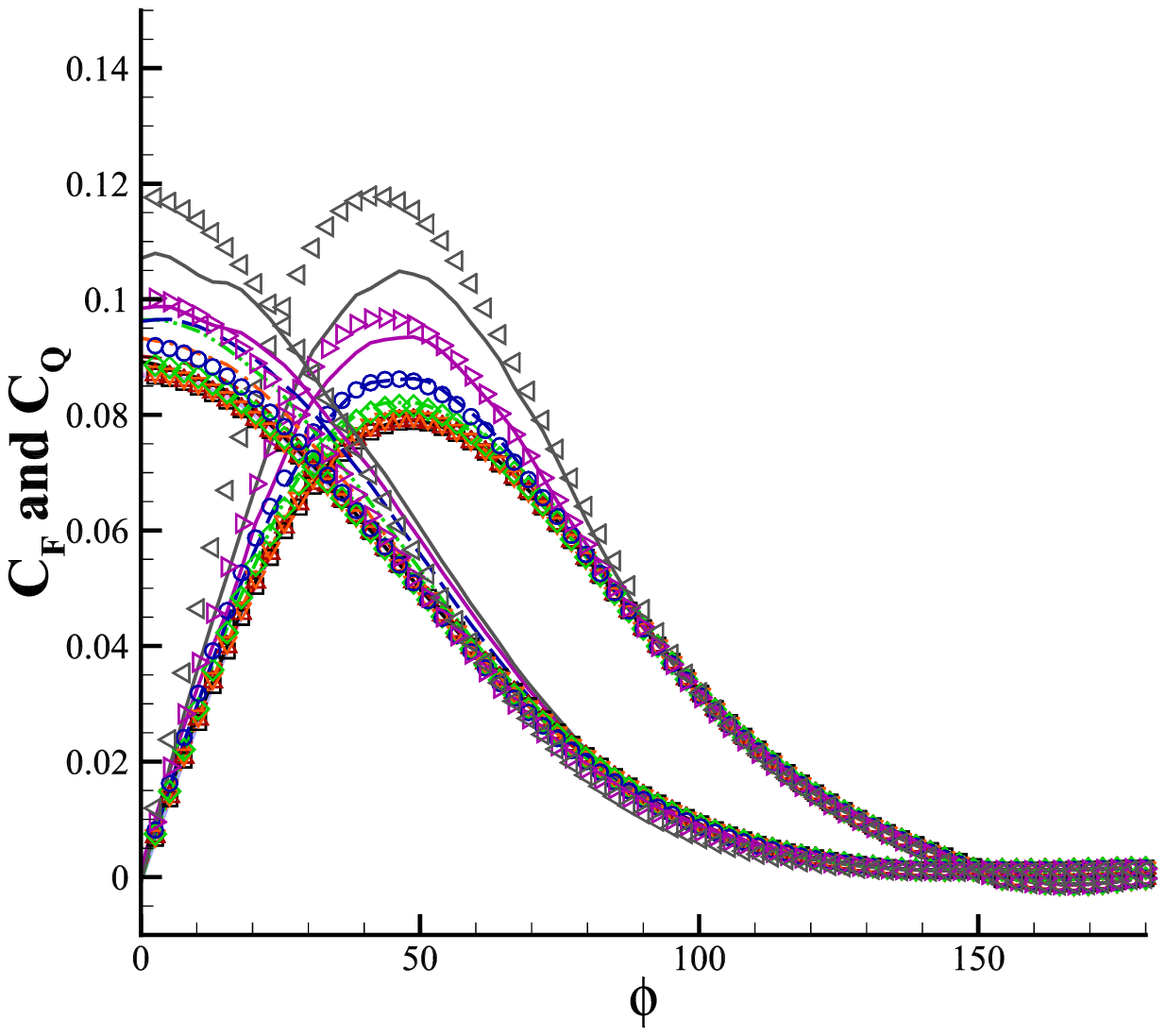}
    	}
    \subfigure[]{
			\includegraphics[width=0.45 \textwidth]{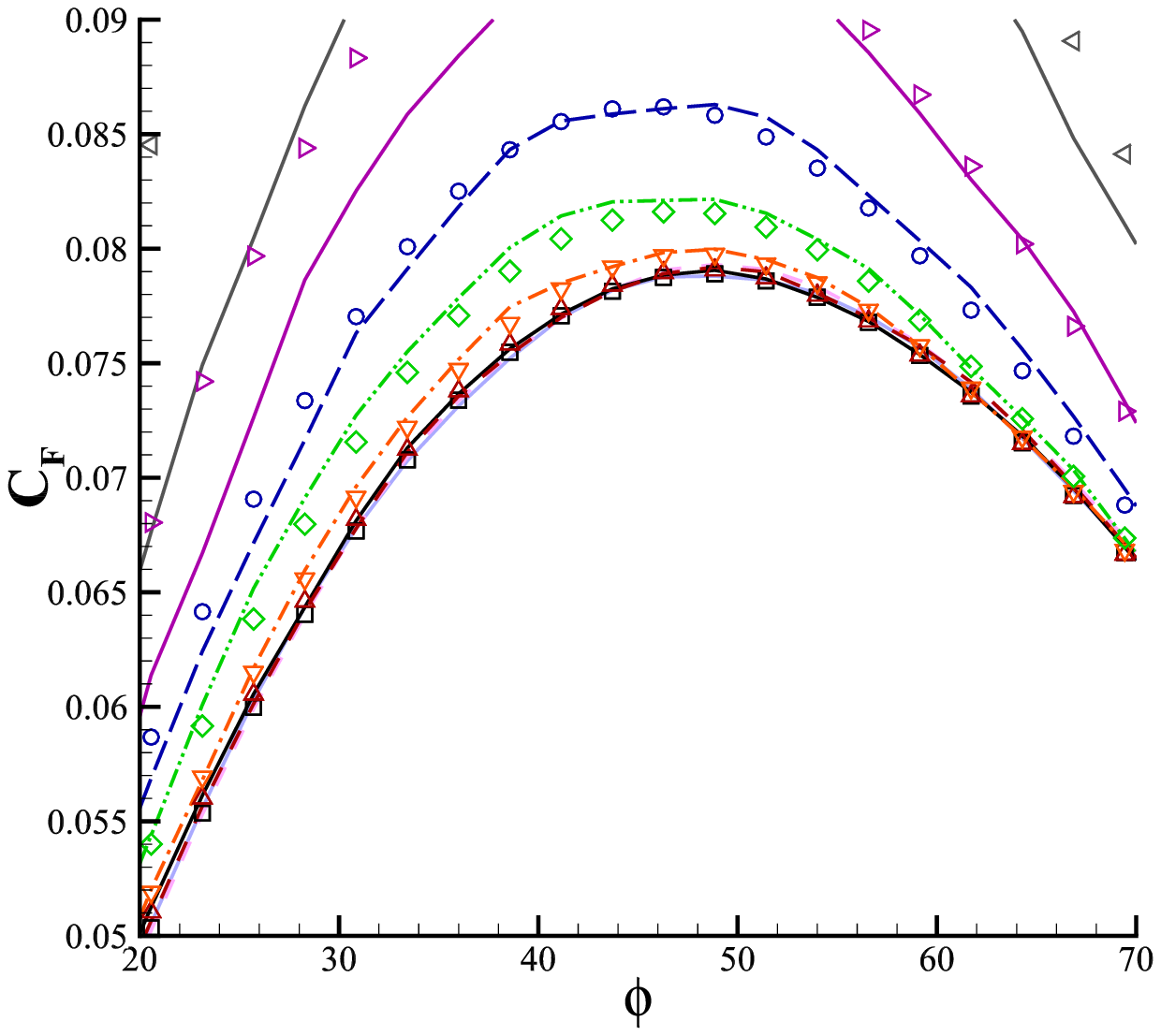}
		}
    \subfigure[]{
    		\includegraphics[width=0.45 \textwidth]{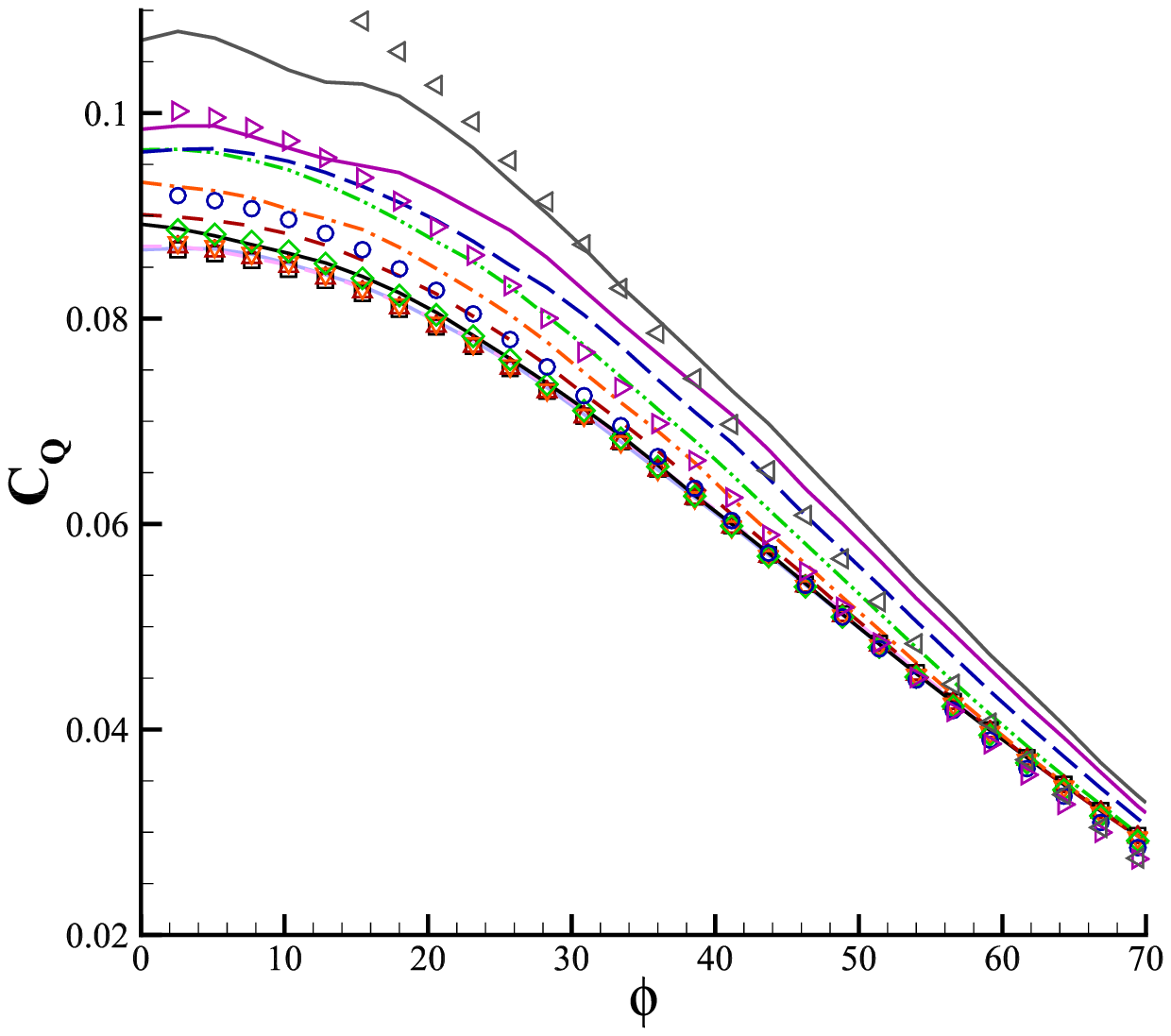}
    	}
	\caption{\label{cylinder-kn0.01_ugks} Comparisons of UGKWP with UGKS on hypersonic cylinder flow at ${\rm{Ma}}_{\infty}=5$, ${\rm{Kn}}_{\infty}=0.01$, ${\rm{Pr}}=1$: (a) Pressure coefficient at the wall $C_P$, (b) shear stress coefficient $C_F$ and heat flux coefficient $C_Q$ at the wall, (c) enlarged view of $C_F$, (d) enlarged view of $C_Q$.}
\end{figure}

\begin{figure}[H]
	\centering
	\subfigure[]{
			\includegraphics[width=0.45 \textwidth]{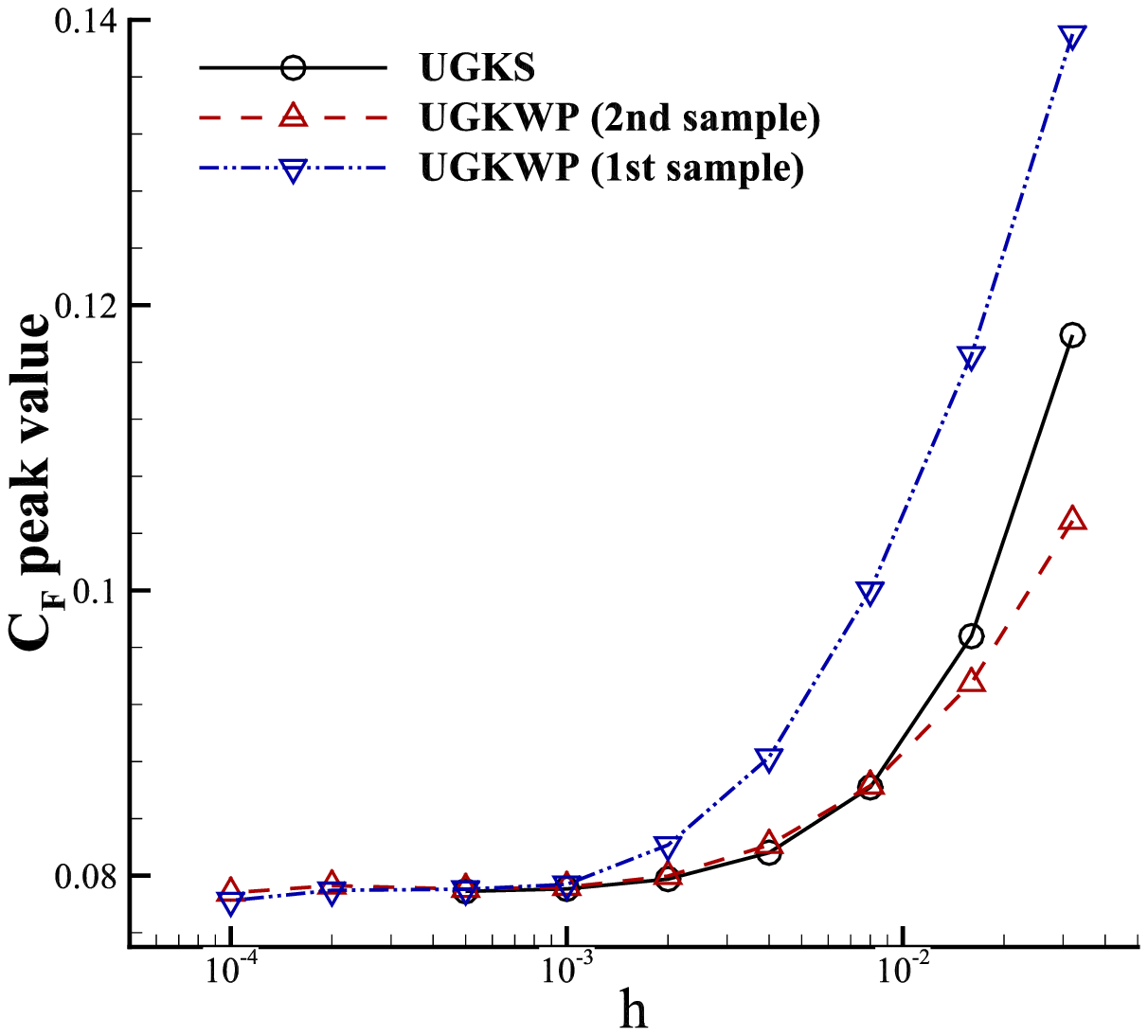}
		}
    \subfigure[]{
    		\includegraphics[width=0.45 \textwidth]{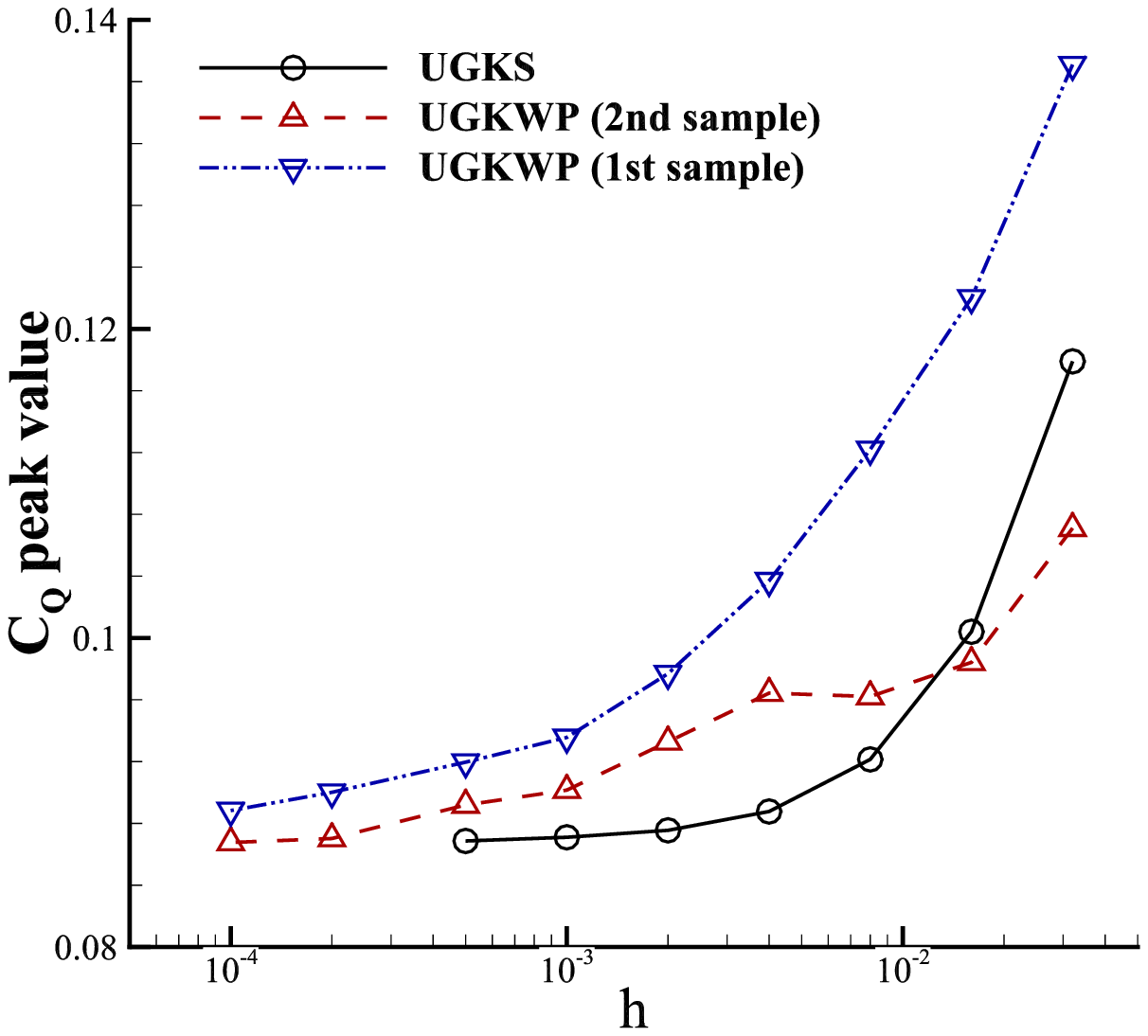}
    	}
	\caption{\label{cylinder-kn0.01line_12} Comparisons among second-order particle sampling UGKWP, first-order particle sampling UGKWP, and UGKS on hypersonic cylinder flow at ${\rm{Ma}}_{\infty}=5$, ${\rm{Kn}}_{\infty}=0.01$, ${\rm{Pr}}=1$: (a) Peak value of shear stress coefficient $C_F$, (b) peak value of heat flux coefficient $C_Q$.}
\end{figure}

The improvement of second-order particle sampling is shown in Figs.~\ref{cylinder-kn0.01_12}, \ref{cylinder-kn0.01_ugks}, and \ref{cylinder-kn0.01line_12}. While $C_P$ exhibits weak mesh sensitivity, $C_F$ and $C_Q$ are highly sensitive to mesh resolution. In Fig.~\ref{cylinder-kn0.01_12}, the second-order sampling (Eq.~\eqref{eq:sample2nd}) in the UGKWP method is compared with the first-order sampling (Eq.~\eqref{eq:sample1st}). For $C_F$, to achieve similar accuracy, the $h_{\rm{2nd}}$ of second-order sampling can be $2$ to $4$ times larger than the $h_{{\rm{1st}}}$ of first-order sampling. For $C_Q$, comparable accuracy is achieved with $h_{\rm{2nd}}$ up to $4$ to $8$ times larger than $h_{\rm{1st}}$. This allows substantial computational savings by using coarser near-wall meshes. In Fig.~\ref{cylinder-kn0.01_ugks}, the UGKWP method with second-order particle sampling is compared with the UGKS. The results for $C_F$ are analogous between the two multiscale methods, while $C_Q$ is more sensitive, showing better performance for the UGKS. This is because the gradients in the UGKS are more reliable without particle noise. Figure~\ref{cylinder-kn0.01line_12} provides a clearer comparison among the three approaches by focusing on the peak values. The improvement in particle sampling brings the UGKWP profile much closer to the UGKS.

\begin{figure}[H]
	\centering
	\subfigure[]{
			\includegraphics[width=0.45 \textwidth]{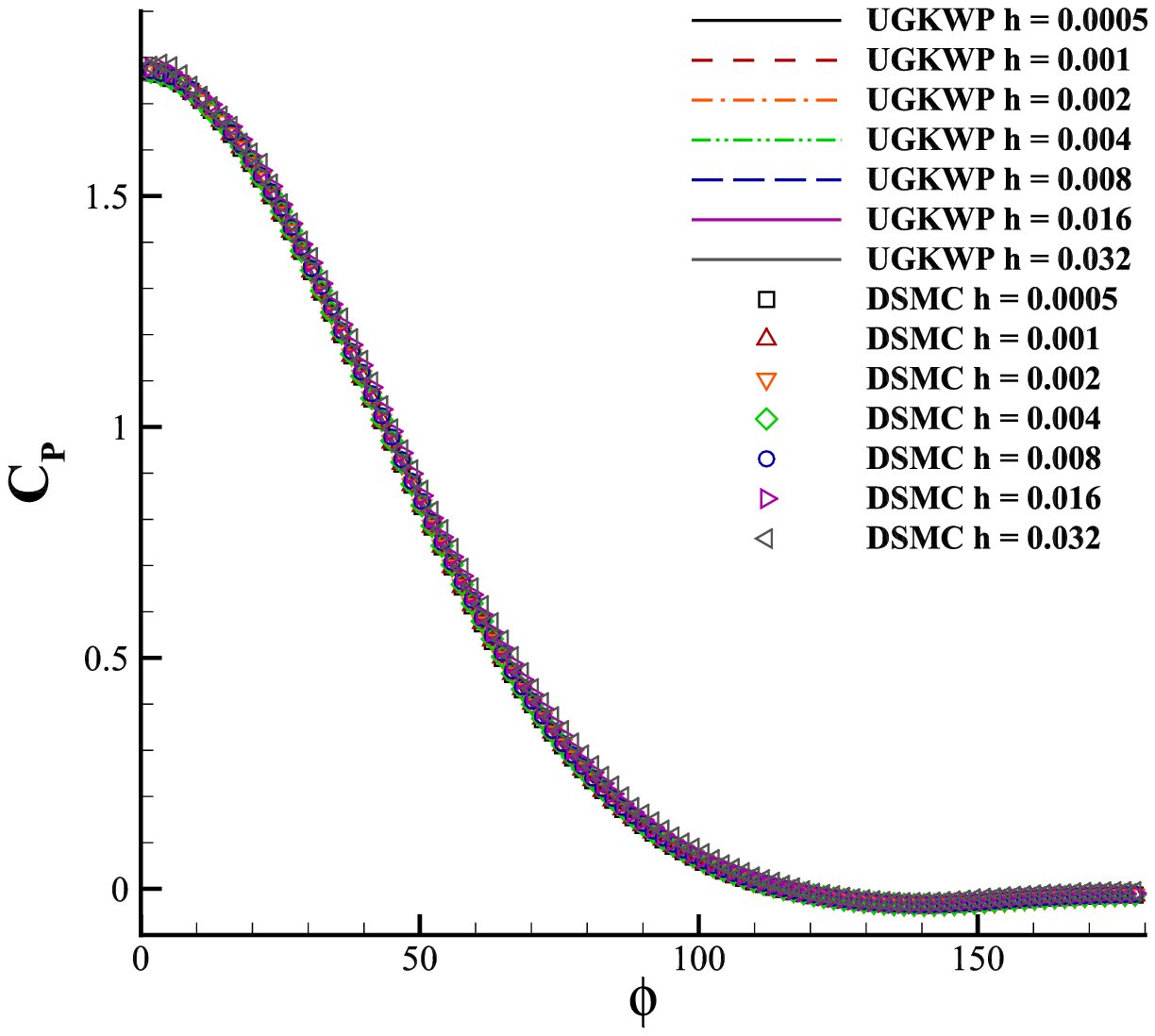}
		}
    \subfigure[]{
    		\includegraphics[width=0.45 \textwidth]{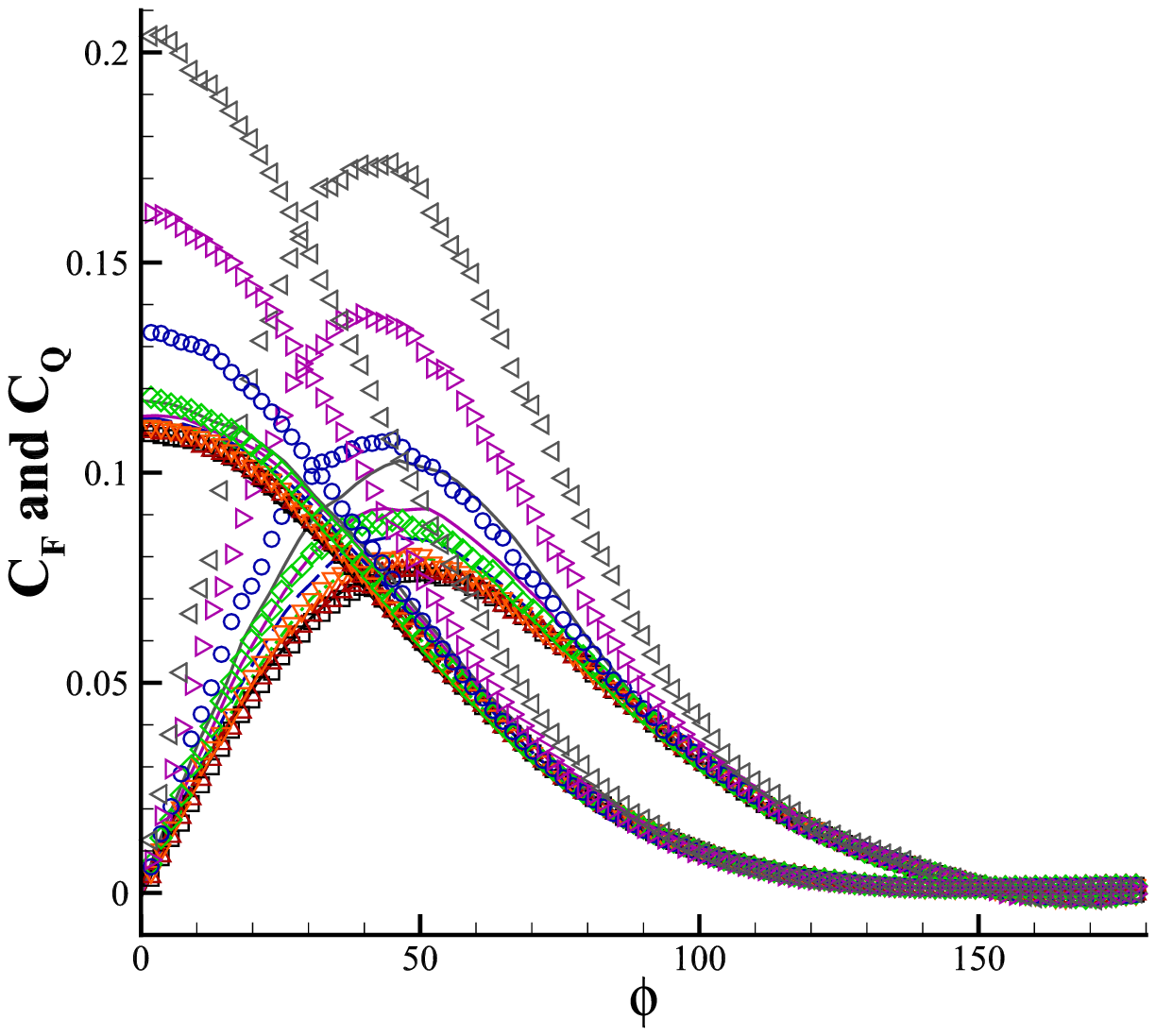}
    	}
    \subfigure[]{
			\includegraphics[width=0.45 \textwidth]{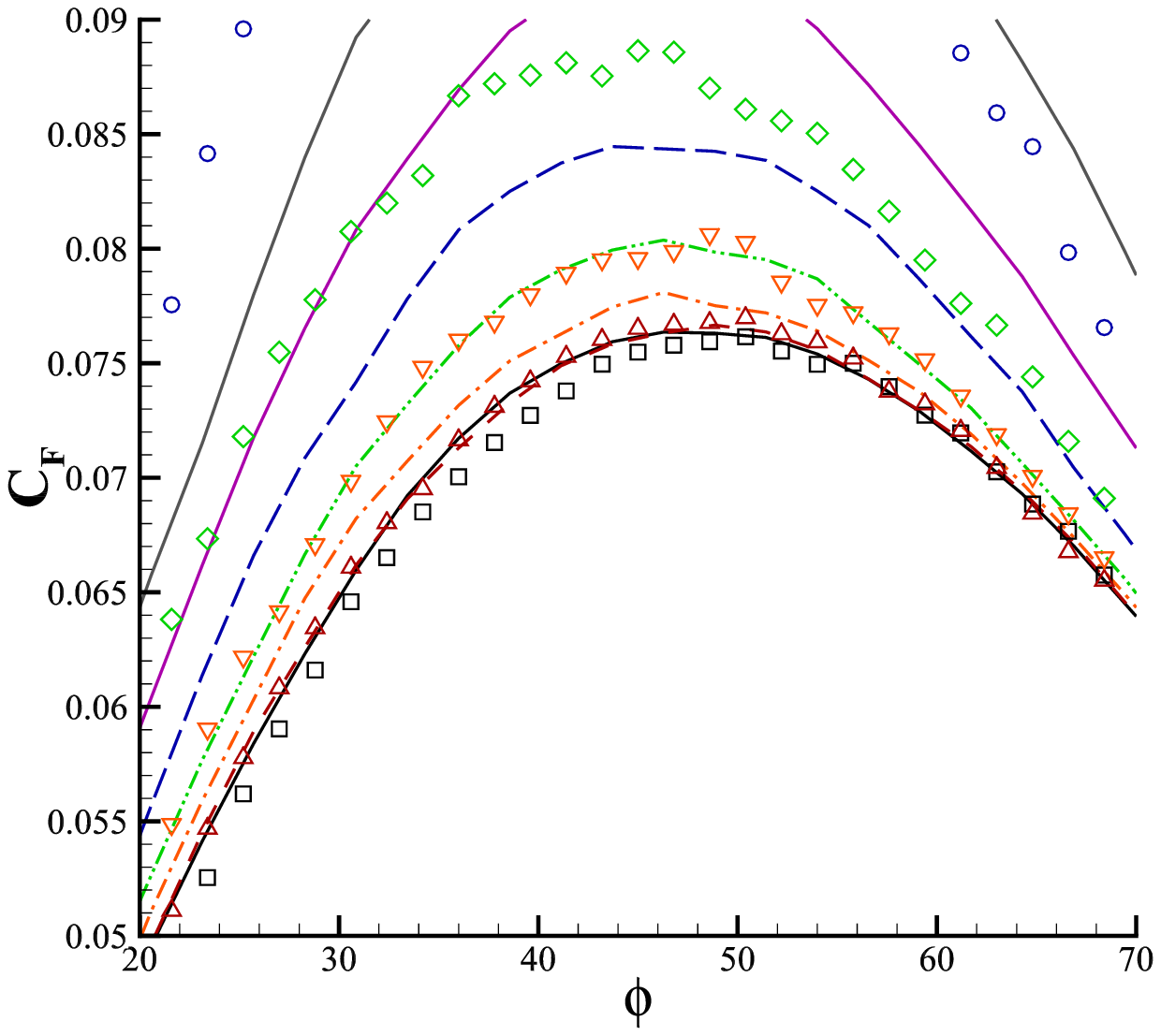}
		}
    \subfigure[]{
    		\includegraphics[width=0.45 \textwidth]{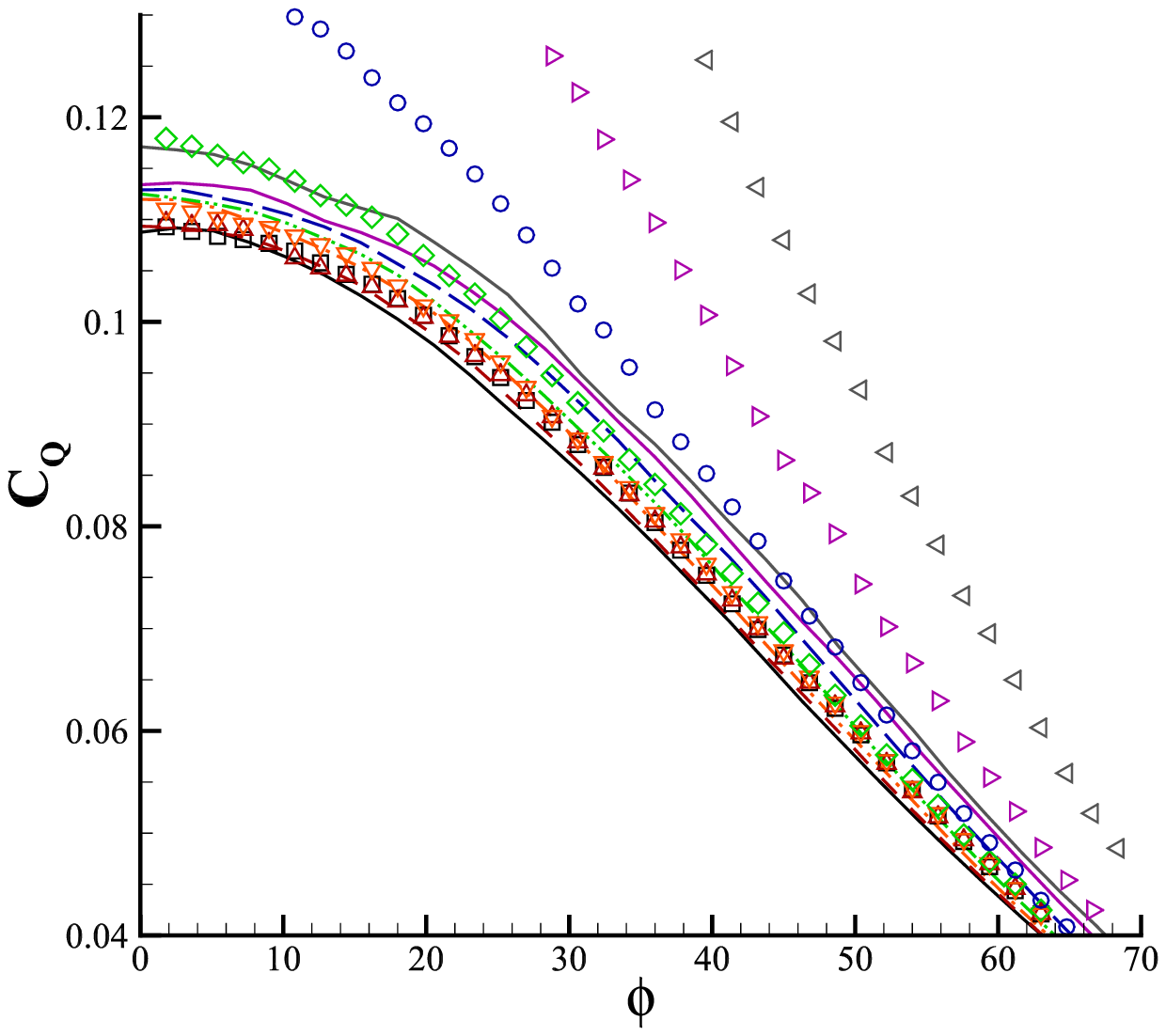}
    	}
	\caption{\label{cylinder-kn0.01_dsmc} Comparisons of UGKWP with DSMC on hypersonic cylinder flow at ${\rm{Ma}}_{\infty}=5$, ${\rm{Kn}}_{\infty}=0.01$, ${\rm{Pr}}=2/3$: (a) Pressure coefficient at the wall $C_P$, (b) shear stress coefficient $C_F$ and heat flux coefficient $C_Q$ at the wall, (c) enlarged view of $C_F$, (d) enlarged view of $C_Q$.}
\end{figure}

\begin{figure}[H]
	\centering
	\subfigure[]{
			\includegraphics[width=0.45 \textwidth]{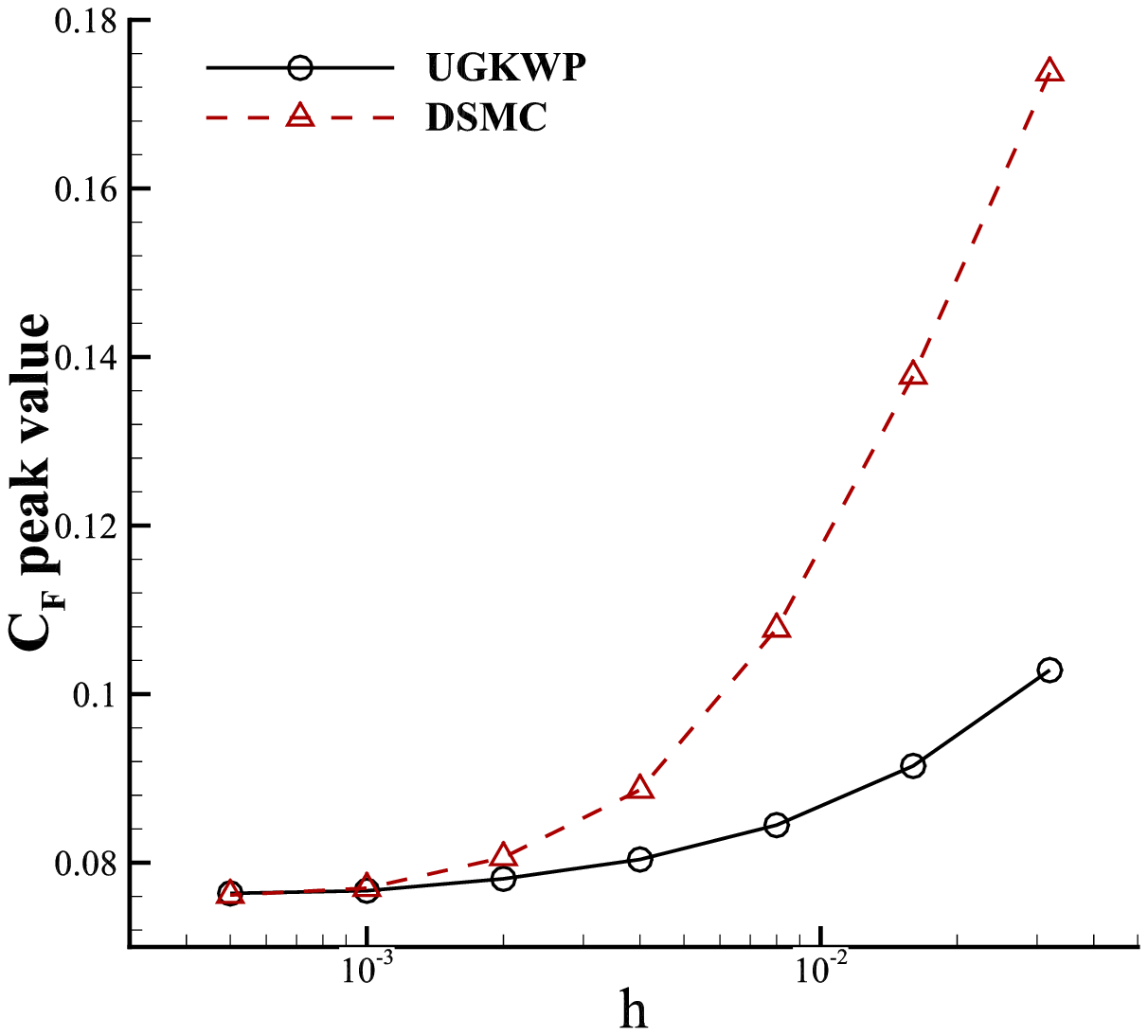}
		}
    \subfigure[]{
    		\includegraphics[width=0.45 \textwidth]{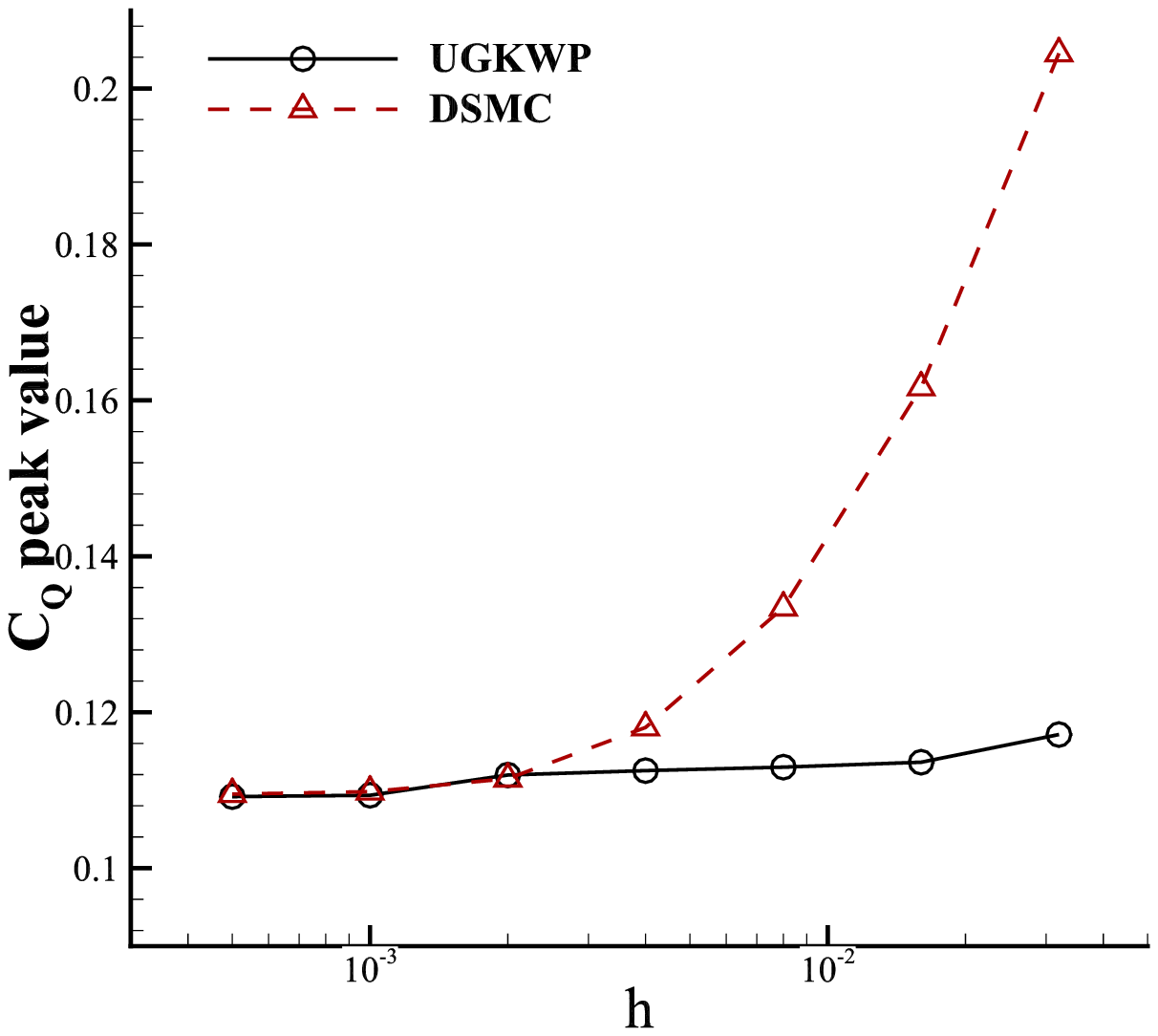}
    	}
	\caption{\label{cylinder-kn0.01line_dsmc} Comparisons of UGKWP with DSMC on hypersonic cylinder flow at ${\rm{Ma}}_{\infty}=5$, ${\rm{Kn}}_{\infty}=0.01$, ${\rm{Pr}}=2/3$: (a) Peak value of shear stress coefficient $C_F$, (b) peak value of heat flux coefficient $C_Q$.}
\end{figure}

\begin{figure}[H]
	\centering
	\subfigure[]{
			\includegraphics[width=0.3 \textwidth]{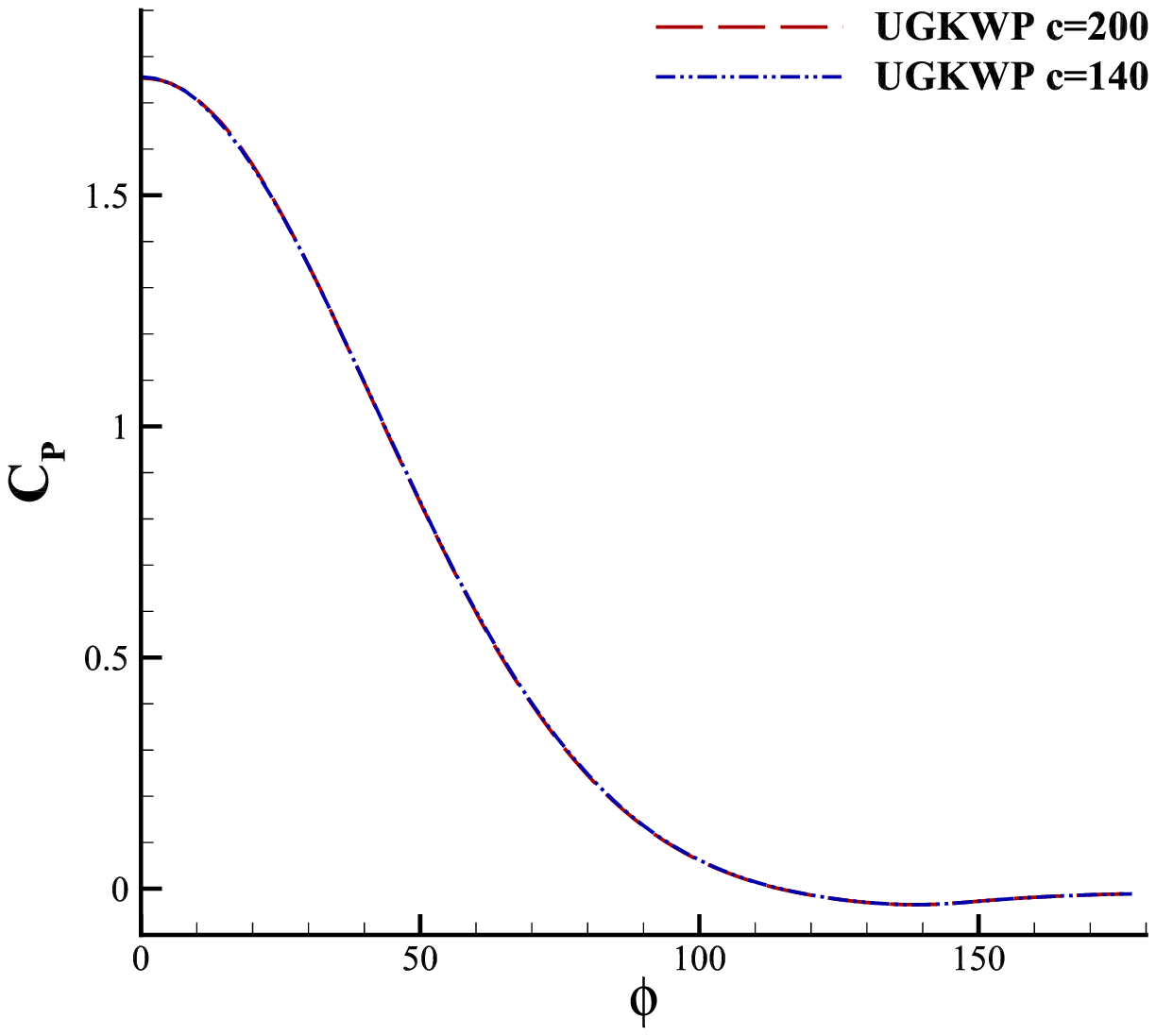}
		}
    \subfigure[]{
    		\includegraphics[width=0.3 \textwidth]{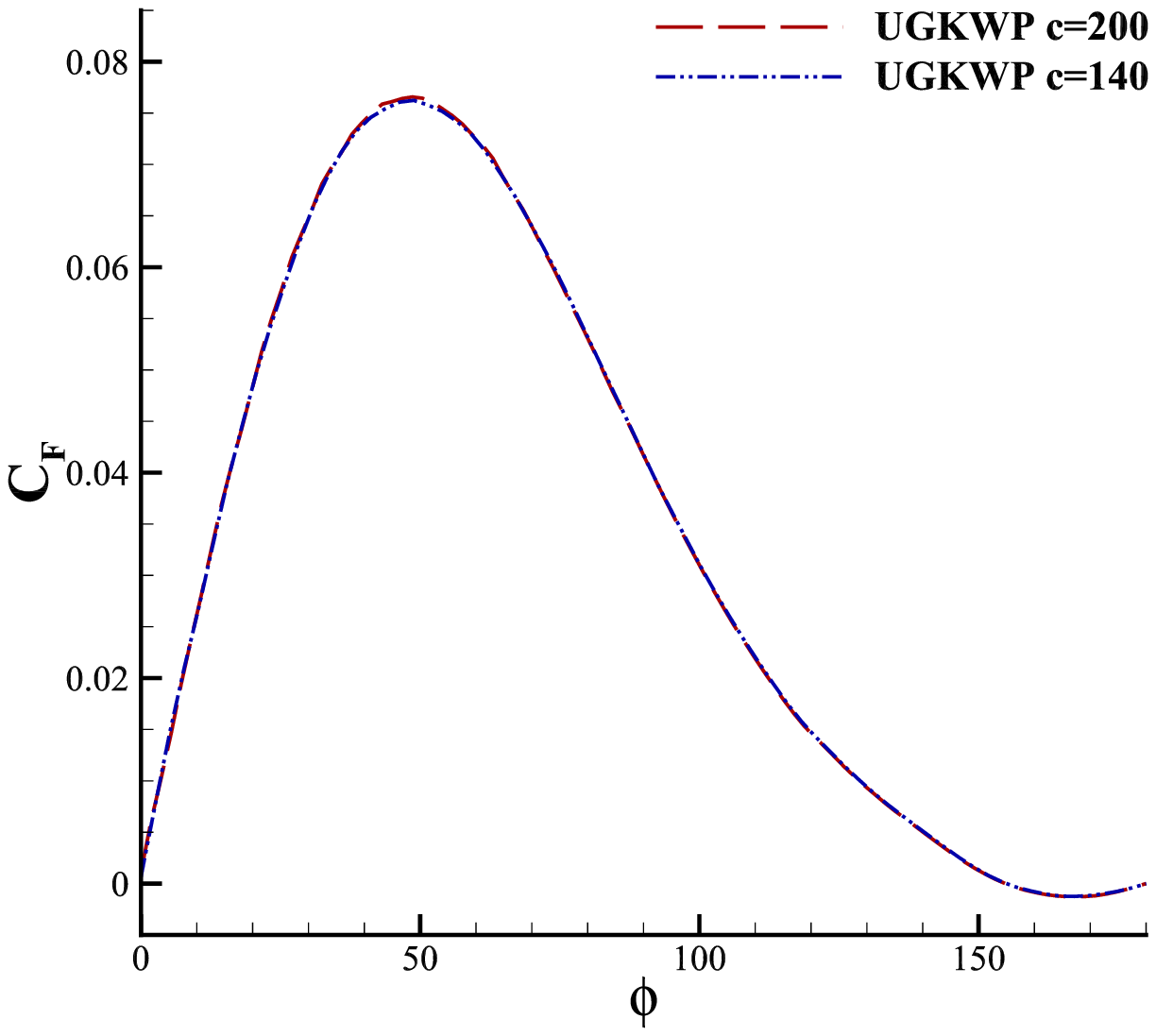}
    	}
    \subfigure[]{
			\includegraphics[width=0.3 \textwidth]{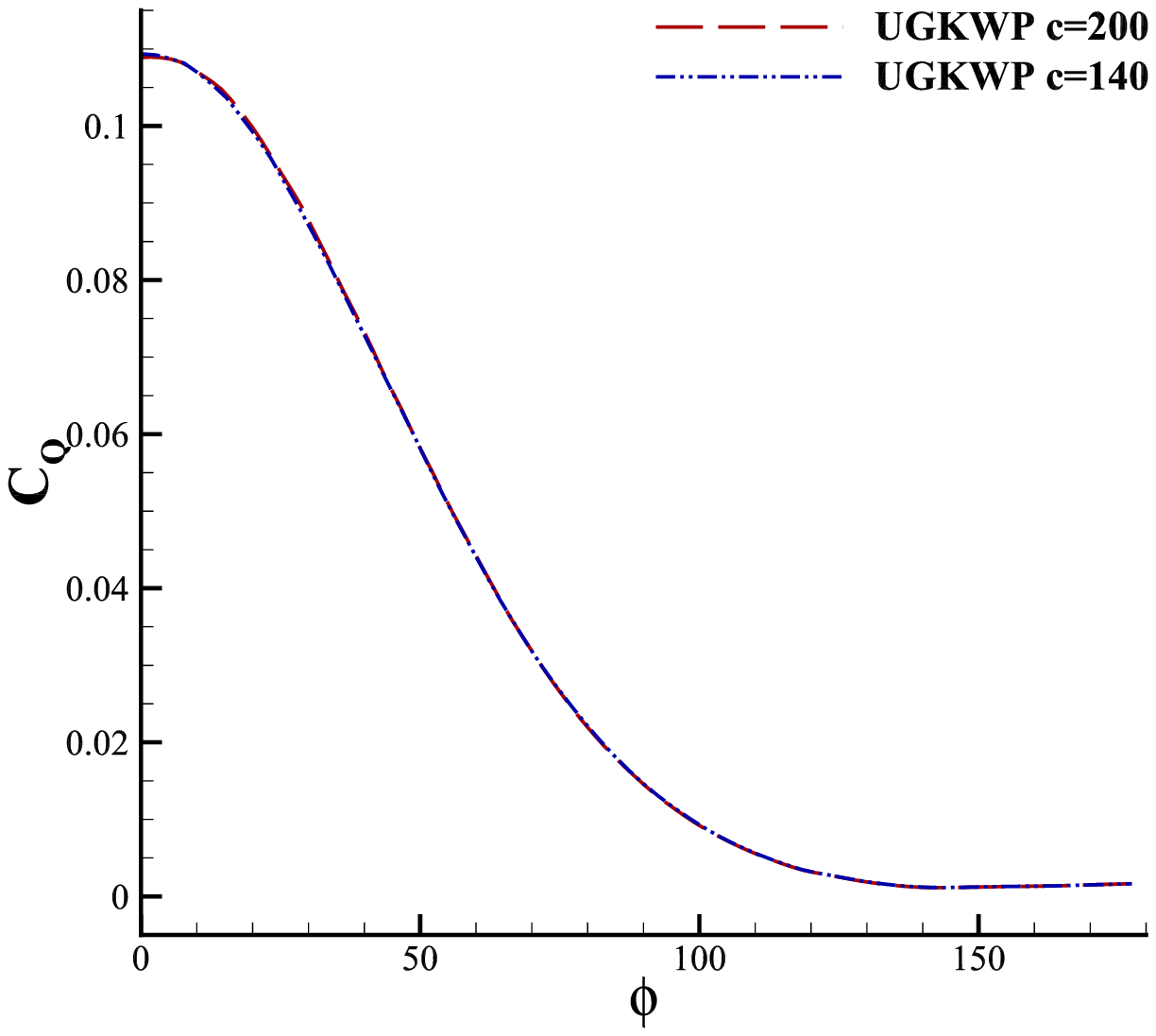}
		}
    \subfigure[]{
    		\includegraphics[width=0.3 \textwidth]{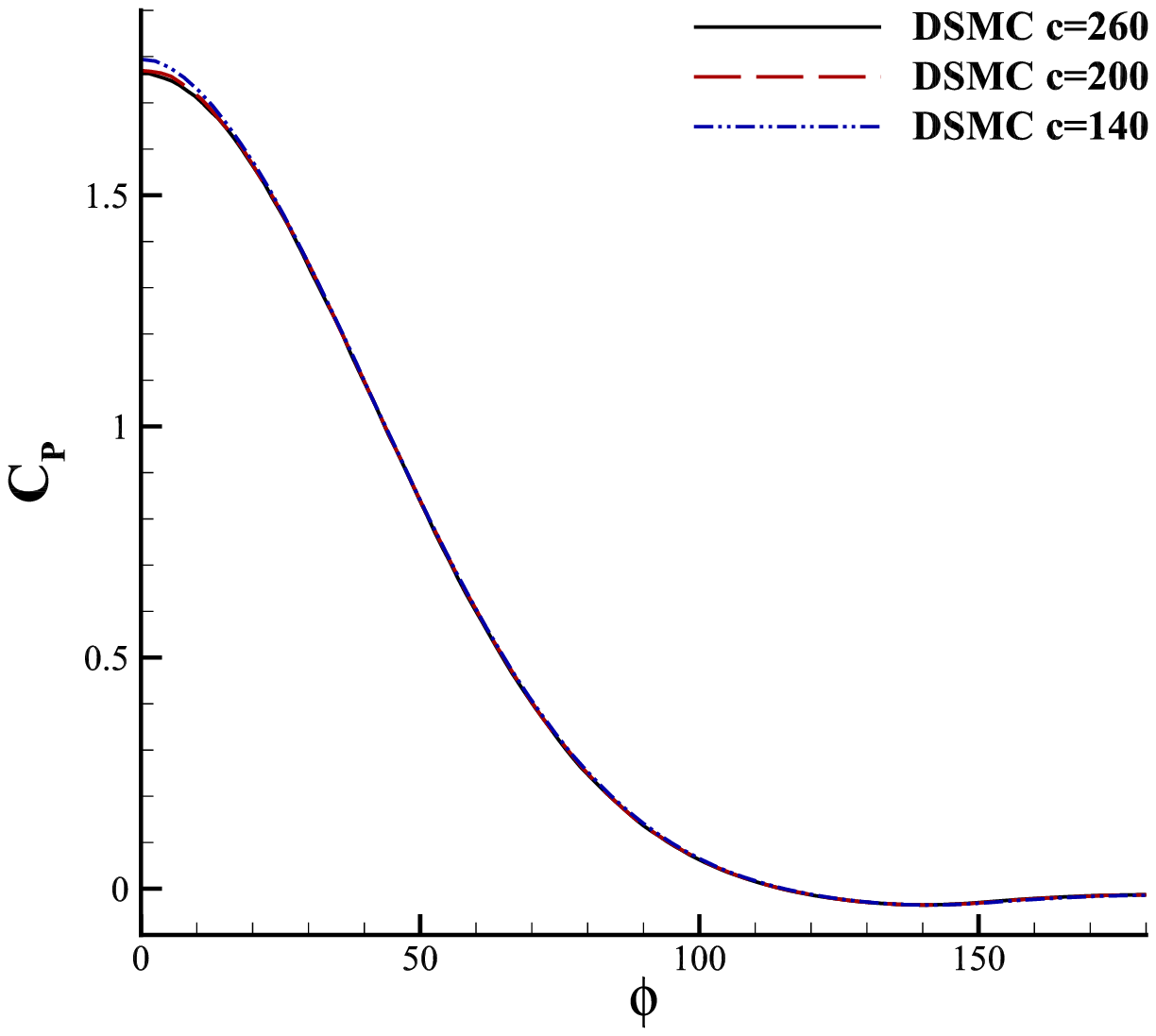}
    	}
    \subfigure[]{
			\includegraphics[width=0.3 \textwidth]{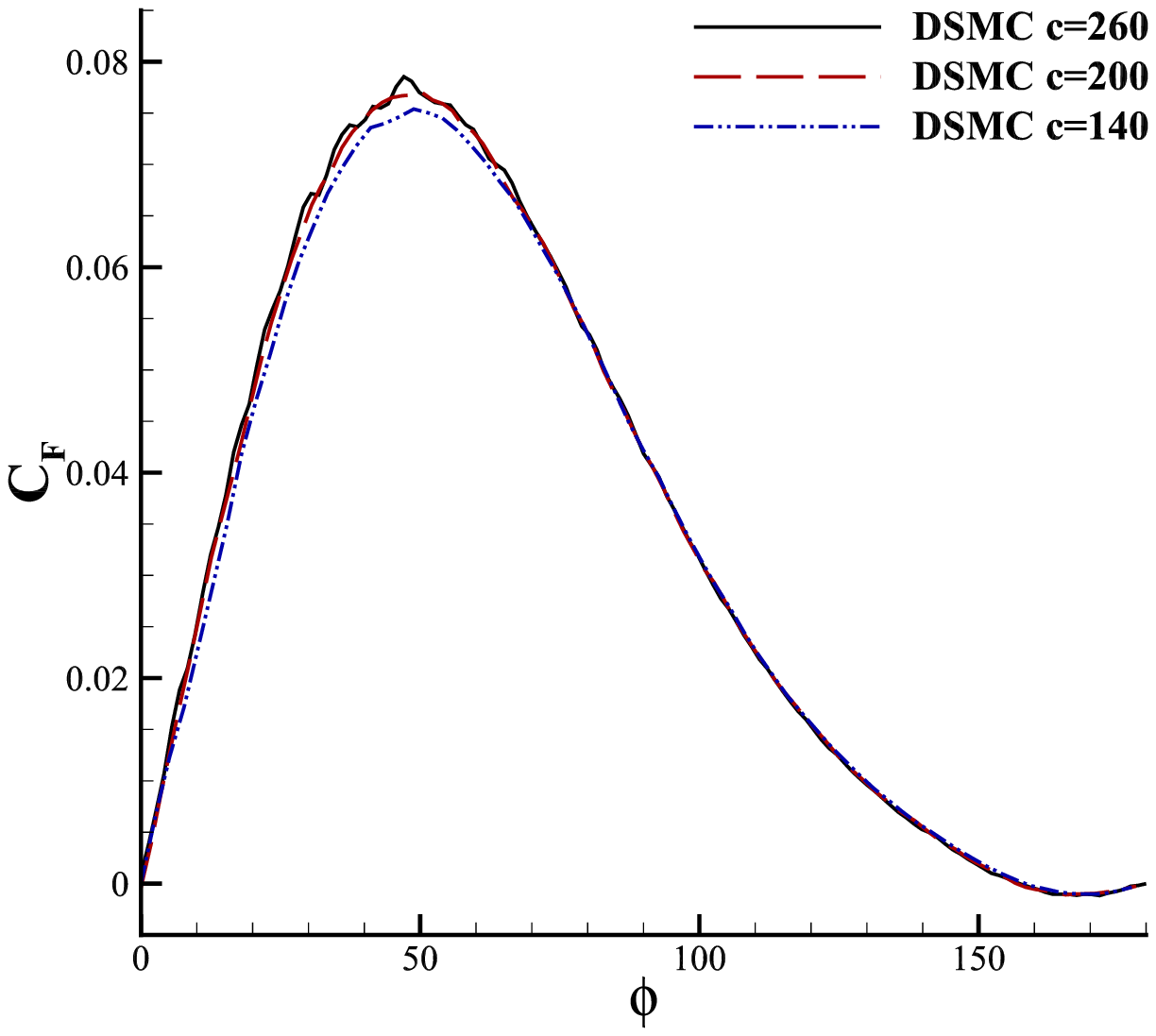}
		}
    \subfigure[]{
    		\includegraphics[width=0.3 \textwidth]{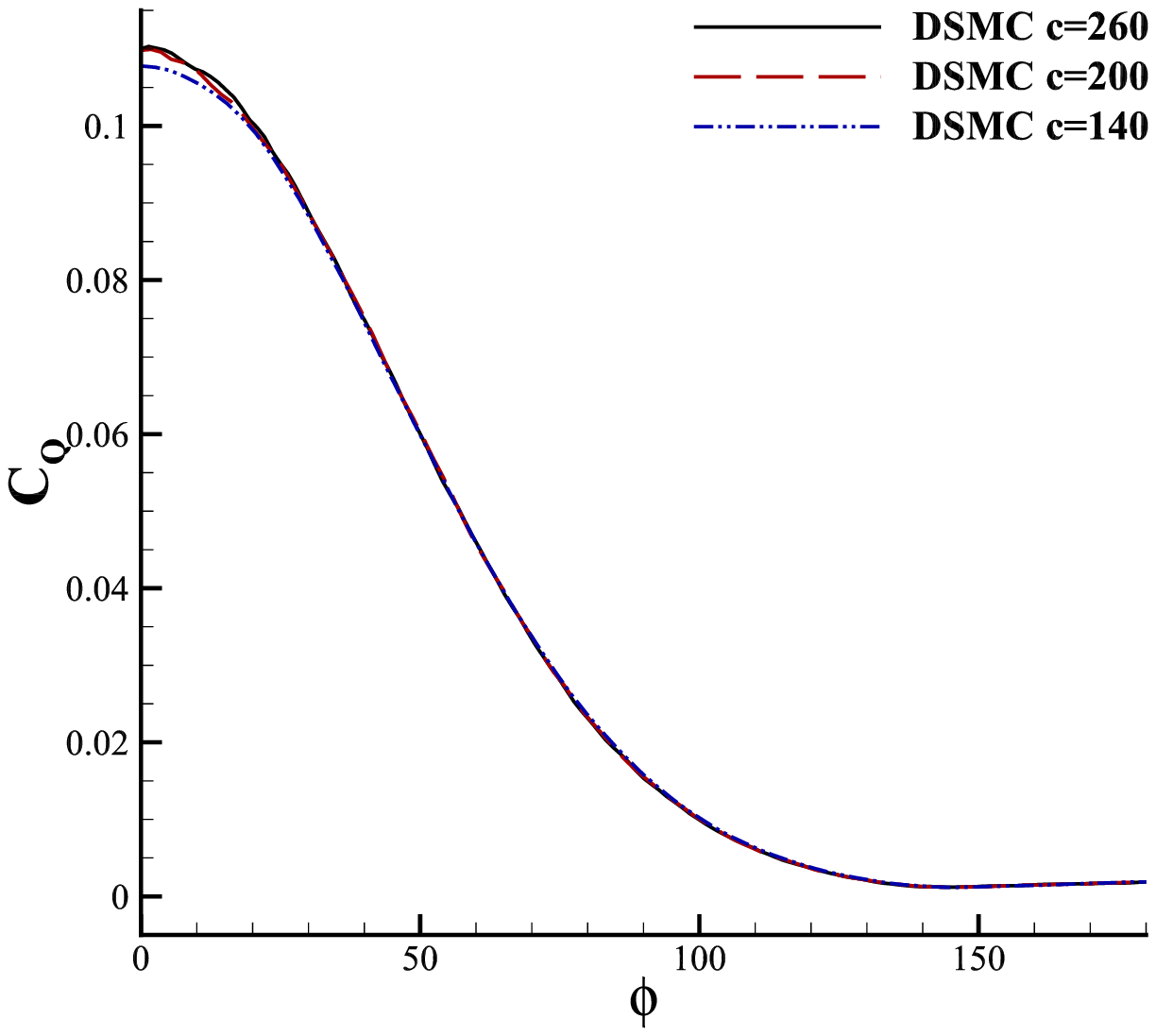}
    	}
	\caption{\label{cylinder-kn0.01_c} Comparisons of cell number $c$ around the circumference: (a) $C_P$ of UGKWP, (b) $C_F$ of UGKWP, (c) $C_Q$ of UGKWP, (d) $C_P$ of DSMC, (e) $C_F$ of DSMC, (f) $C_Q$ of DSMC.}
\end{figure}

Furthermore, the multiscale UGKWP method is compared with the single-scale DSMC method. For this comparison, ${\rm{Pr}}=2/3$ is adopted. As shown in Figs.~\ref{cylinder-kn0.01_dsmc} and \ref{cylinder-kn0.01line_dsmc}, the UGKWP method shows significantly better mesh-independent behavior. At $h=0.032$, the $C_F$ peak value predicted by the UGKWP method deviates by about $35\%$ relative to the finest-mesh solution, whereas for DSMC the deviation is about $130\%$. The $C_Q$ peak value predicted by the UGKWP method deviates by less than $6\%$, while for DSMC it exceeds $85\%$. The deviation of DSMC rises rapidly when $h$ and $\Delta t$ increase beyond the microscopic scales $\lambda$ and $\tau$ (see Refs.~\cite{dsmc1,dsmc2,dsmc3,dsmc4} for details). Apart from $h$, the UGKWP method also requires fewer cells around the circumference, as shown in Fig.~\ref{cylinder-kn0.01_c}. Based on the results of Fig.~\ref{cylinder-kn0.01_c}, in the mesh-independence study with respect to $h$, $c=140$ is set for the UGKWP method and $c=200$ is set for the DSMC method.

\begin{figure}[H]
	\centering
	\subfigure[]{
			\includegraphics[width=0.45 \textwidth]{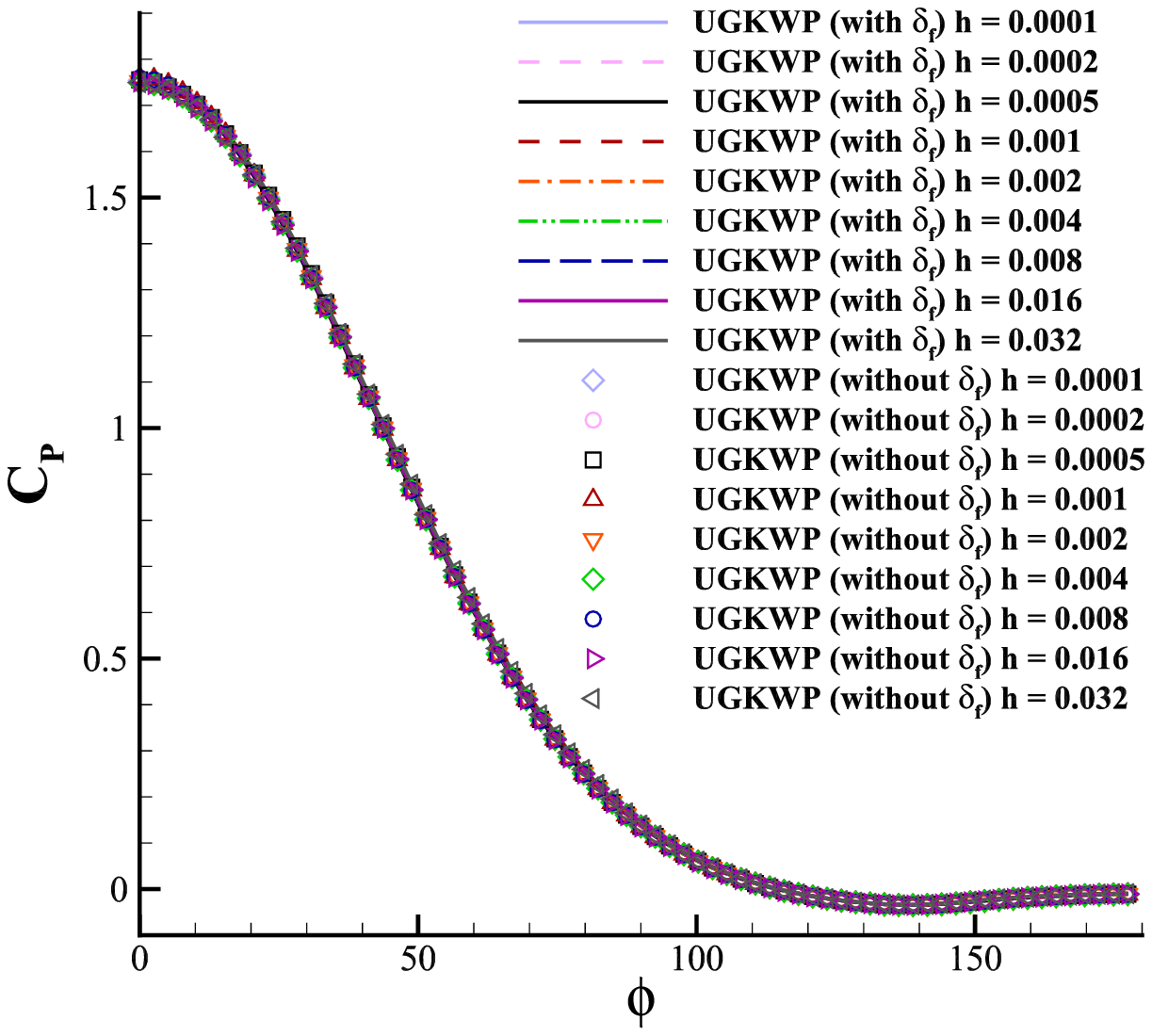}
		}
    \subfigure[]{
    		\includegraphics[width=0.45 \textwidth]{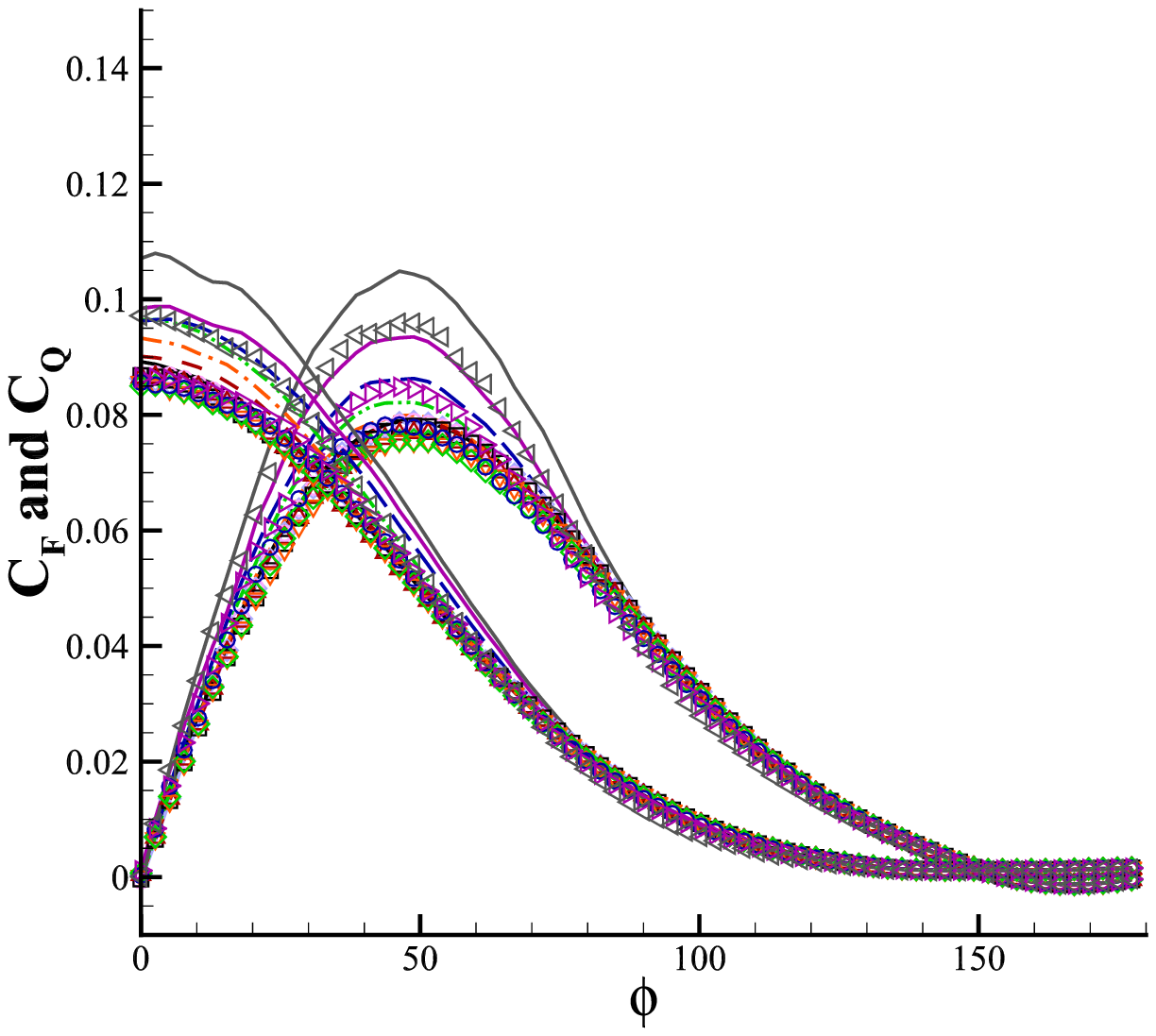}
    	}
    \subfigure[]{
			\includegraphics[width=0.45 \textwidth]{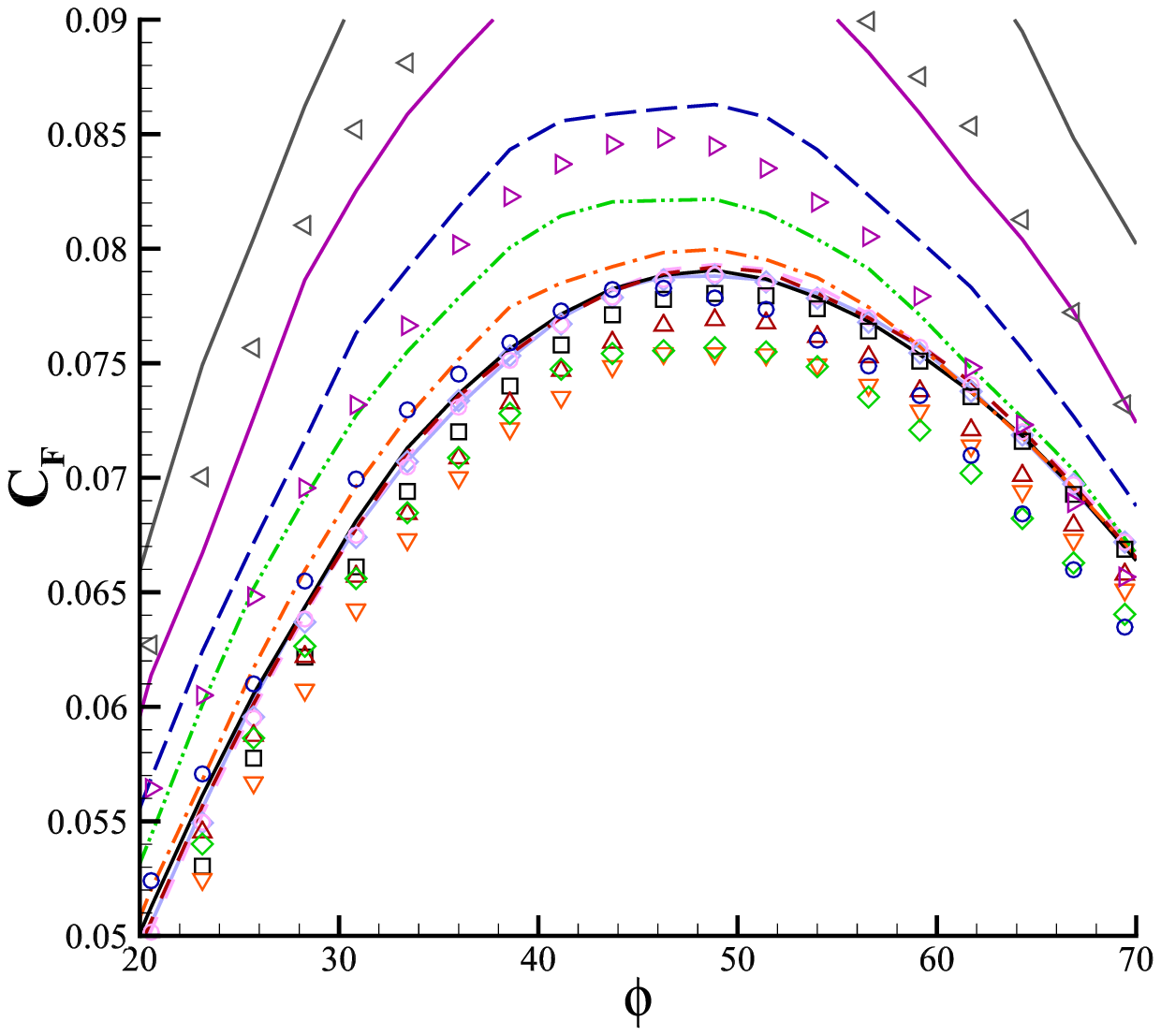}
		}
    \subfigure[]{
    		\includegraphics[width=0.45 \textwidth]{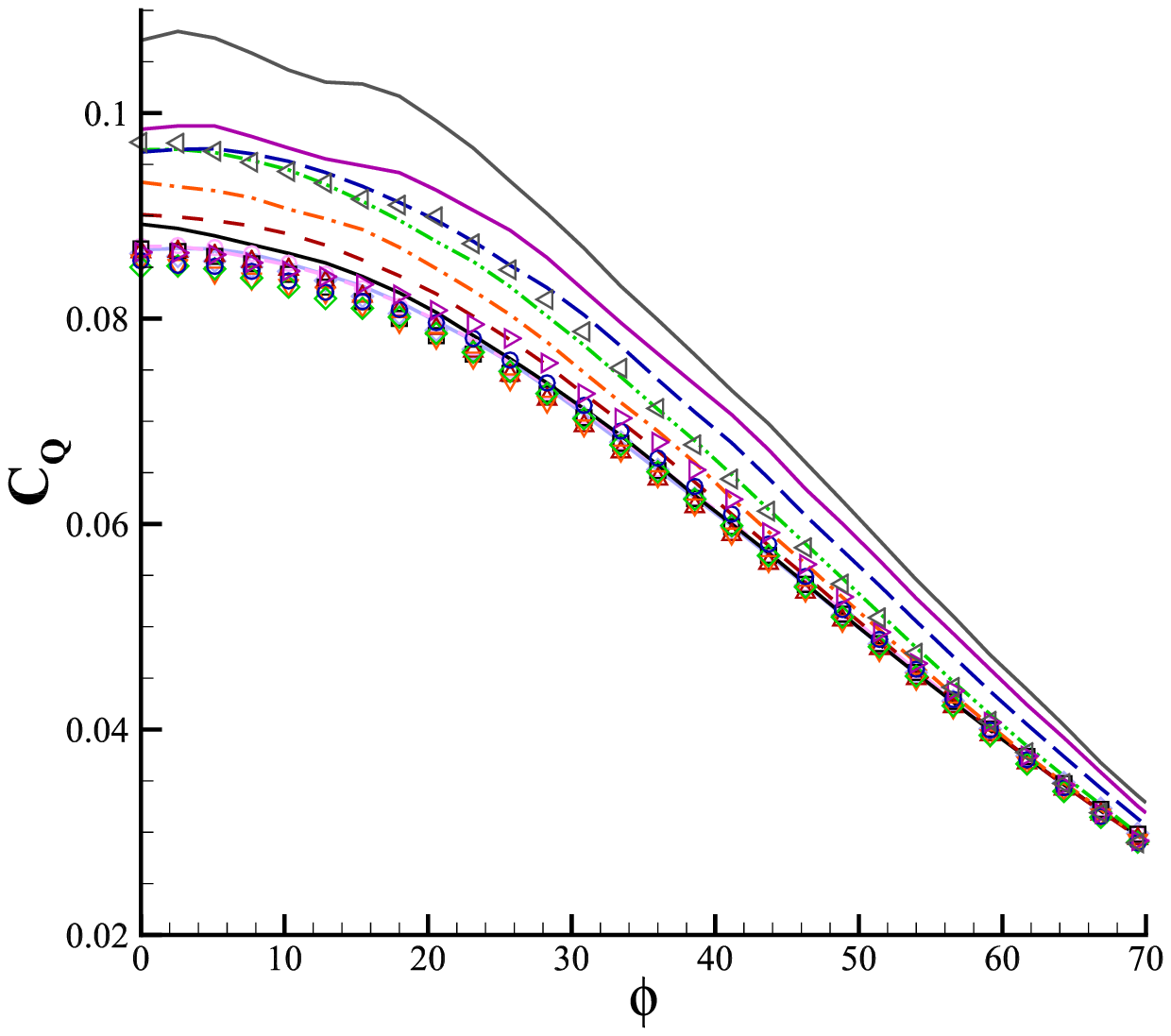}
    	}
	\caption{\label{cylinder-kn0.01_no6} Effect of the $\delta_f$ term on UGKWP solutions on hypersonic cylinder flow at ${\rm{Ma}}_{\infty}=5$, ${\rm{Kn}}_{\infty}=0.01$, ${\rm{Pr}}=1$: (a) Pressure coefficient at the wall $C_P$, (b) shear stress coefficient $C_F$ and heat flux coefficient $C_Q$ at the wall, (c) enlarged view of $C_F$, (d) enlarged view of $C_Q$.}
\end{figure}

\begin{figure}[H]
	\centering
	\subfigure[]{
			\includegraphics[width=0.45 \textwidth]{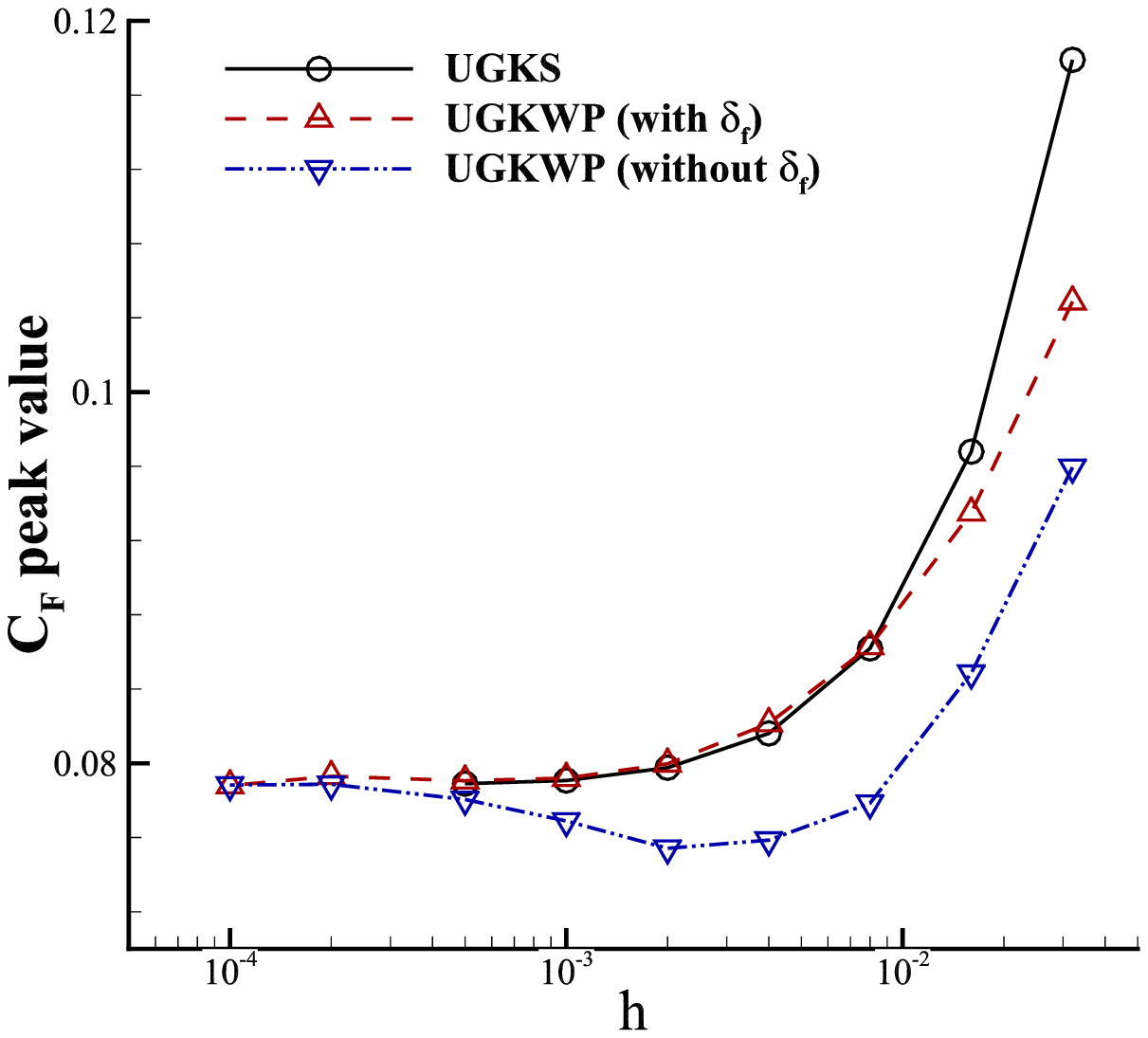}
		}
    \subfigure[]{
    		\includegraphics[width=0.45 \textwidth]{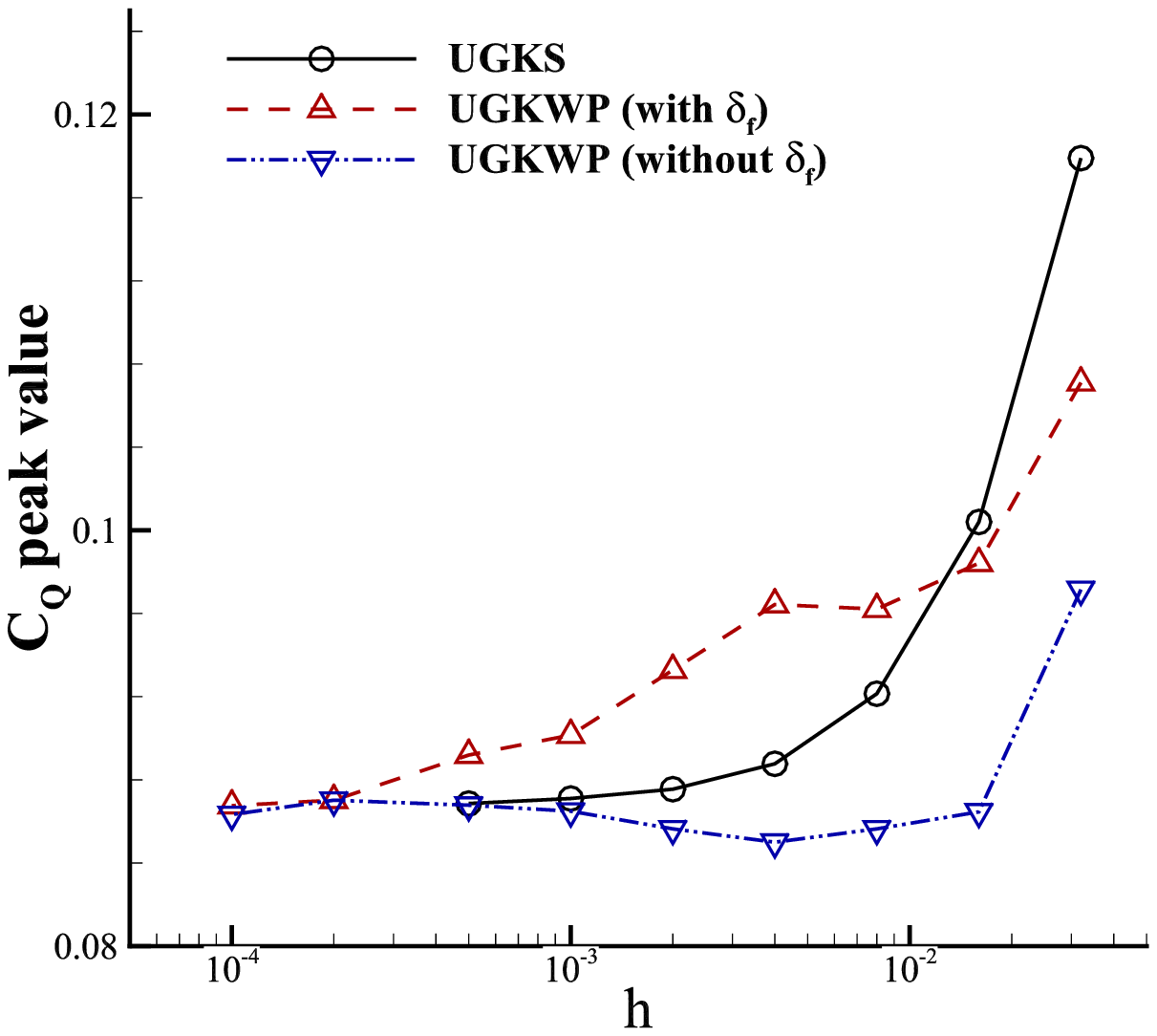}
    	}
	\caption{\label{cylinder-kn0.01line_6} Effect of the $\delta_f$ term on UGKWP solutions on hypersonic cylinder flow at ${\rm{Ma}}_{\infty}=5$, ${\rm{Kn}}_{\infty}=0.01$, ${\rm{Pr}}=1$: (a) Peak value of shear stress coefficient $C_F$, (b) peak value of heat flux coefficient $C_Q$.}
\end{figure}

Regarding the modification in $\boldsymbol{F}^{fr,wave}$, the performance is tested with and without the Chapman--Enskog expansion term ($\delta_f$ term in Eq.~\eqref{eq:ffrwave}). As shown in Figs.~\ref{cylinder-kn0.01_no6} and \ref{cylinder-kn0.01line_6}, when $h$ increases, the $C_F$ and $C_Q$ profiles of UGKWP without the $\delta_f$ term first decrease and then increase again. This occurs because the viscous contribution associated with the Chapman--Enskog expansion is omitted, whose influence is larger than that of the coarser mesh. Consequently, introducing this term yields more reliable results.

\begin{figure}[H]
	\centering
	\subfigure[]{
			\includegraphics[width=0.45 \textwidth]{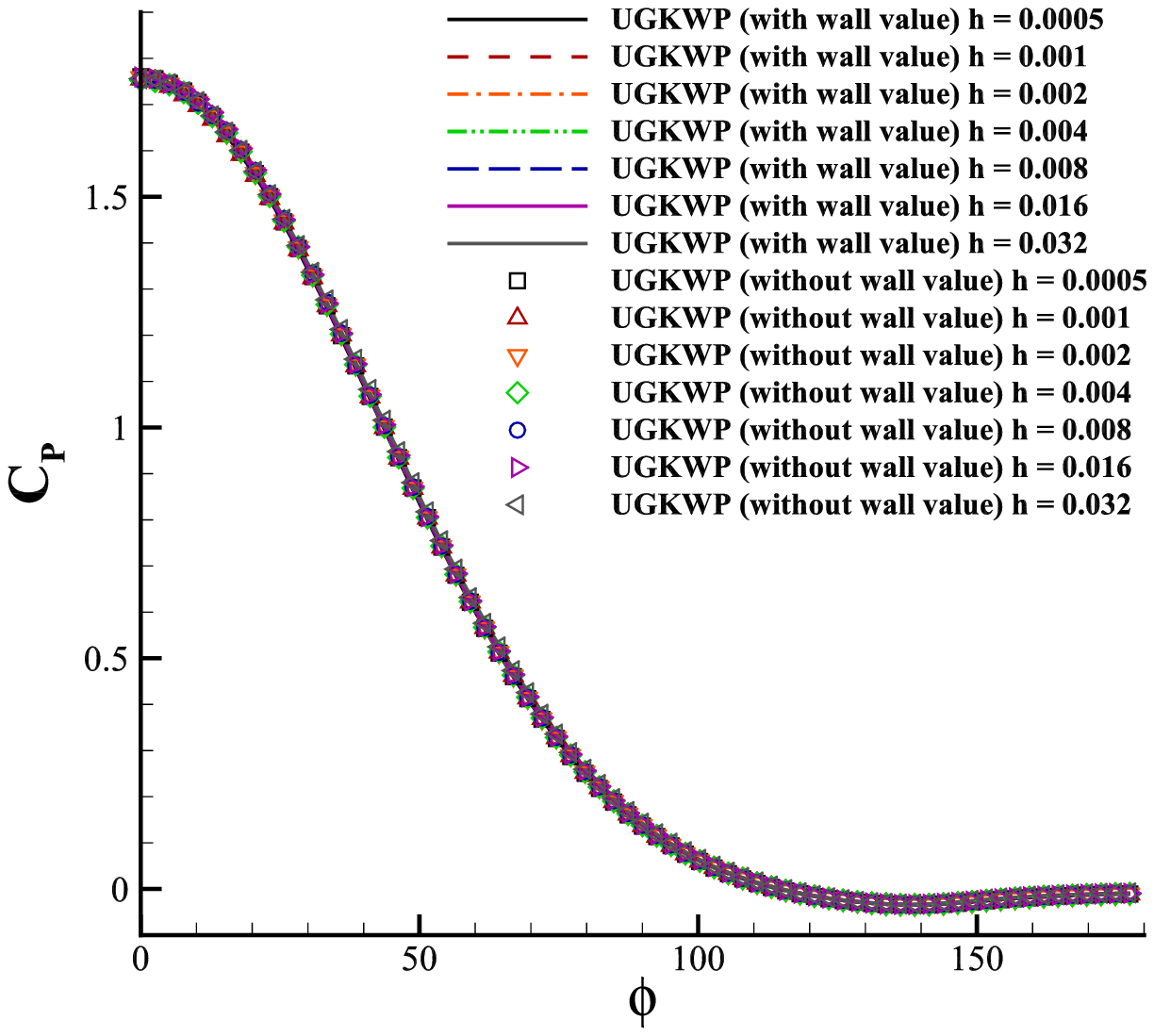}
		}
    \subfigure[]{
    		\includegraphics[width=0.45 \textwidth]{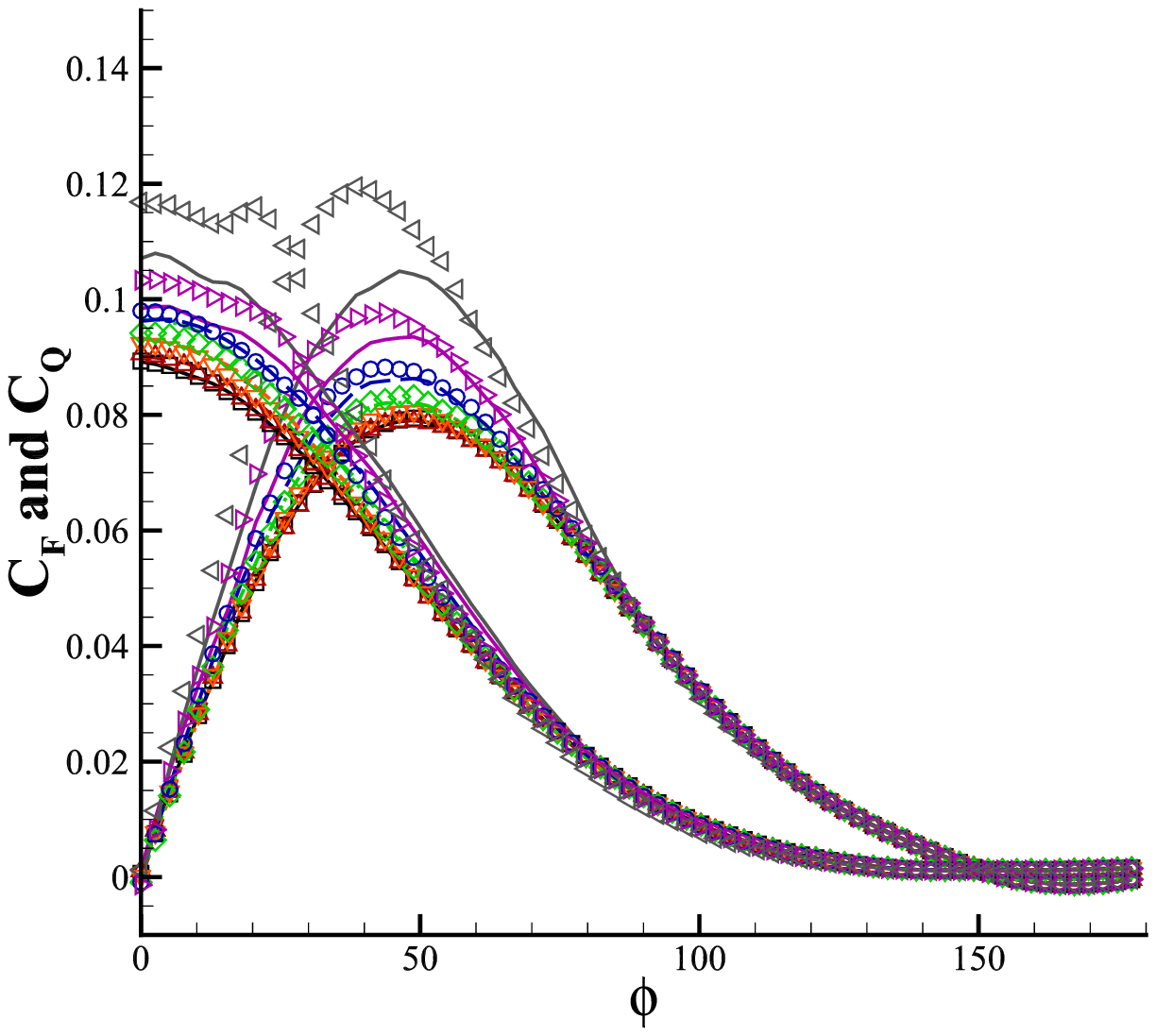}
    	}
    \subfigure[]{
			\includegraphics[width=0.45 \textwidth]{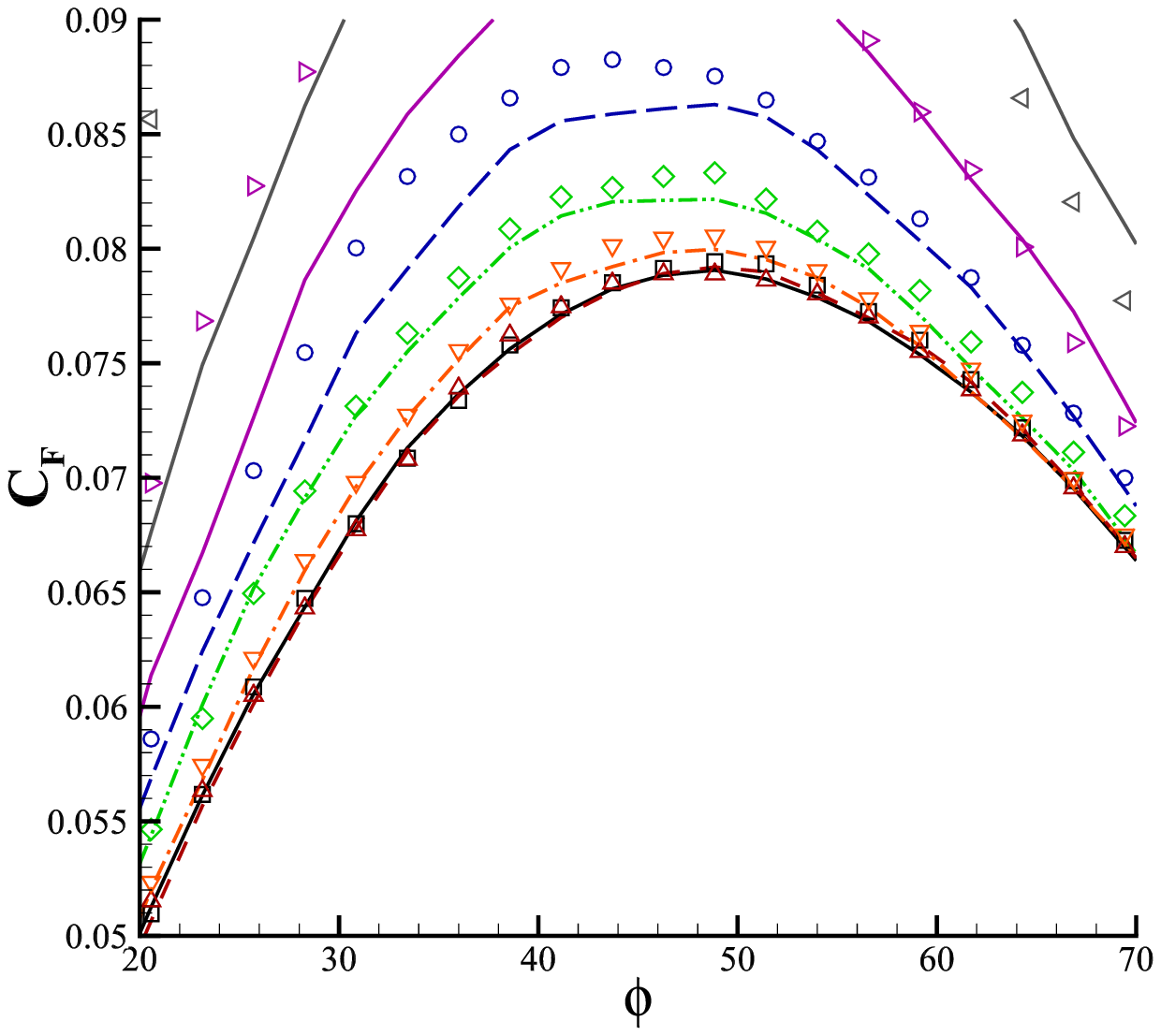}
		}
    \subfigure[]{
    		\includegraphics[width=0.45 \textwidth]{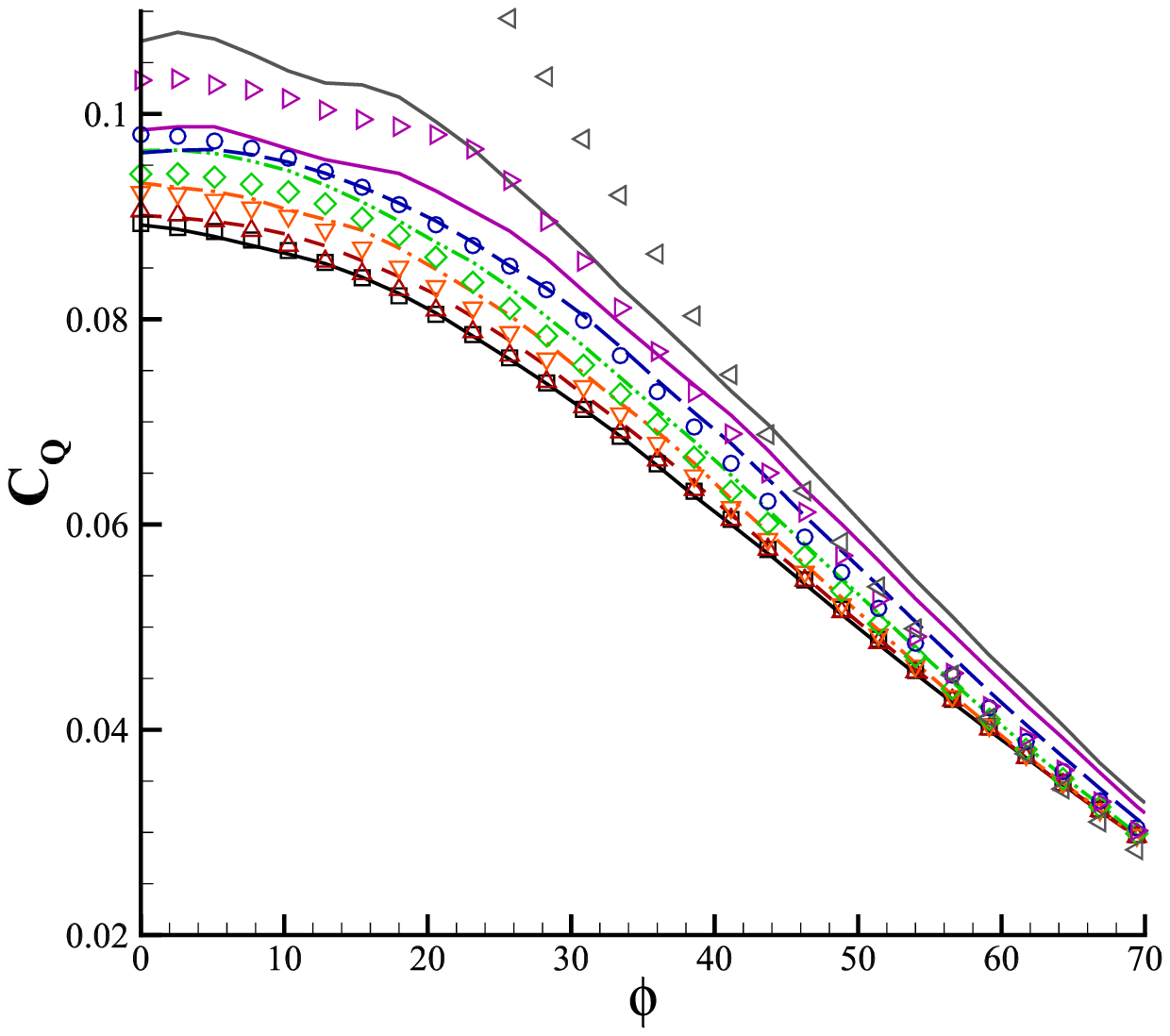}
    	}
	\caption{\label{cylinder-kn0.01_wallGrad} Comparisons of UGKWP method with and without wall values in gradient calculations on hypersonic cylinder flow at ${\rm{Ma}}_{\infty}=5$, ${\rm{Kn}}_{\infty}=0.01$, ${\rm{Pr}}=1$: (a) Pressure coefficient at the wall $C_P$, (b) shear stress coefficient $C_F$ and heat flux coefficient $C_Q$ at the wall, (c) enlarged view of $C_F$, (d) enlarged view of $C_Q$.}
\end{figure}

The modification in the gradient calculation, specifically whether the wall value is considered (Eq.~\eqref{eq:wls-wall}) or not, is also tested as shown in Fig.~\ref{cylinder-kn0.01_wallGrad}. The influence is negligible at low $h$ values, but accuracy is improved at large $h$ values.

\begin{figure}[H]
	\centering
	\subfigure[]{
			\includegraphics[width=0.3 \textwidth]{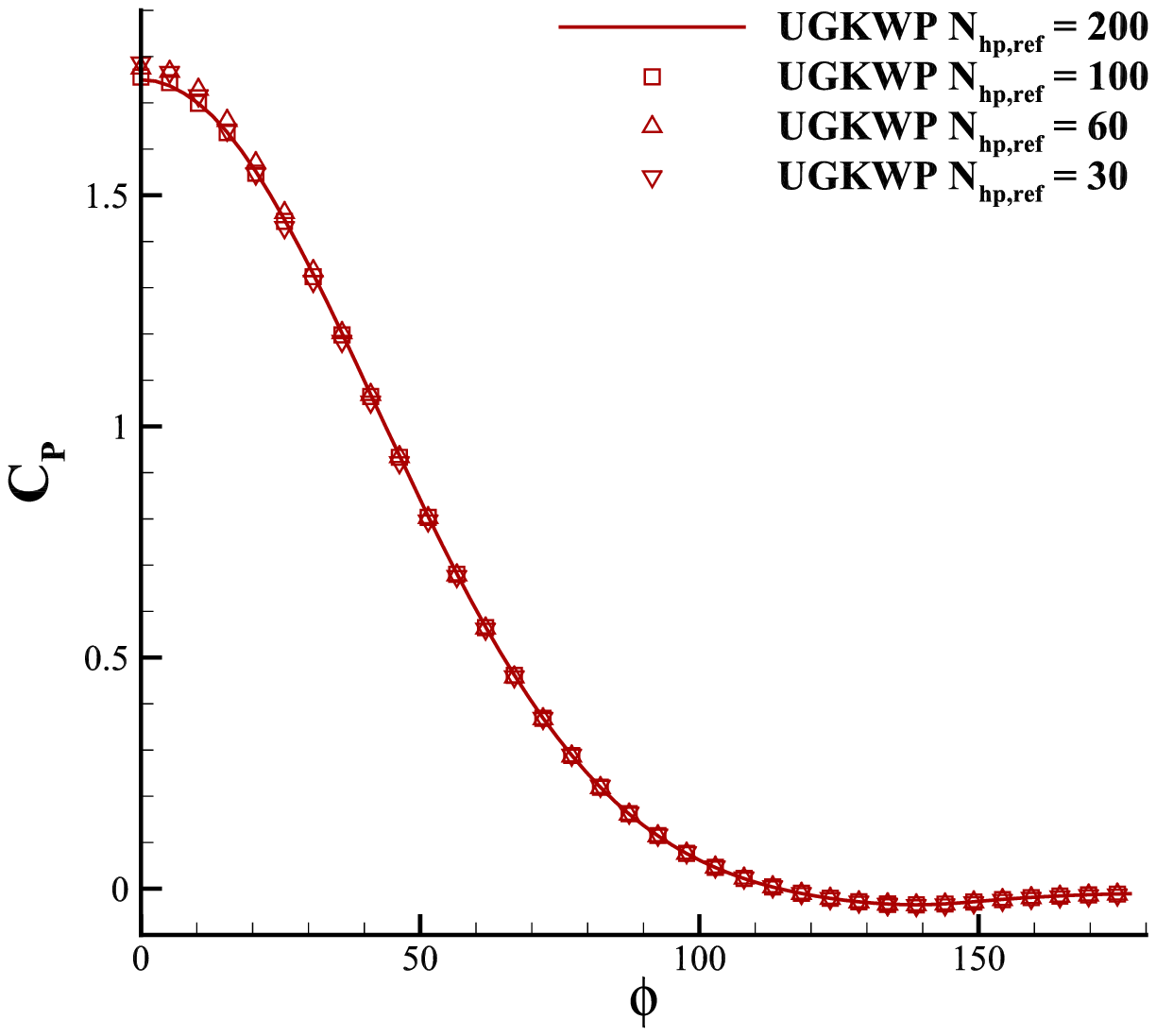}
		}
    \subfigure[]{
    		\includegraphics[width=0.3 \textwidth]{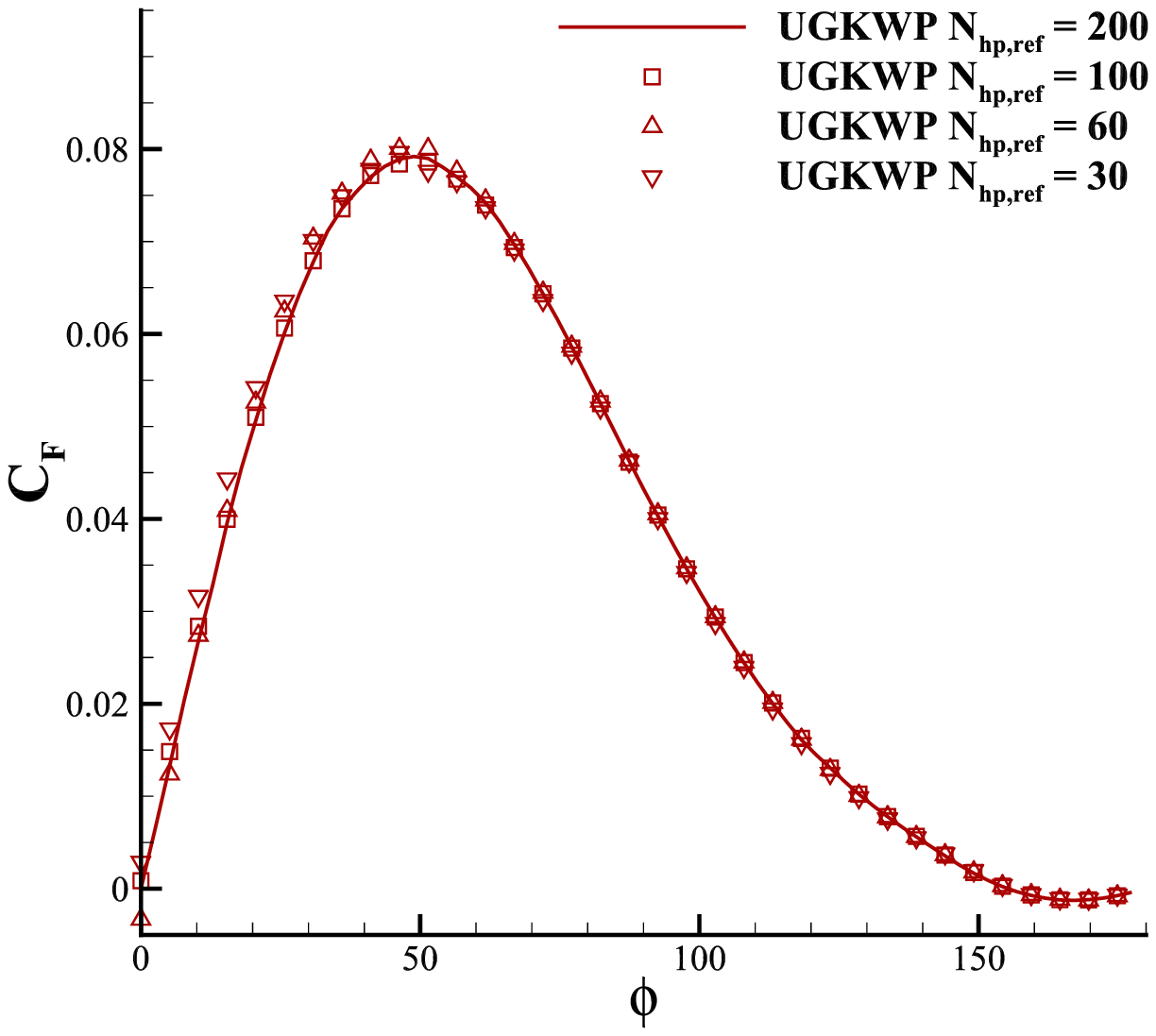}
    	}
    \subfigure[]{
			\includegraphics[width=0.3 \textwidth]{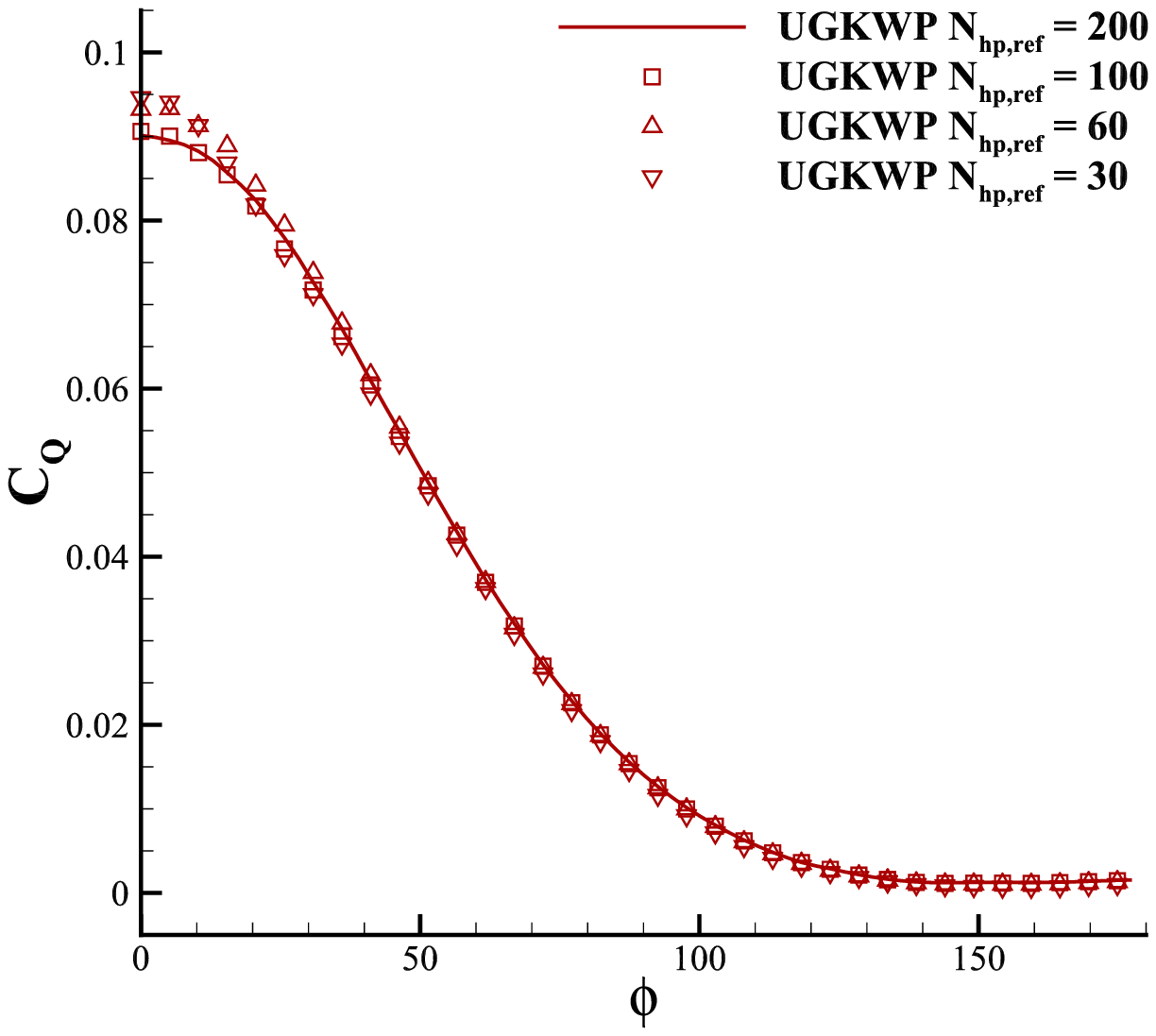}
		}
    \subfigure[]{
    		\includegraphics[width=0.3 \textwidth]{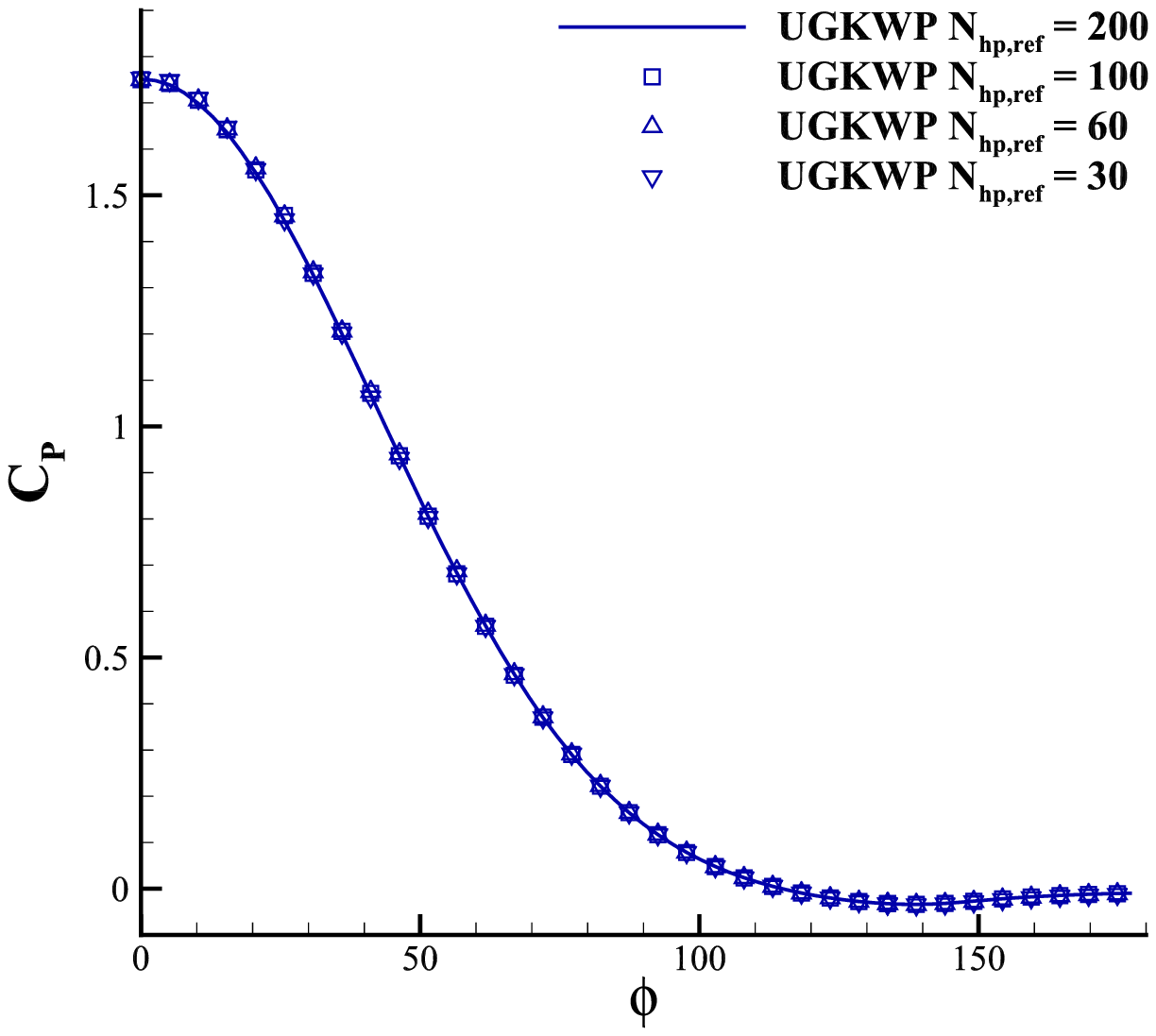}
    	}
    \subfigure[]{
			\includegraphics[width=0.3 \textwidth]{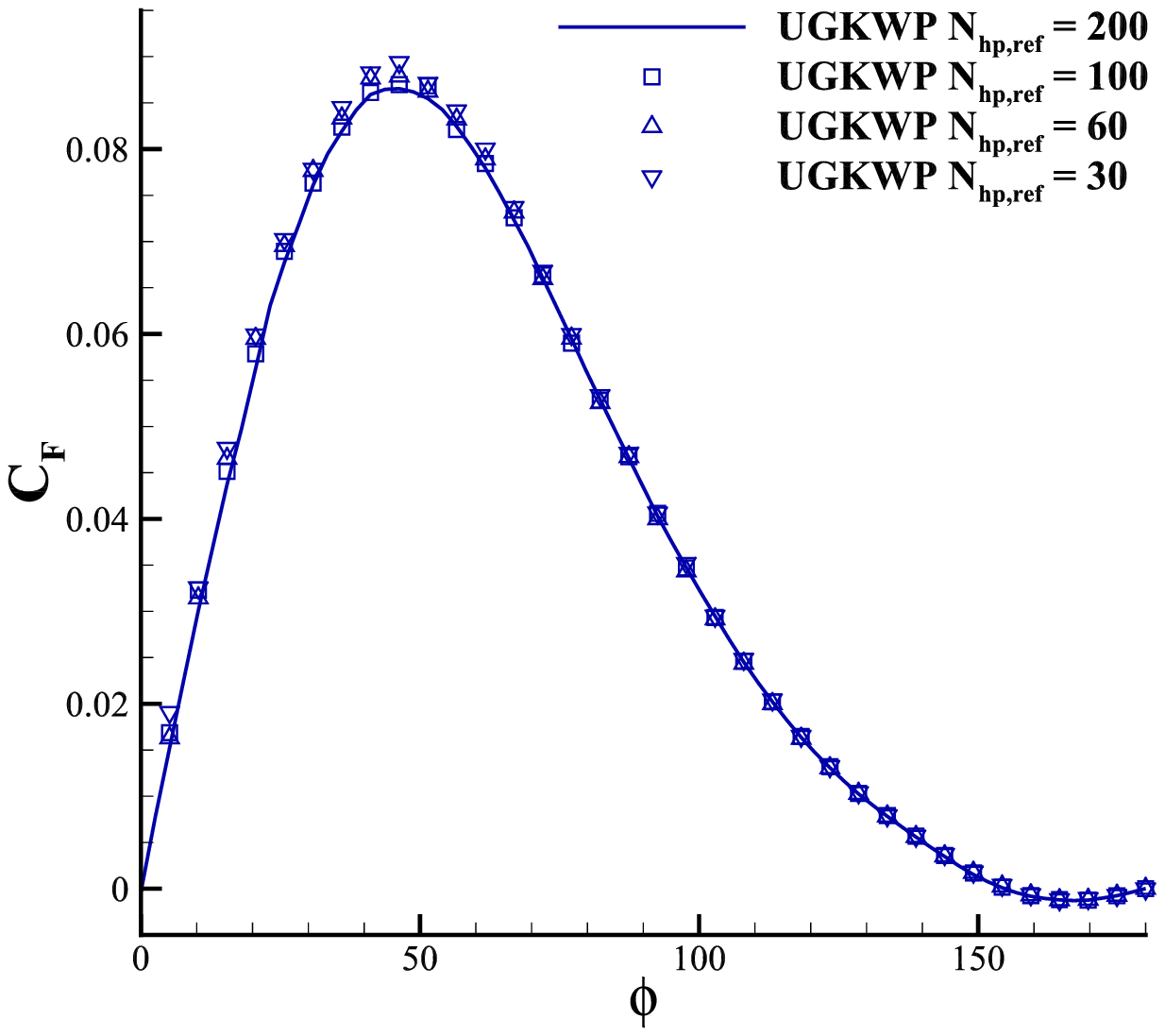}
		}
    \subfigure[]{
    		\includegraphics[width=0.3 \textwidth]{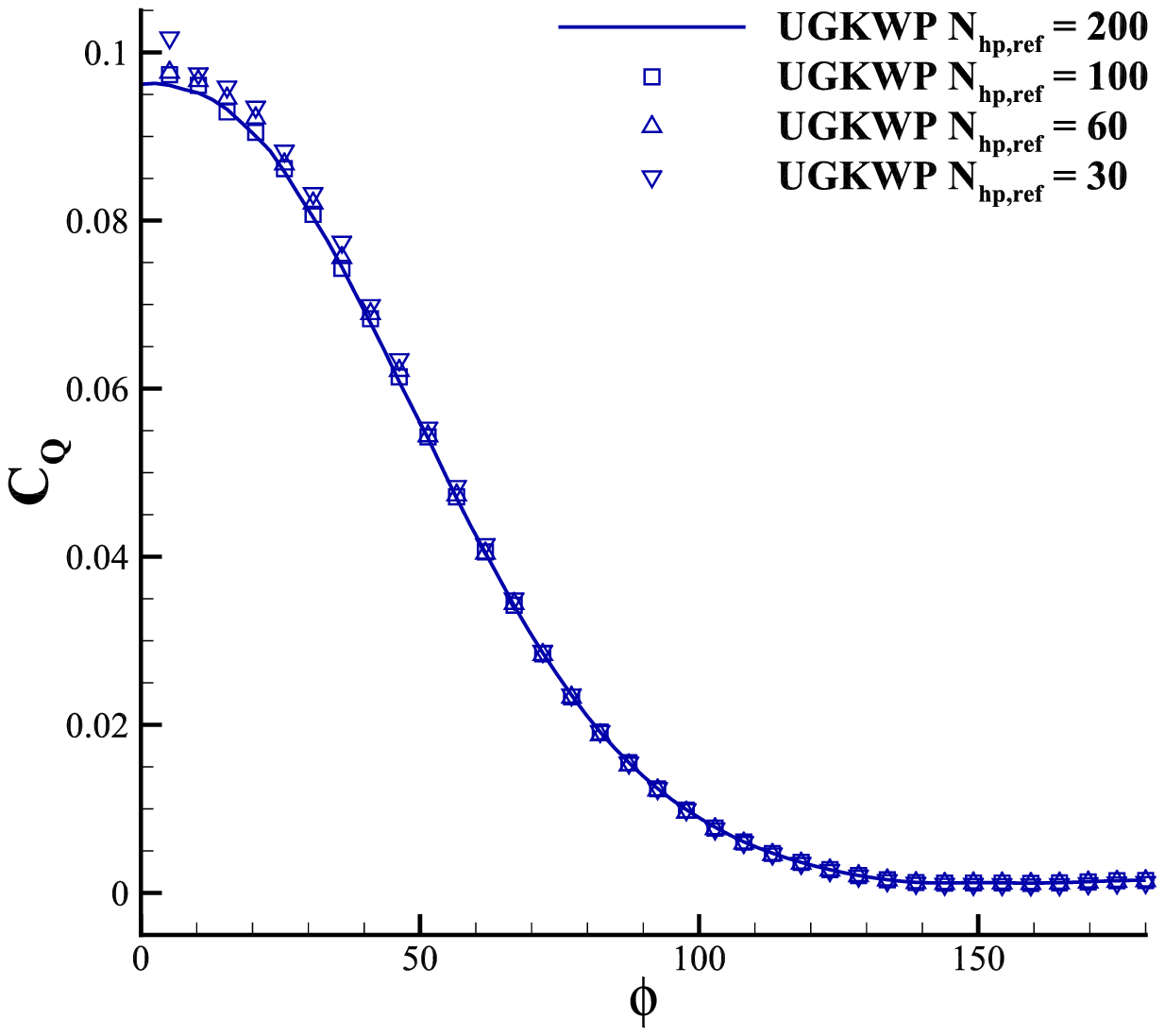}
    	}
	\caption{\label{cylinder-kn0.01_nref} $N^{hp,{\rm{ref}}}$ independence study on hypersonic cylinder flow at ${\rm{Ma}}_{\infty}=5$, ${\rm{Kn}}_{\infty}=0.01$, ${\rm{Pr}}=1$: (a) $C_P$ at $h=0.001$, (b) $C_F$ at $h=0.001$, (c) $C_Q$ at $h=0.001$, (d) $C_P$ at $h=0.008$, (e) $C_F$ at $h=0.008$, (f) $C_Q$ at $h=0.008$.}
\end{figure}

Additionally, the $N^{hp,{\rm{ref}}}$ independence is studied for $h=0.001$ and $h=0.008$ cases. As shown in Fig.~\ref{cylinder-kn0.01_nref}, the performances at both large $h$ and small $h$ are similar. Good consistency is achieved at $N^{hp,{\rm{ref}}}=200$ and $N^{hp,{\rm{ref}}}=100$. The deviation increases when $N^{hp,{\rm{ref}}}$ is further reduced.

\begin{figure}[H]
	\centering
	\subfigure[]{
			\includegraphics[width=0.45 \textwidth]{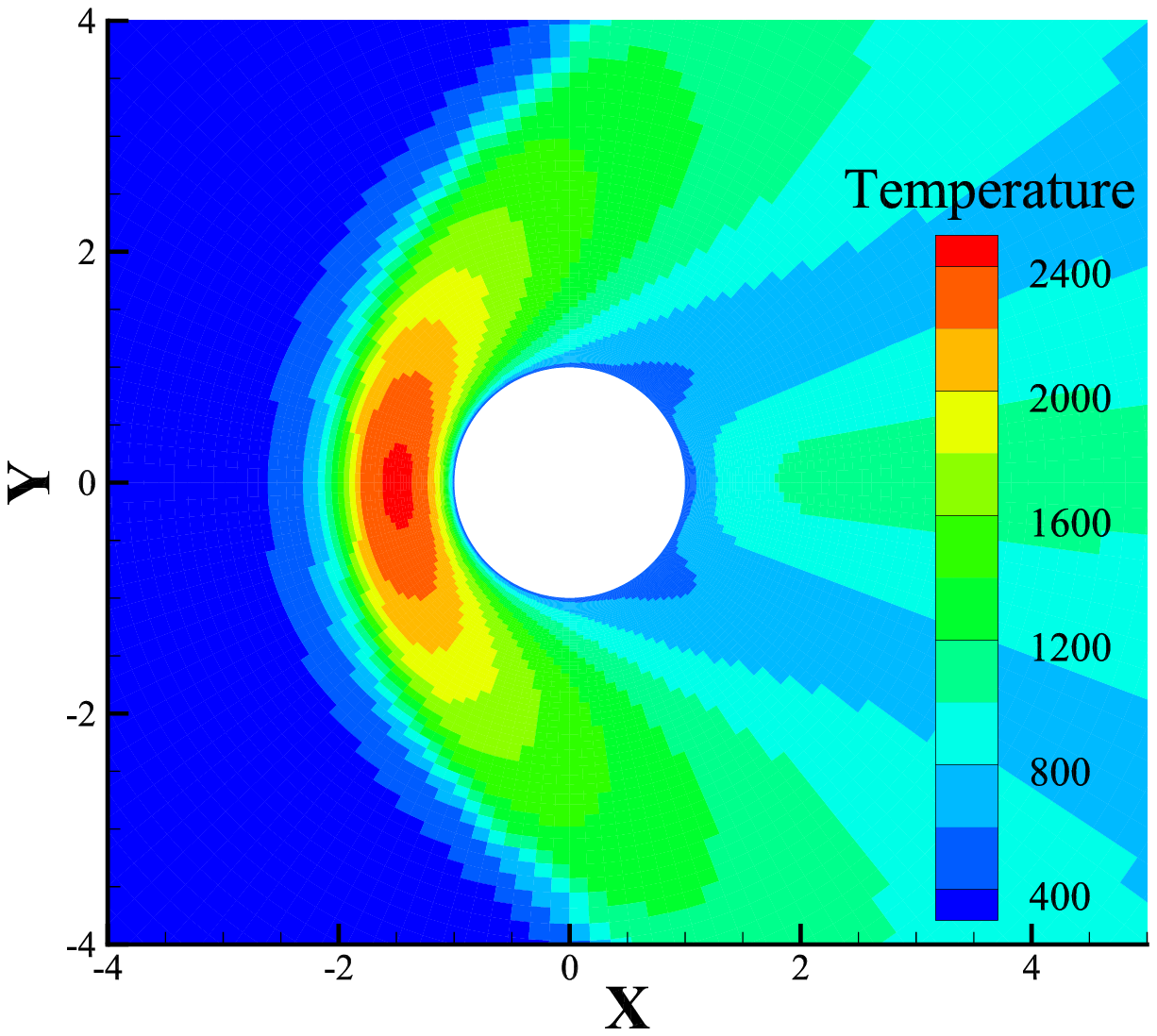}
		}
    \subfigure[]{
    		\includegraphics[width=0.45 \textwidth]{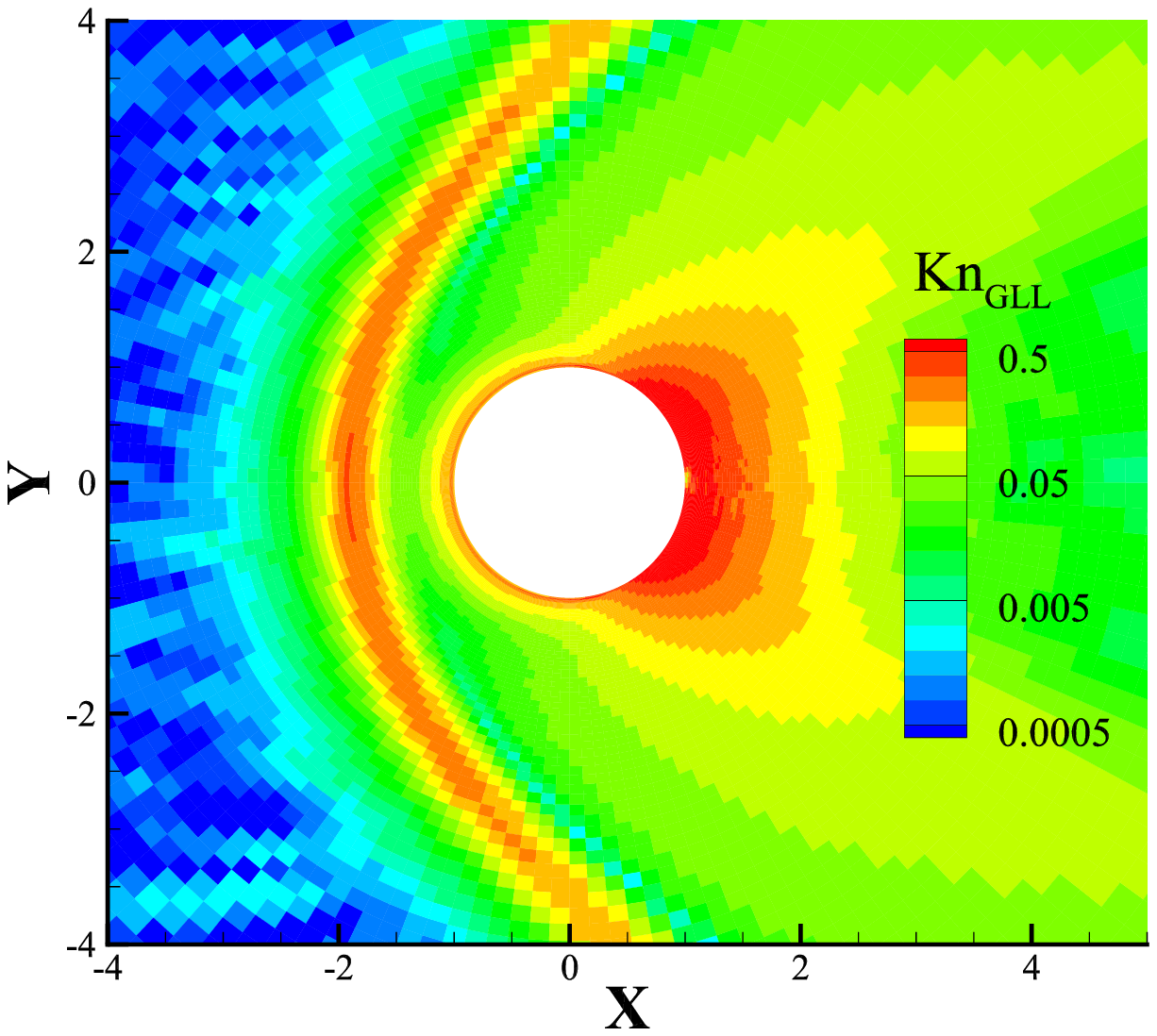}
    	}
	\caption{\label{cylinder-kn0.1-contour} Contours of hypersonic cylinder flow at ${\rm{Ma}}_{\infty}=5$, ${\rm{Kn}}_{\infty}=0.1$, ${\rm{Pr}}=1$: (a) Temperature, (b) $\rm{Kn_{GLL}}$.}
\end{figure}

\begin{figure}[H]
	\centering
	\subfigure[]{
			\includegraphics[width=0.45 \textwidth]{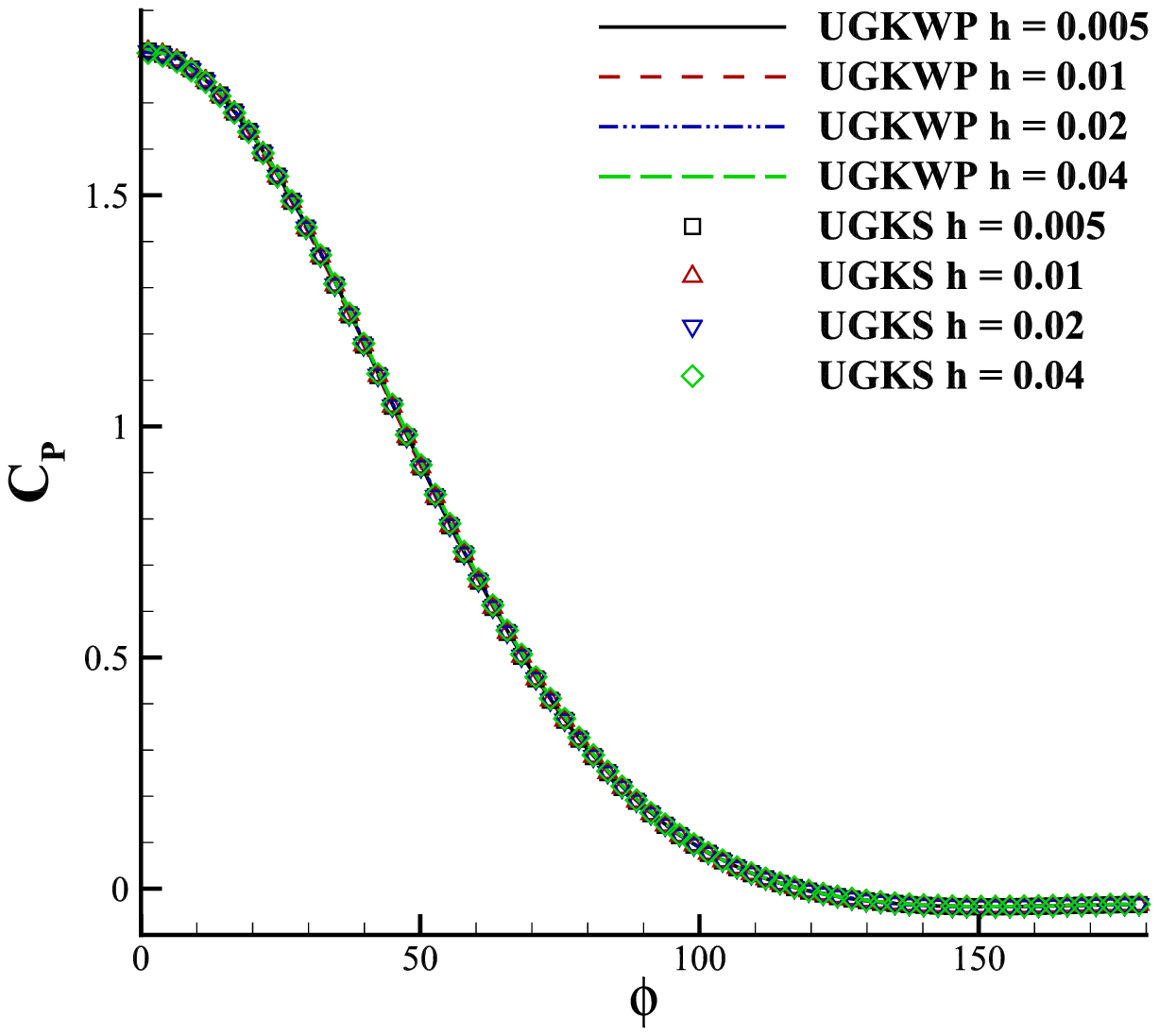}
		}
    \subfigure[]{
    		\includegraphics[width=0.45 \textwidth]{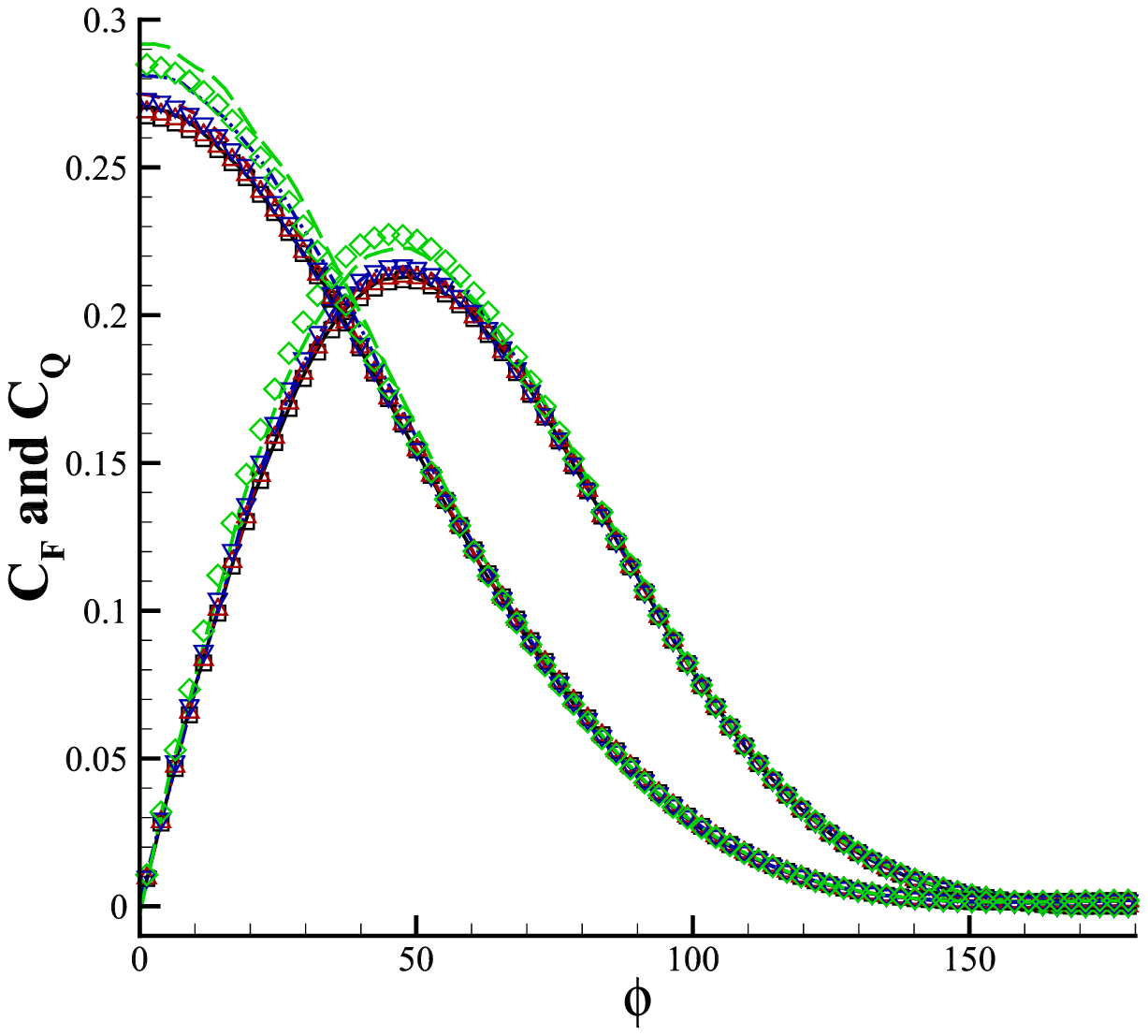}
    	}
    \subfigure[]{
			\includegraphics[width=0.45 \textwidth]{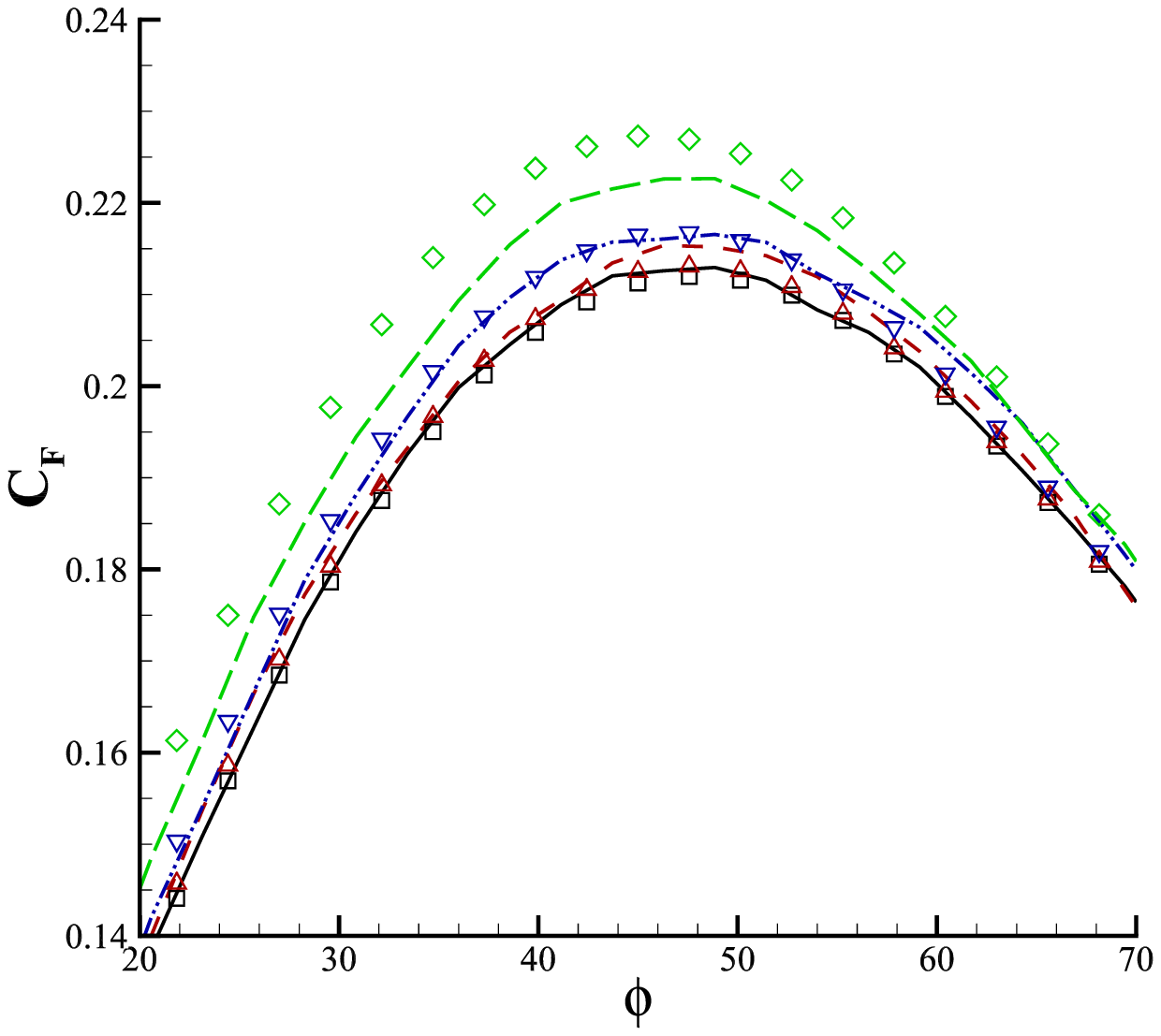}
		}
    \subfigure[]{
    		\includegraphics[width=0.45 \textwidth]{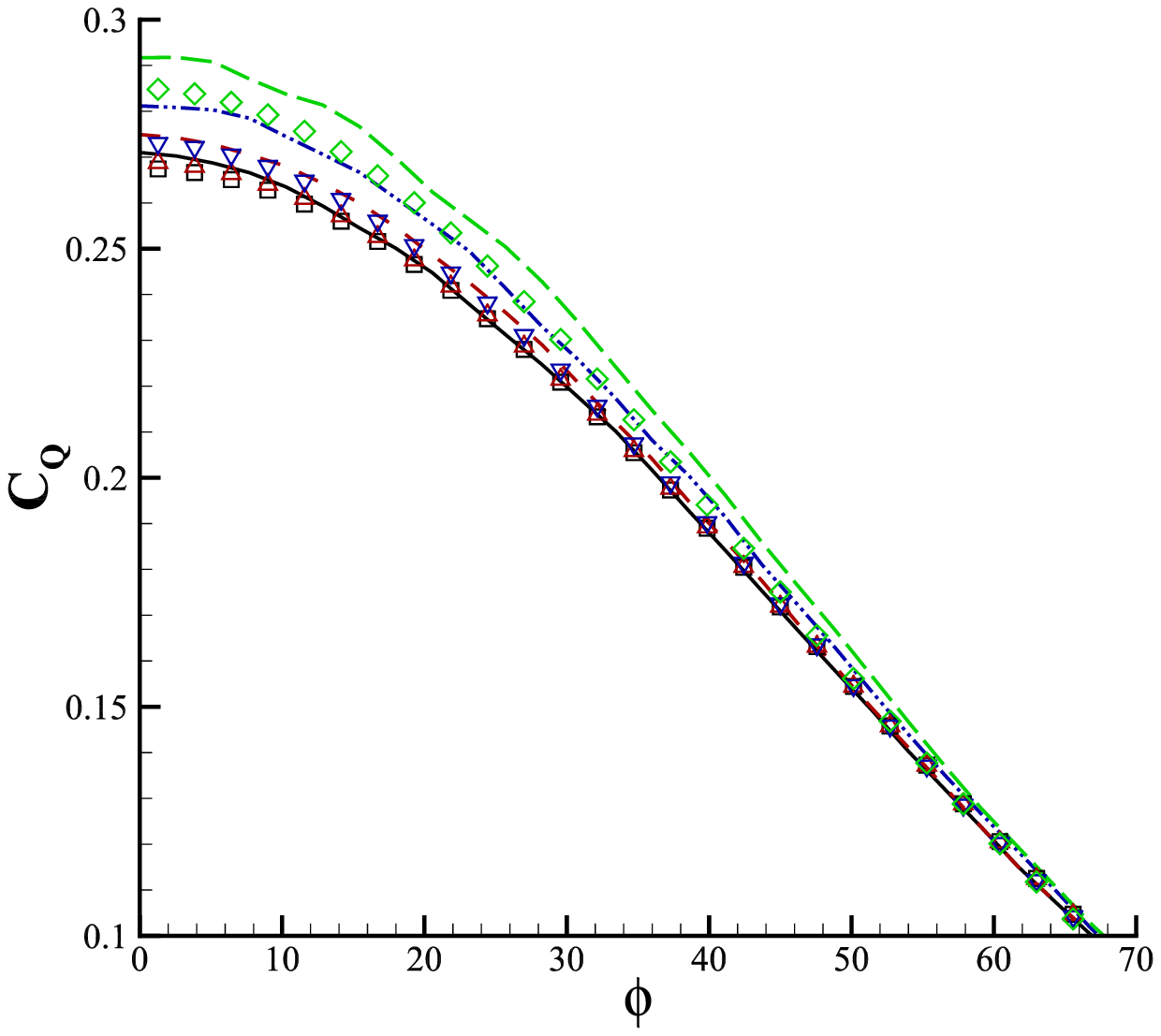}
    	}
	\caption{\label{cylinder-kn0.1_ugks} Comparisons of UGKWP with UGKS on hypersonic cylinder flow at ${\rm{Ma}}_{\infty}=5$, ${\rm{Kn}}_{\infty}=0.1$, ${\rm{Pr}}=1$: (a) Pressure coefficient at the wall $C_P$, (b) shear stress coefficient $C_F$ and heat flux coefficient $C_Q$ at the wall, (c) enlarged view of $C_F$, (d) enlarged view of $C_Q$.}
\end{figure}

A more rarefied case is conducted by increasing ${\rm{Kn}}_{\infty}$ to $0.1$. The temperature and $\rm{Kn_{GLL}}$ contours are shown in Fig.~\ref{cylinder-kn0.1-contour}. The shock wave is much thicker than in the ${\rm{Kn}}_{\infty}=0.01$ case, and the value of $\rm{Kn_{GLL}}$ is much larger. As shown in Fig.~\ref{cylinder-kn0.1_ugks}, the results of the UGKWP method match well with those of the UGKS.

\subsection{Hypersonic flow around a blunt cone at ${\rm{Ma}}_{\infty}=10.15$}\label{sec:cone}
\begin{figure}[H]
	\centering
	\subfigure[]{
			\includegraphics[width=0.45 \textwidth]{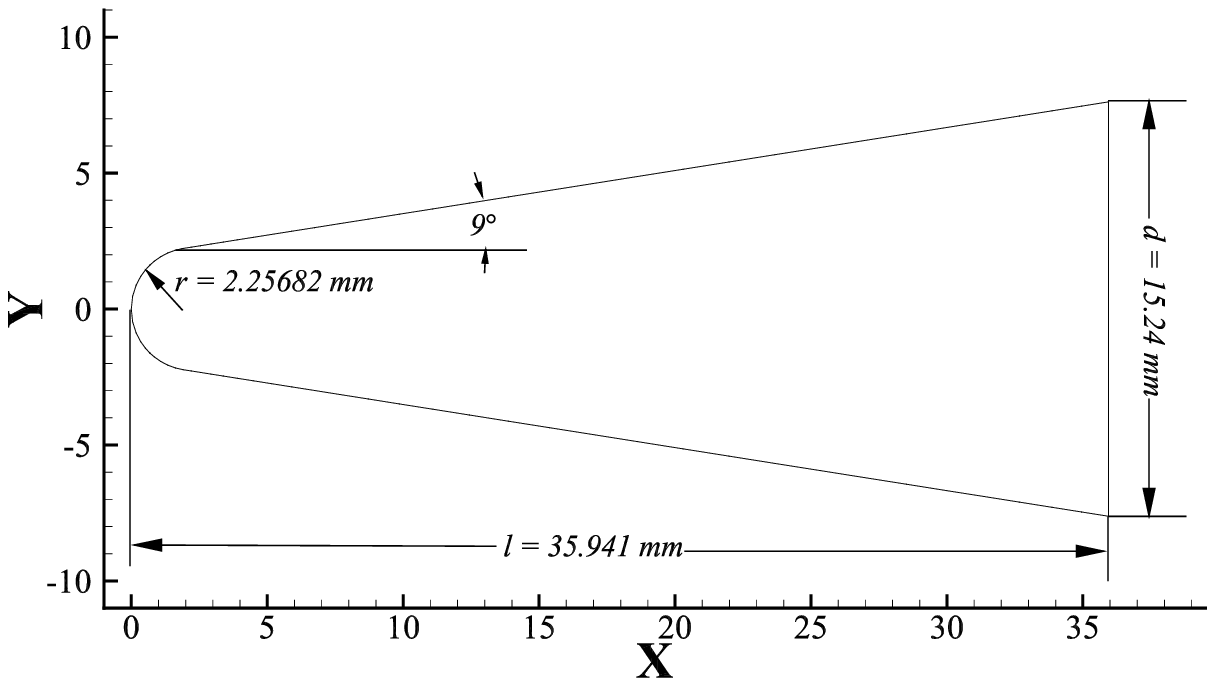}
		}
    \subfigure[]{
    		\includegraphics[width=0.45 \textwidth]{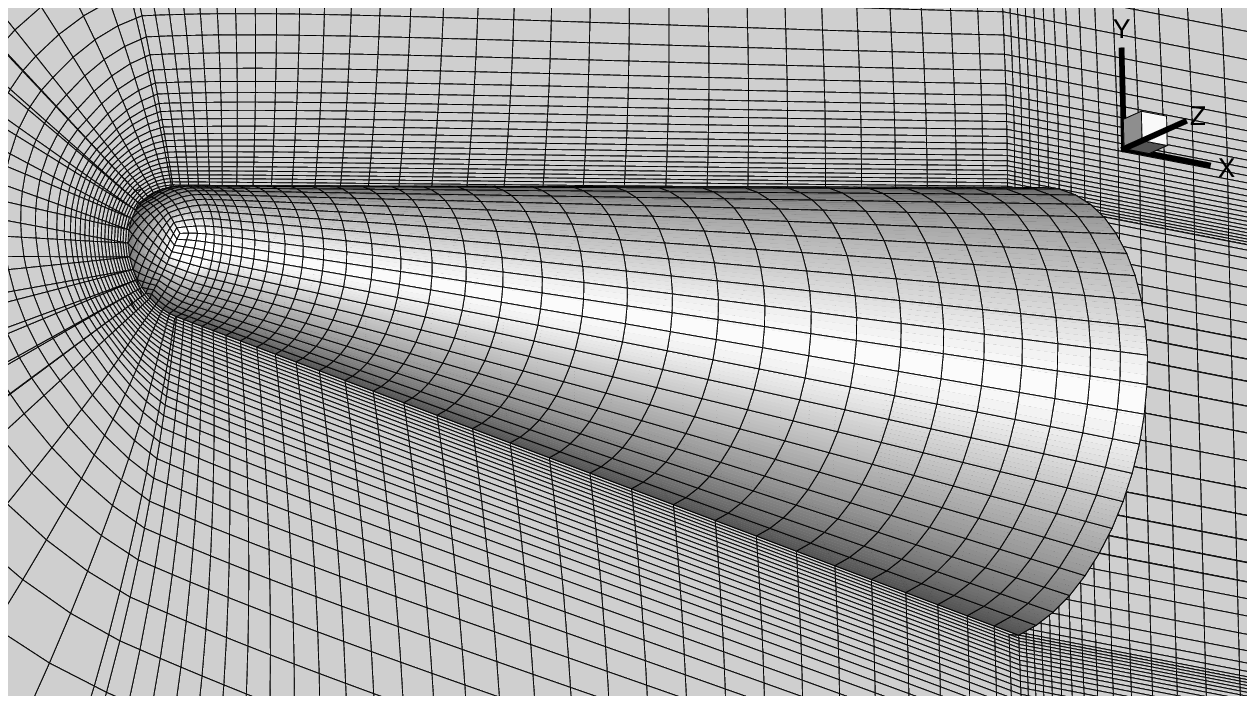}
    	}
	\caption{\label{cone-n2-mesh} (a) Geometric shape (dimensions in millimeters) and (b) mesh of the blunt cone ($N_2$, ${\rm{Ma}}_{\infty}=10.15$, ${\rm{Kn}}_{\infty,{\rm{HS}}}=0.065$, $H\approx93{\rm{km}}$).}
\end{figure}

\begin{figure}[H]
	\centering
	\subfigure[]{
			\includegraphics[width=0.22 \textwidth]{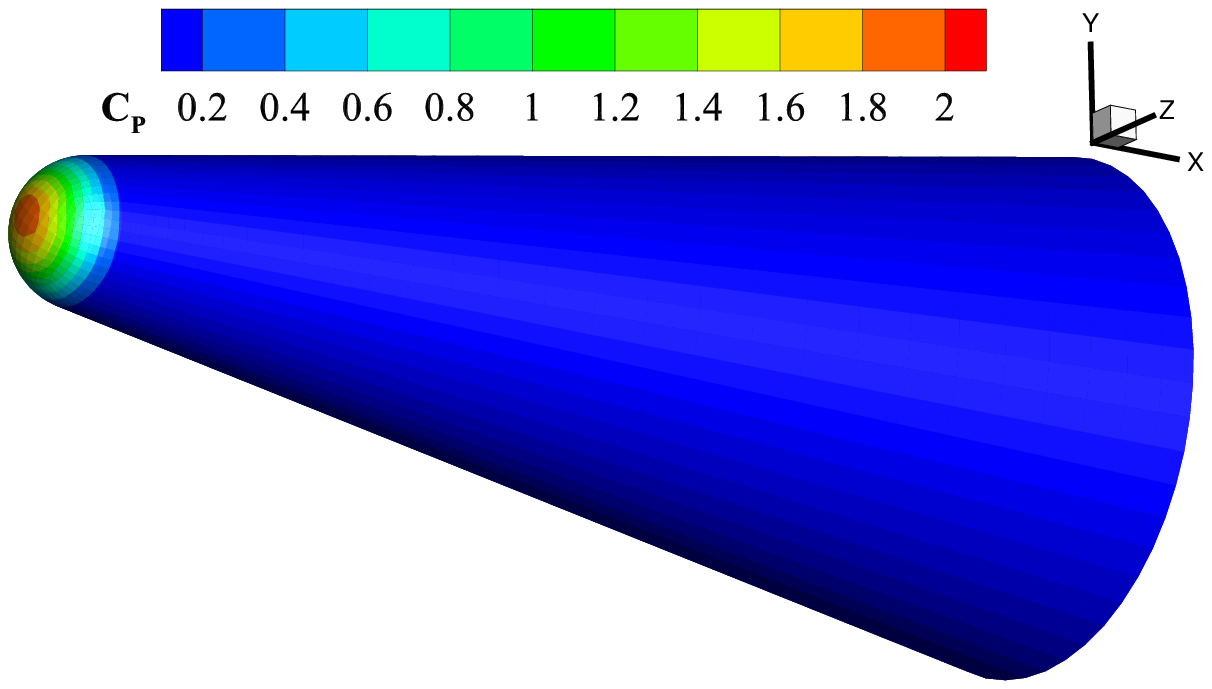}
		}
    \subfigure[]{
    		\includegraphics[width=0.22 \textwidth]{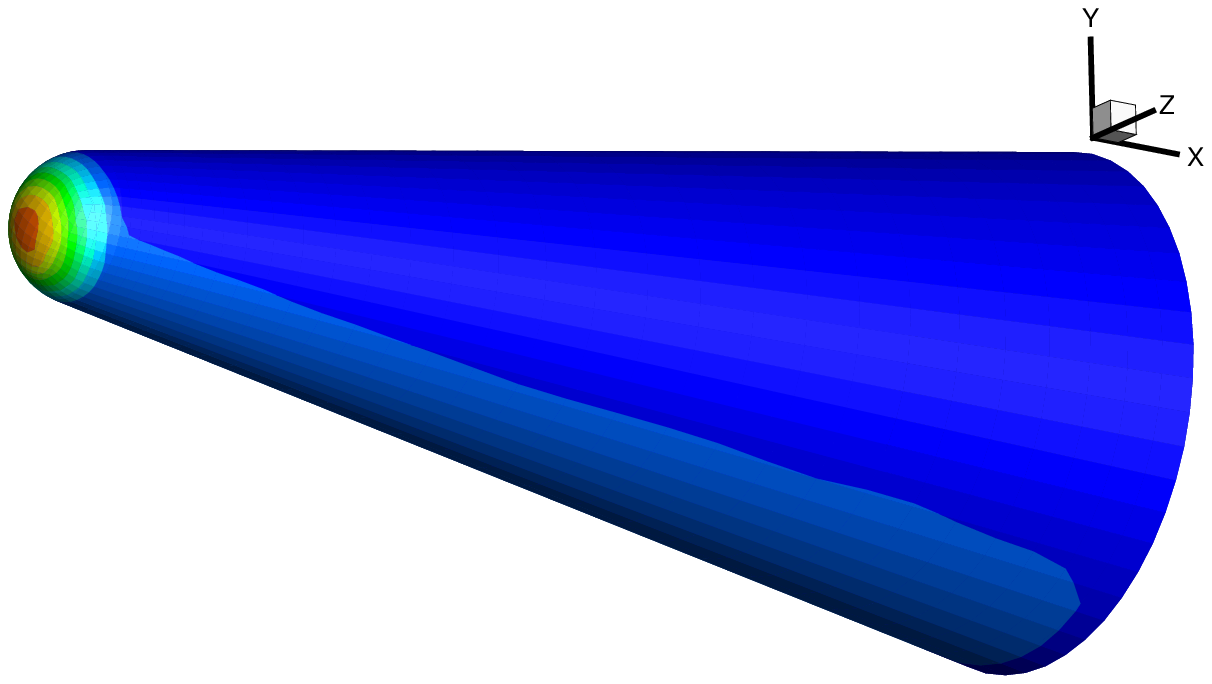}
    	}
    \subfigure[]{
			\includegraphics[width=0.22 \textwidth]{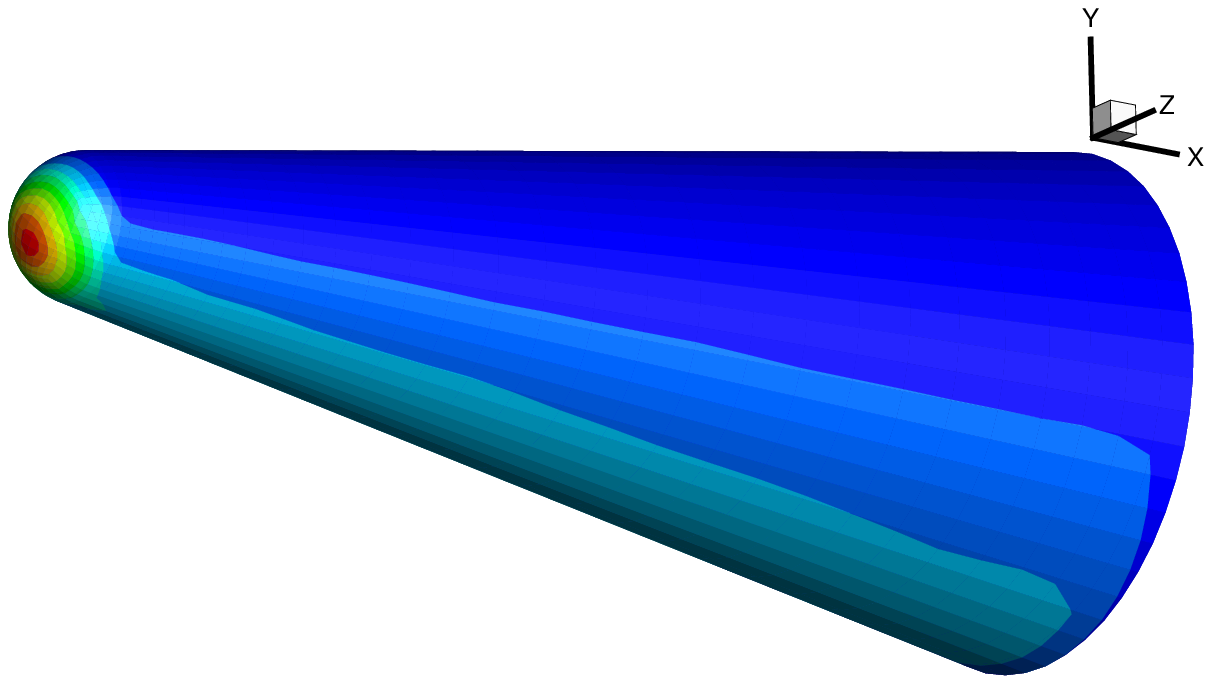}
		}
    \subfigure[]{
    		\includegraphics[width=0.22 \textwidth]{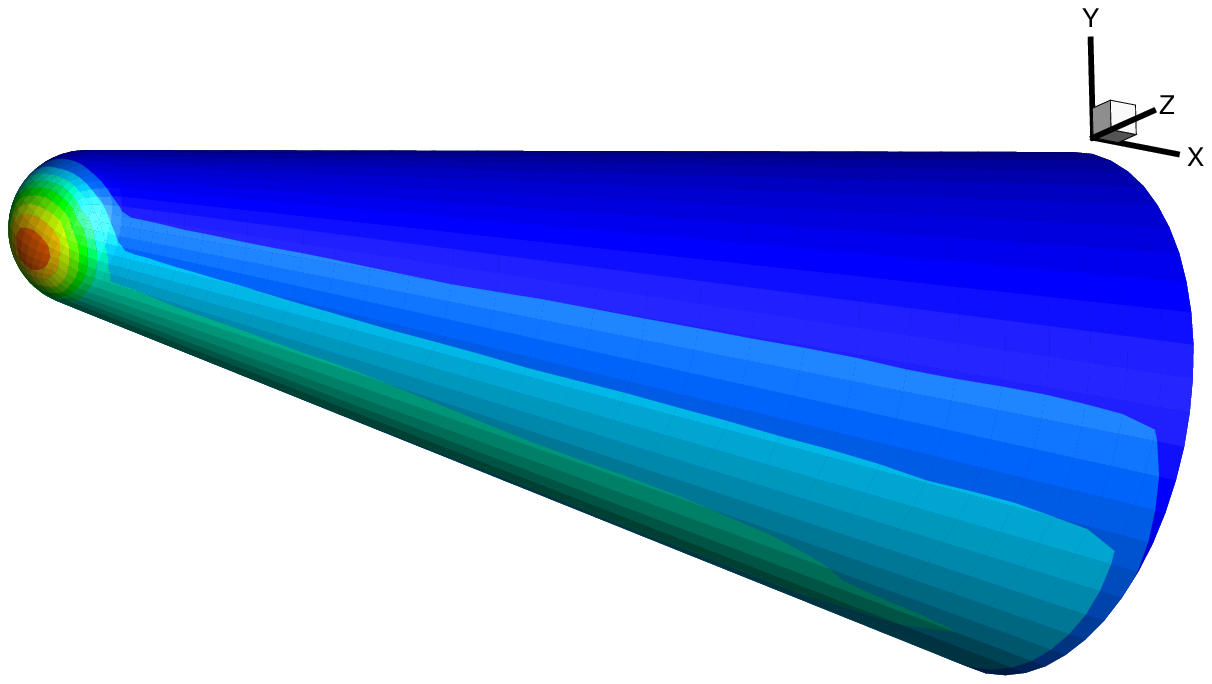}
    	}
	\caption{\label{cone-n2-cp} $C_P$ contours on the surface of hypersonic flow over a blunt cone at four angles of attack ($N_2$, ${\rm{Ma}}_{\infty}=10.15$, ${\rm{Kn}}_{{\rm{HS}},\infty}=0.065$, $H\approx93{\rm{km}}$): (a) $0^{\circ}$, (b) $10^{\circ}$, (c) $20^{\circ}$, (d) $25^{\circ}$.}
\end{figure}

\begin{figure}[H]
	\centering
	\subfigure[]{
			\includegraphics[width=0.22 \textwidth]{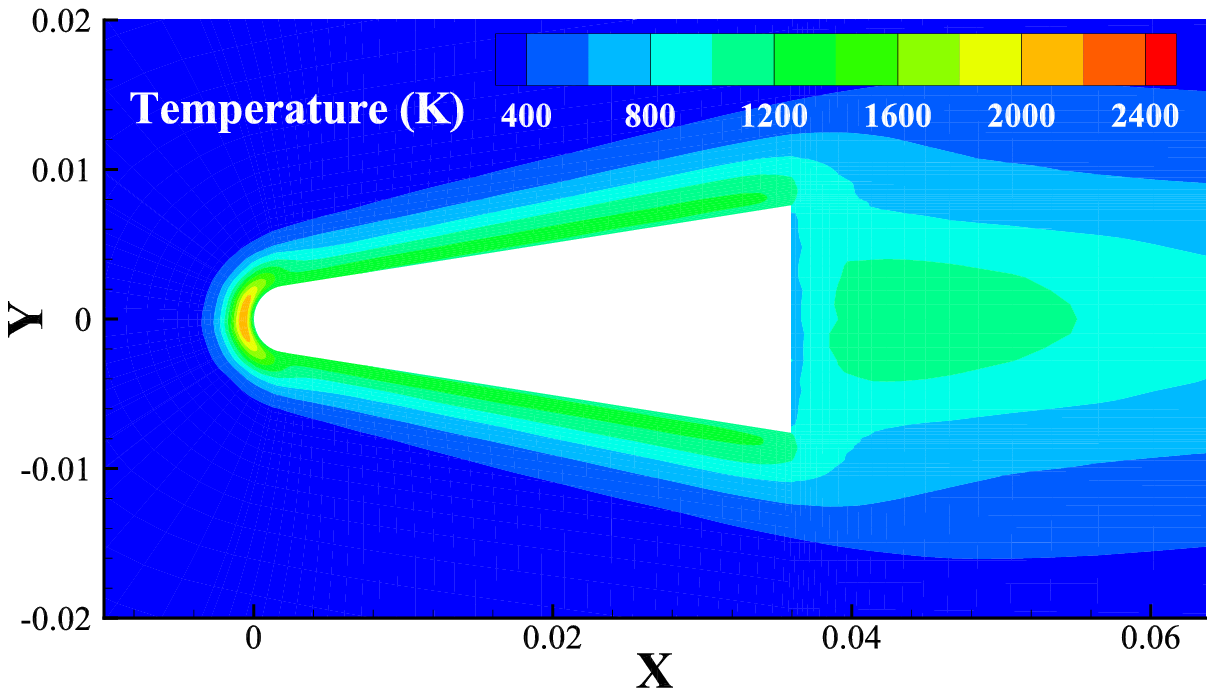}
		}
    \subfigure[]{
    		\includegraphics[width=0.22 \textwidth]{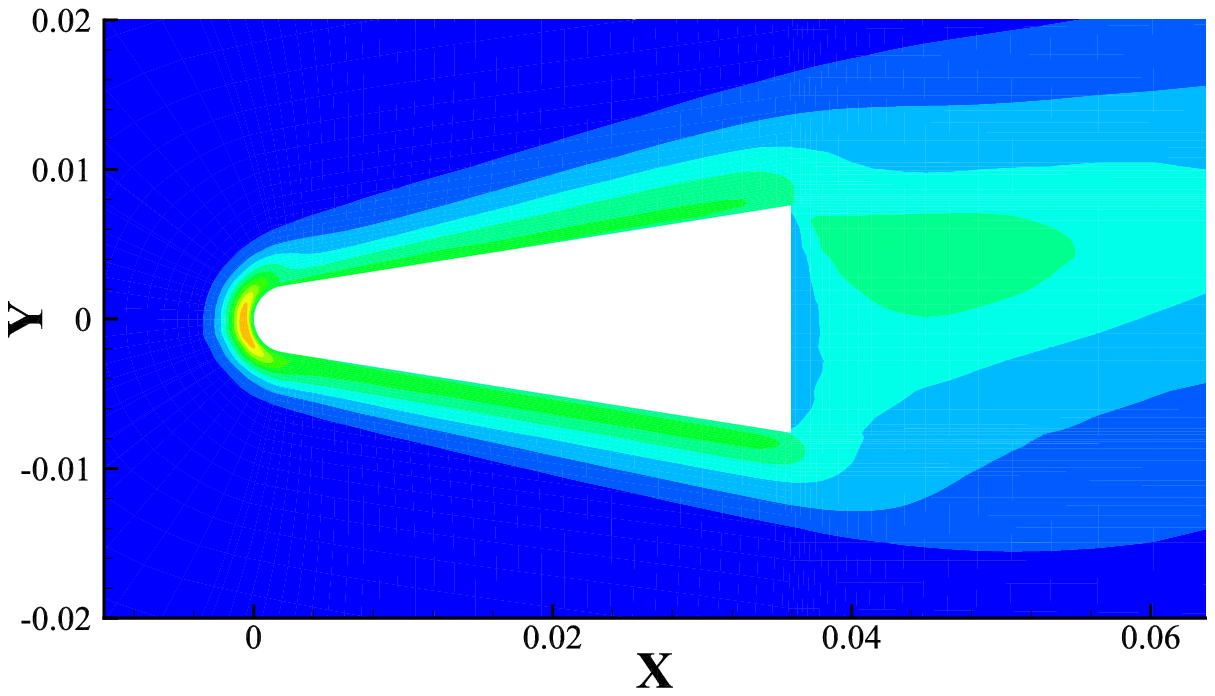}
    	}
    \subfigure[]{
			\includegraphics[width=0.22 \textwidth]{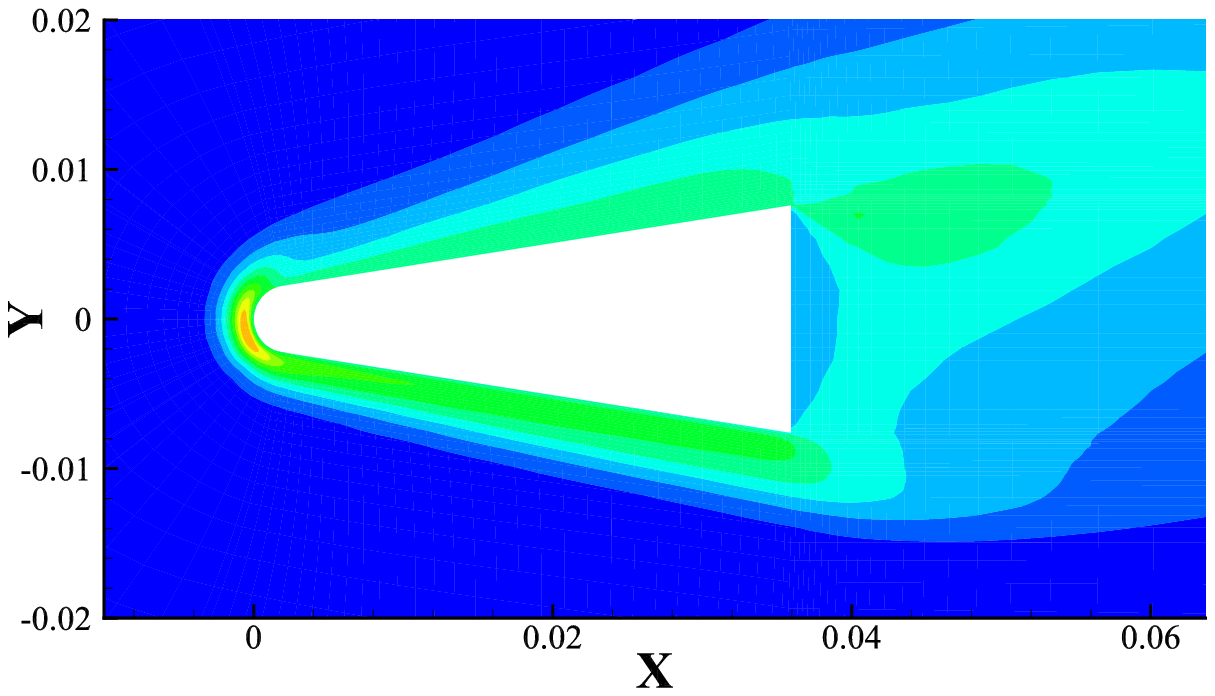}
		}
    \subfigure[]{
    		\includegraphics[width=0.22 \textwidth]{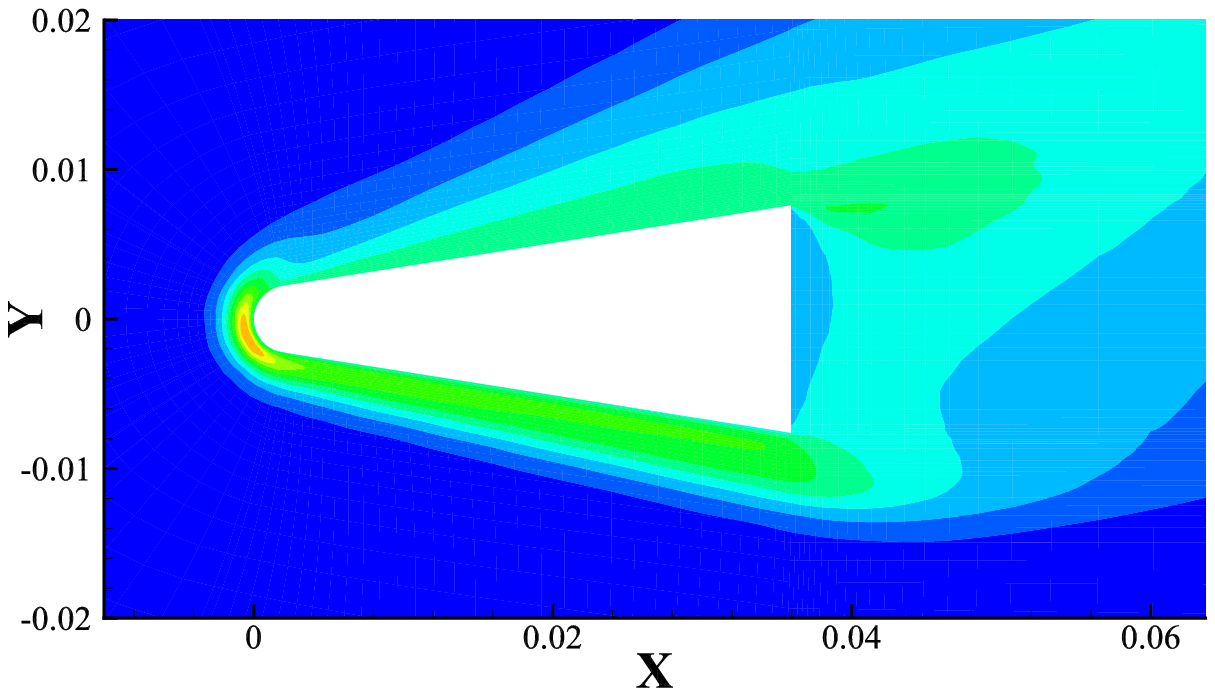}
    	}
	\caption{\label{cone-n2-t} Temperature contours at the symmetry plane of hypersonic flow over a blunt cone at four angles of attack ($N_2$, ${\rm{Ma}}_{\infty}=10.15$, ${\rm{Kn}}_{{\rm{HS}},\infty}=0.065$, $H\approx93{\rm{km}}$, dimensions in meters): (a) $0^{\circ}$, (b) $10^{\circ}$, (c) $20^{\circ}$, (d) $25^{\circ}$.}
\end{figure}

\begin{figure}[H]
	\centering
	\subfigure[]{
			\includegraphics[width=0.30 \textwidth]{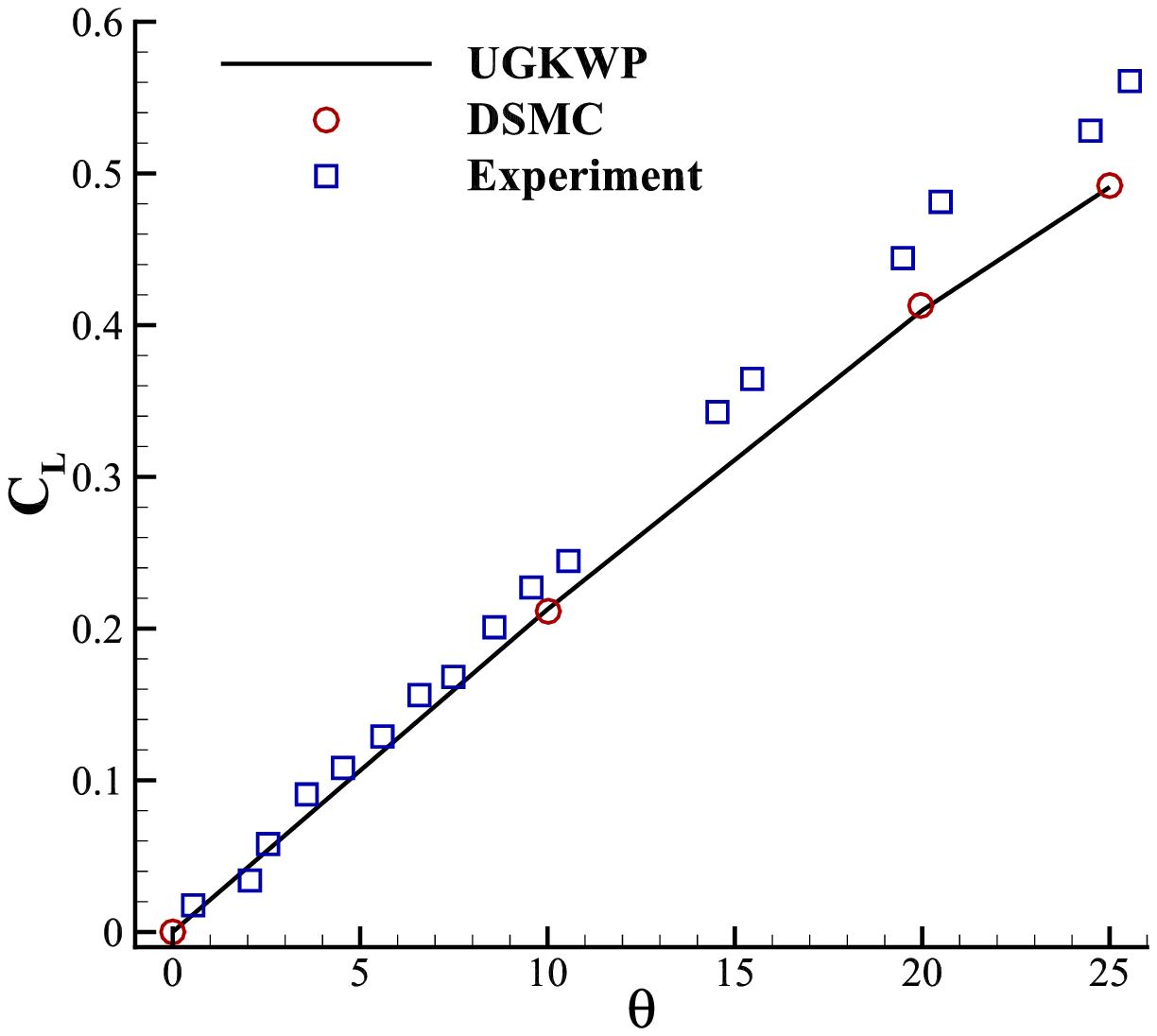}
		}
    \subfigure[]{
    		\includegraphics[width=0.30 \textwidth]{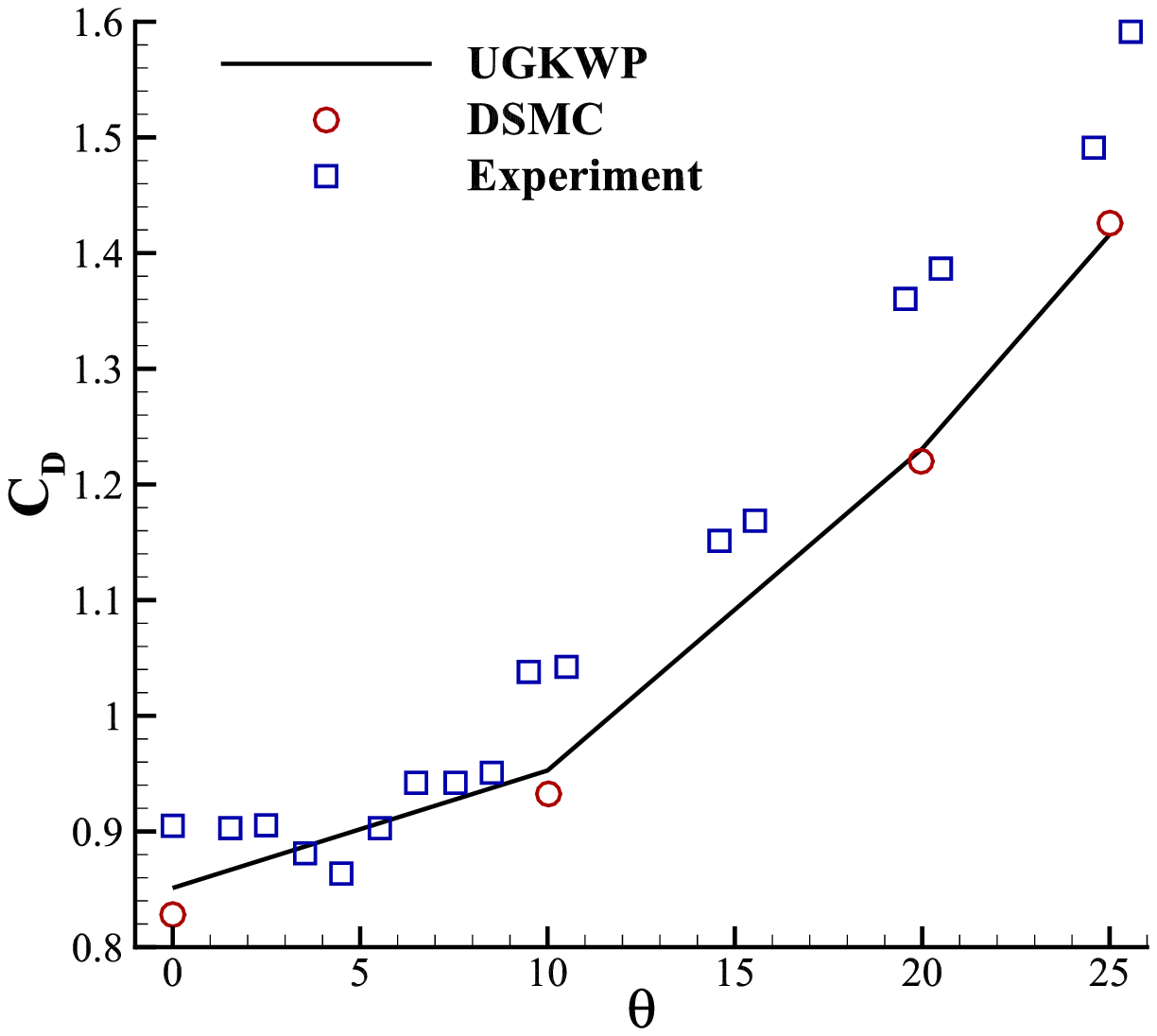}
    	}
    \subfigure[]{
			\includegraphics[width=0.30 \textwidth]{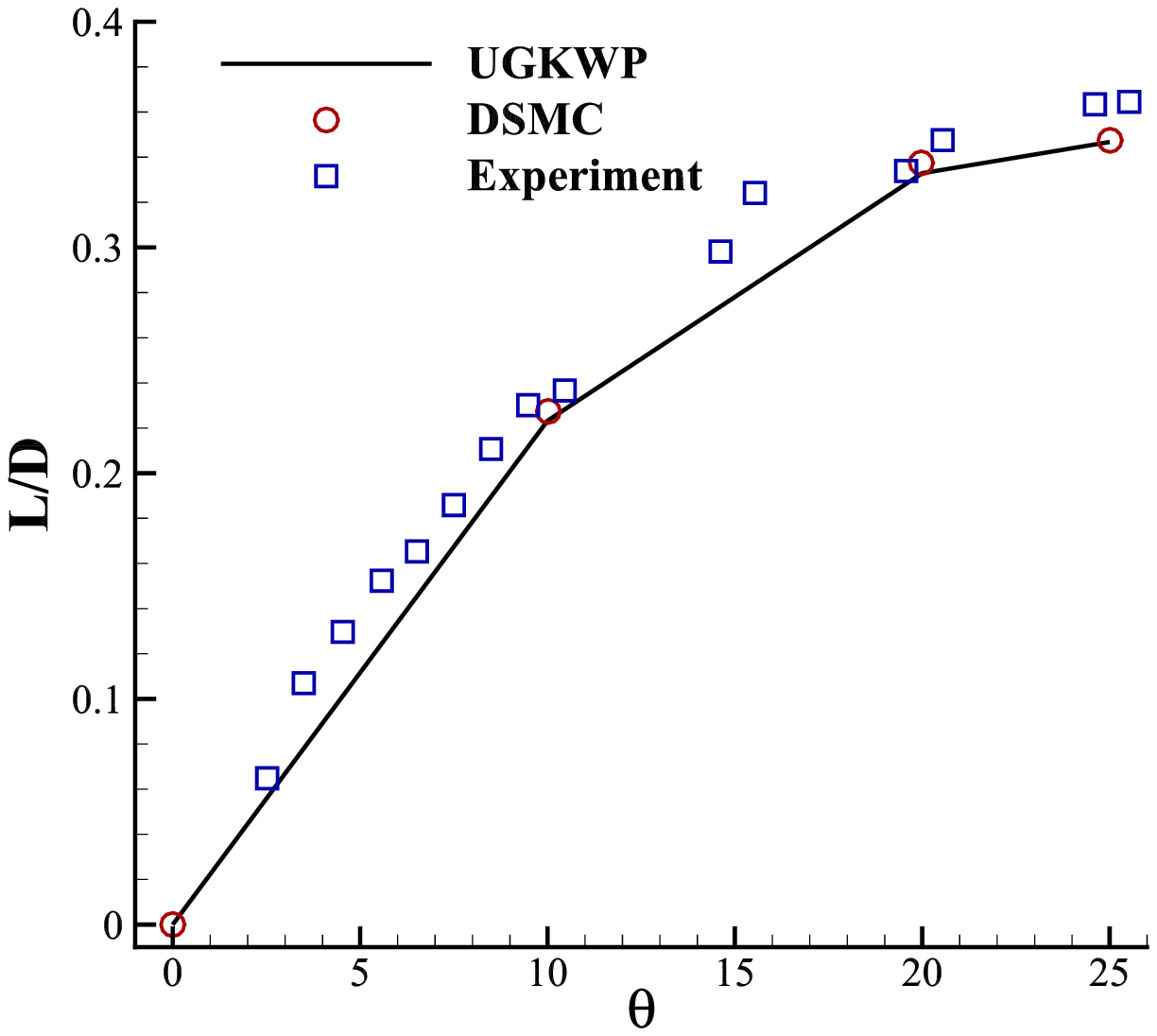}
		}
	\caption{\label{cone-n2-aero} Aerodynamic force coefficients of hypersonic flow over a blunt cone ($N_2$, ${\rm{Ma}}_{\infty}=10.15$, ${\rm{Kn}}_{{\rm{HS}},\infty}=0.065$, $H\approx93{\rm{km}}$): (a) $C_L$, (b) $C_D$, (c) $L/D$.}
\end{figure}

In this section, three-dimensional cases are considered to assess the mesh independence of the UGKWP method for overall aerodynamic force coefficients, including the lift coefficient ($C_L$), drag coefficient ($C_D$), and lift-to-drag ratio ($L/D$). The results are also compared with the DSMC method. The geometric shape of the blunt cone is taken from Refs.~\cite{cone-exp,cone-dsmc,cone-zr}, as shown in Fig.~\ref{cone-n2-mesh} along with the mesh. When calculating aerodynamic force coefficients and farfield dimensionless quantities, the reference length is set to the base diameter, the reference area is set to the base area, and the moment reference center is set to the origin. First, the experimental condition~\cite{cone-exp} is selected for validation of the solver and surface mesh: ${\rm{Ma}}_{\infty}=10.15$, ${\rm{Kn}}_{{\rm{HS}},\infty}=0.065$, $T_{\infty}=143.5{\rm{K}}$, and $T_{w}=600{\rm{K}}$. The subscript ``${\rm{HS}}$'' in ${\rm{Kn}}_{{\rm{HS}},\infty}$ indicates that the hard sphere (HS) model is used to calculate $\lambda_{\rm{HS}}$. Equation~\eqref{eq:lambda} is used with $\omega=0.5$. The gas is nitrogen ($N_2$) with $D=2$ and ${\rm{Pr}}=0.72$, and the viscosity coefficient is calculated using the power-law model in the VHS model with $\omega=0.74$. According to Ref.~\cite{h1976}, if the base diameter of the actual vehicle is $0.4{\rm{m}}$, the corresponding altitude is approximately $93{\rm{km}}$. For the numerical simulation, the height of the first layer mesh is set to $0.0001{\rm{m}}$ and $N^{hp,{\rm{ref}}}$ is set to $100$. The $C_P$ contours at the surface and temperature contours at the symmetry plane are shown in Figs.~\ref{cone-n2-cp} and \ref{cone-n2-t}, respectively. For the aerodynamic force coefficients at different angles of attack, the UGKWP results match well with the DSMC~\cite{cone-dsmc} and experimental data~\cite{cone-exp}, as shown in Fig.~\ref{cone-n2-aero}.

\begin{figure}[H]
	\centering
	\subfigure[]{
			\includegraphics[width=0.45 \textwidth]{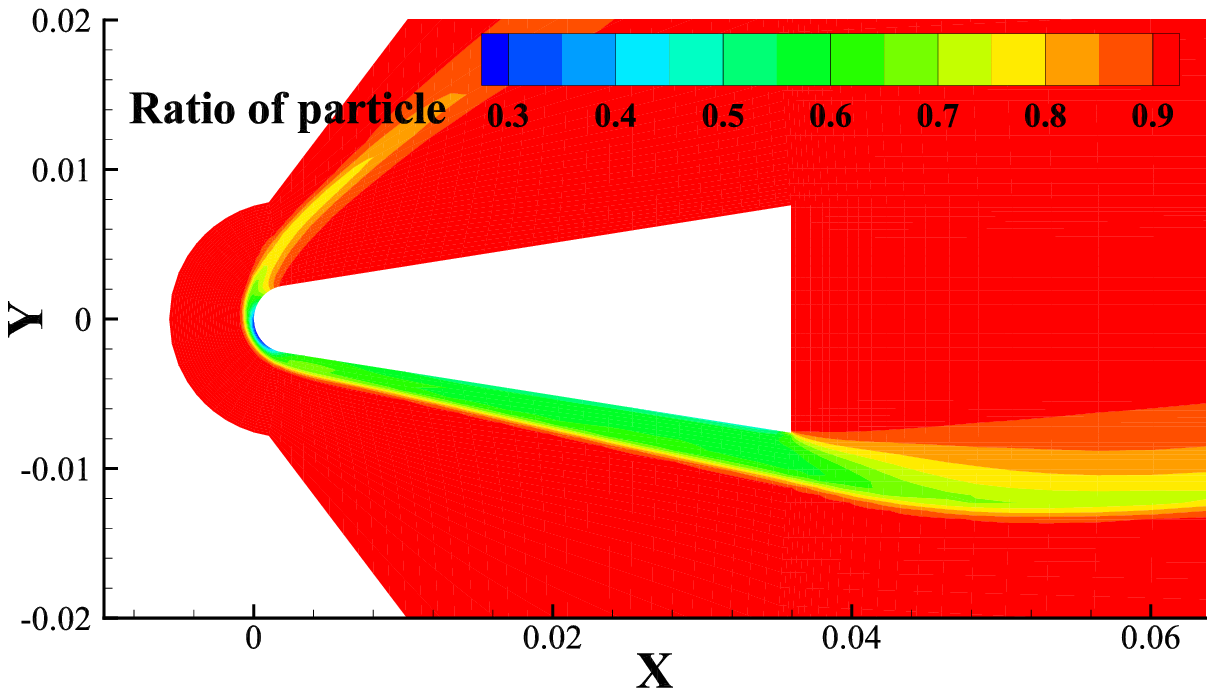}
		}
    \subfigure[]{
    		\includegraphics[width=0.45 \textwidth]{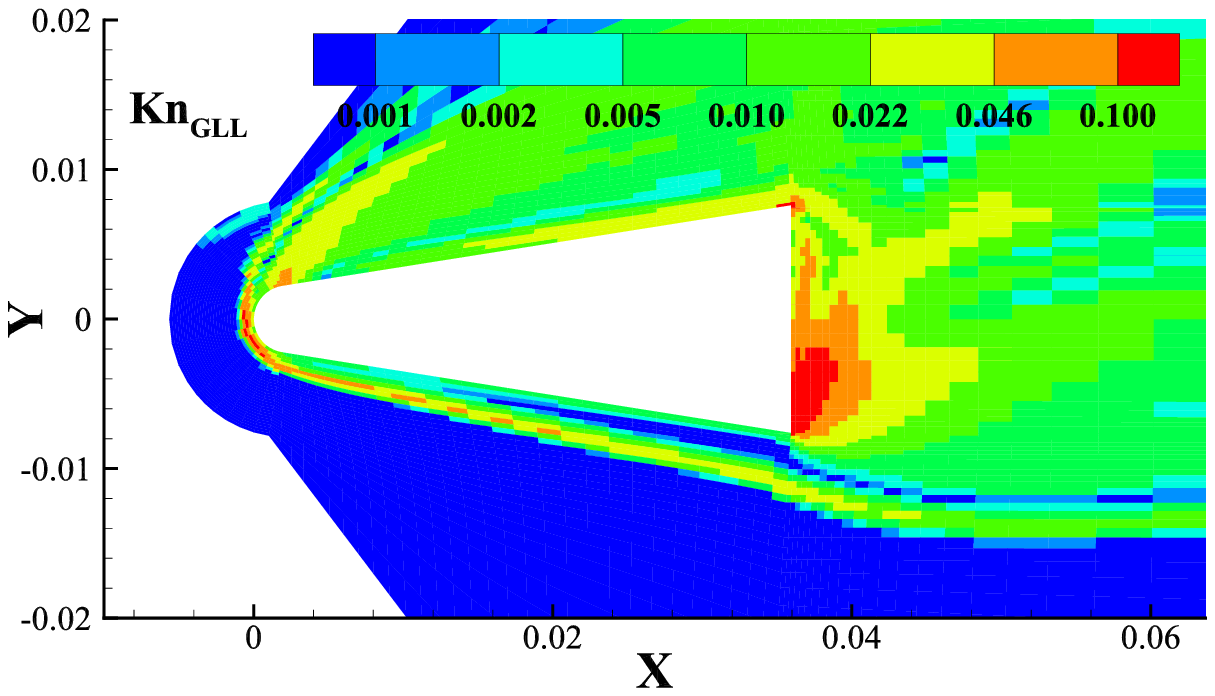}
    	}
	\caption{\label{cone-ar-scale} Contours at the symmetry plane of hypersonic flow over a blunt cone ($Ar$, ${\rm{Ma}}_{\infty}=10.15$, ${\rm{Kn}}_{{\rm{HS}},\infty}=0.002$, $H\approx70{\rm{km}}$, angle of attack $25^{\circ}$, $h=0.0002{\rm{m}}$): (a) Ratio of particles, (b) $\rm{Kn_{GLL}}$.}
\end{figure}

\begin{figure}[H]
	\centering
	\subfigure[]{
			\includegraphics[width=0.22 \textwidth]{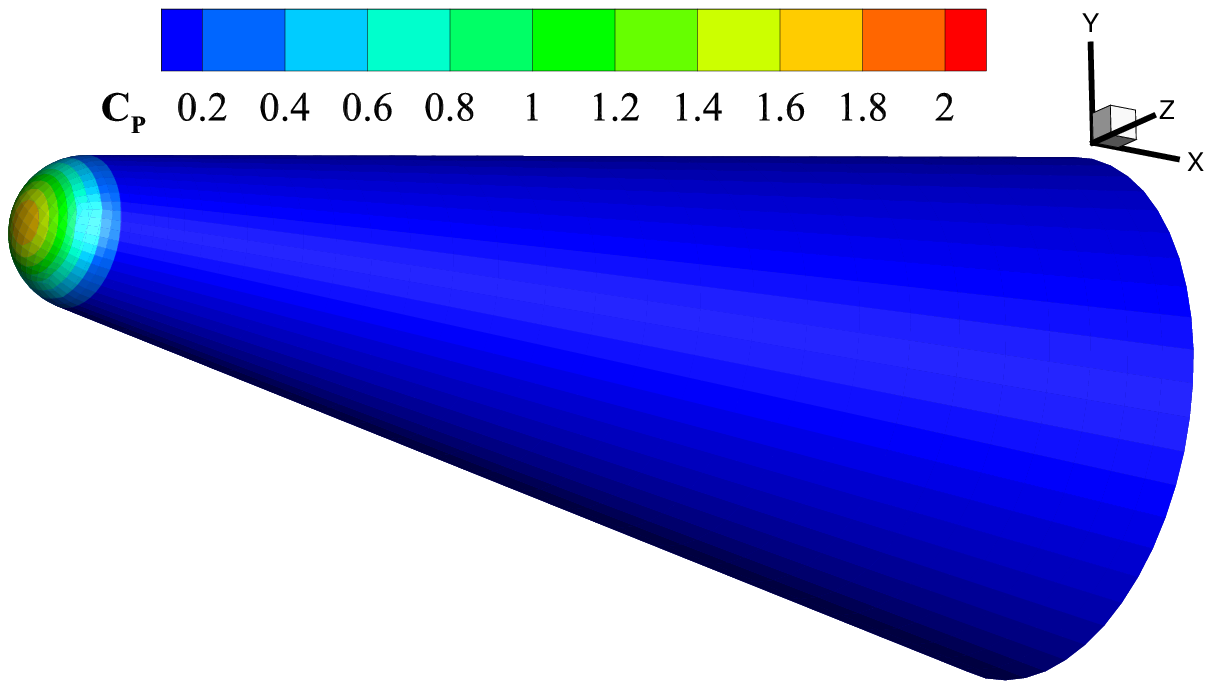}
		}
    \subfigure[]{
    		\includegraphics[width=0.22 \textwidth]{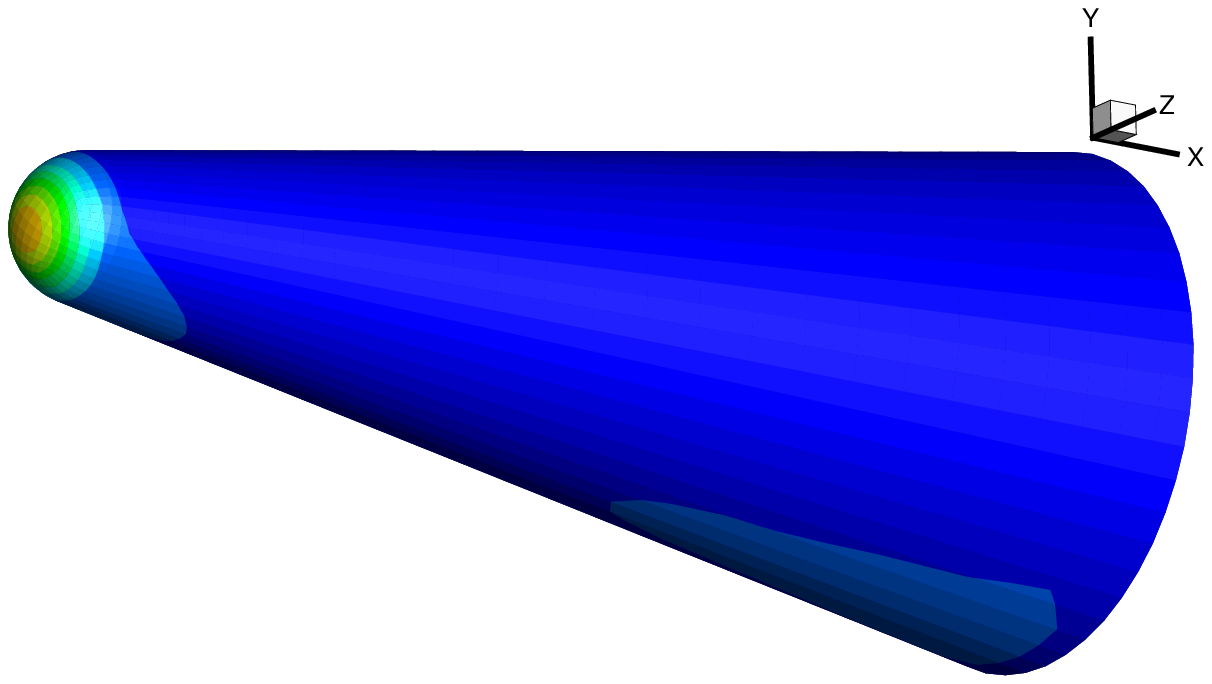}
    	}
    \subfigure[]{
			\includegraphics[width=0.22 \textwidth]{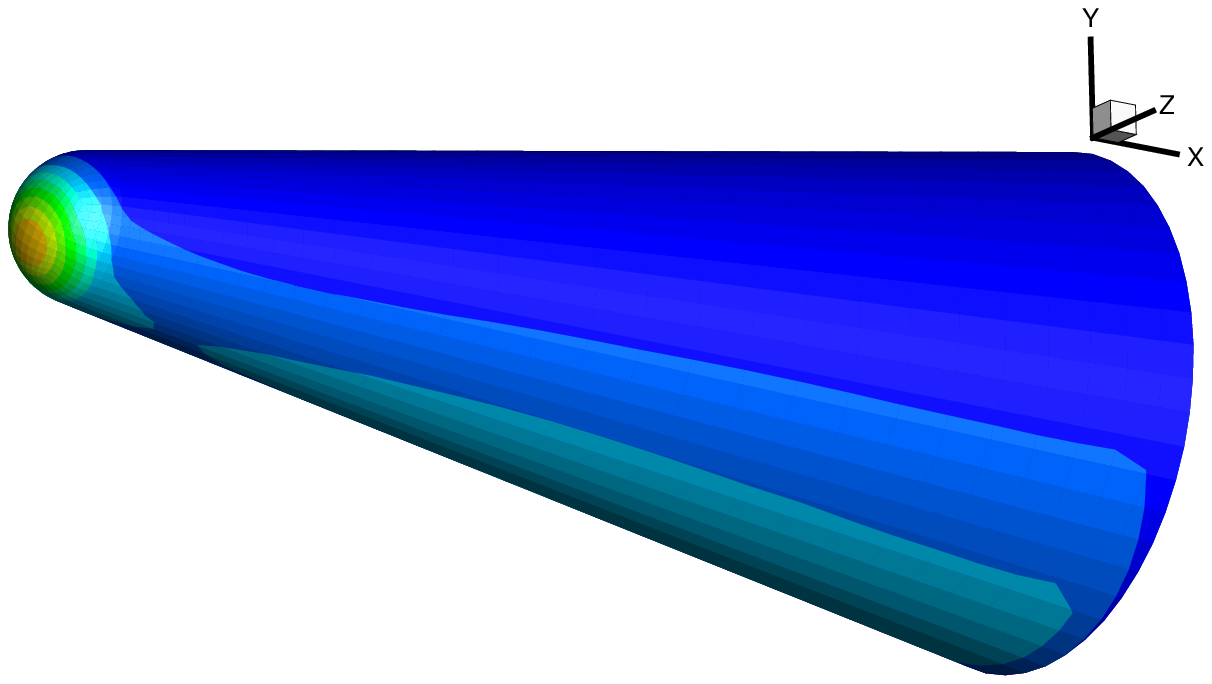}
		}
    \subfigure[]{
    		\includegraphics[width=0.22 \textwidth]{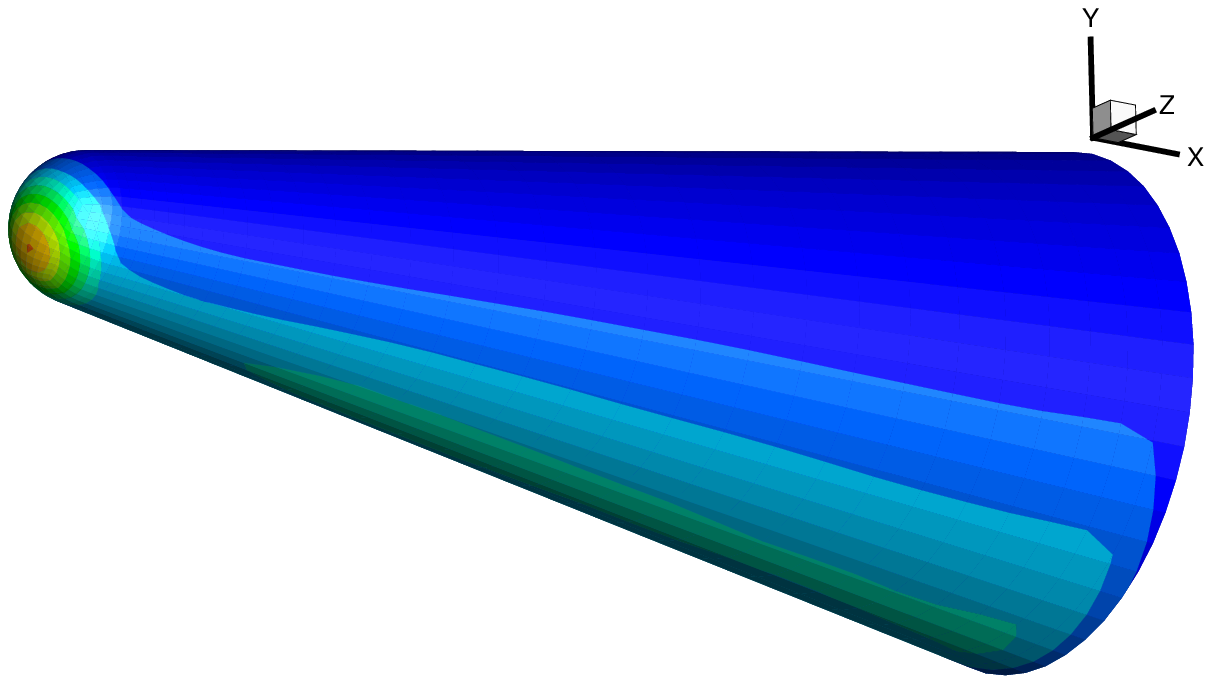}
    	}
	\caption{\label{cone-ar-cp} $C_P$ contours on the surface of hypersonic flow over a blunt cone at four angles of attack ($Ar$, ${\rm{Ma}}_{\infty}=10.15$, ${\rm{Kn}}_{{\rm{HS}},\infty}=0.002$, $H\approx70{\rm{km}}$): (a) $0^{\circ}$, (b) $10^{\circ}$, (c) $20^{\circ}$, (d) $25^{\circ}$.}
\end{figure}

\begin{figure}[H]
	\centering
	\subfigure[]{
			\includegraphics[width=0.22 \textwidth]{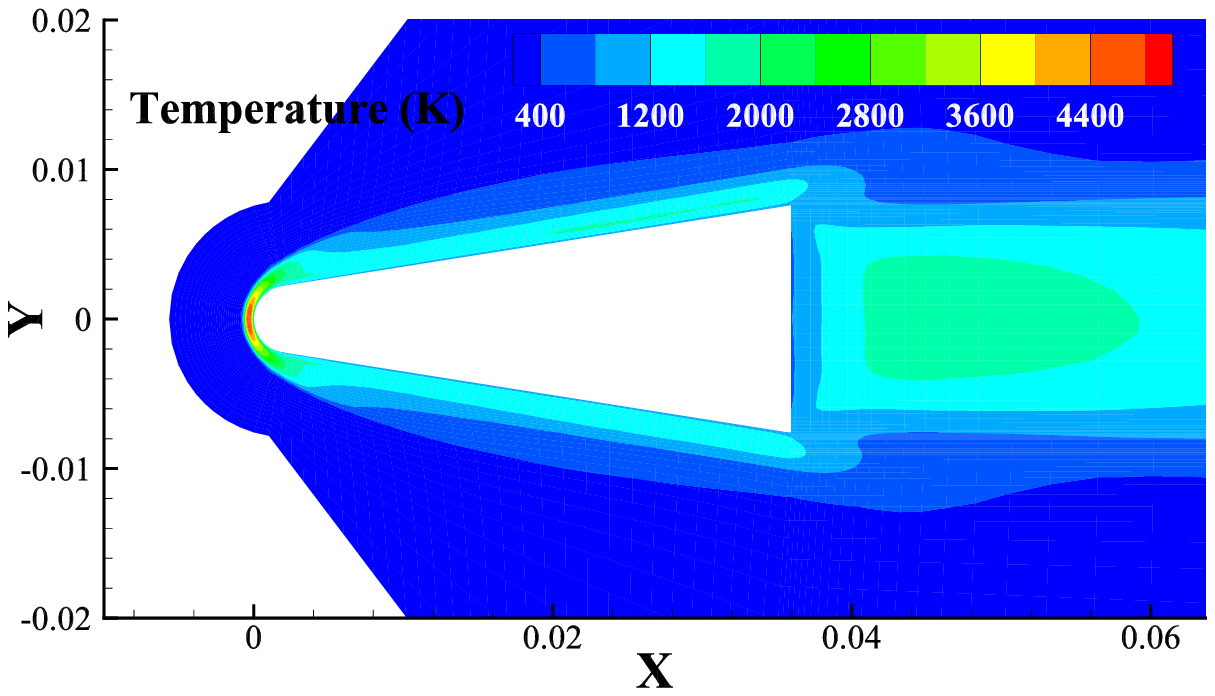}
		}
    \subfigure[]{
    		\includegraphics[width=0.22 \textwidth]{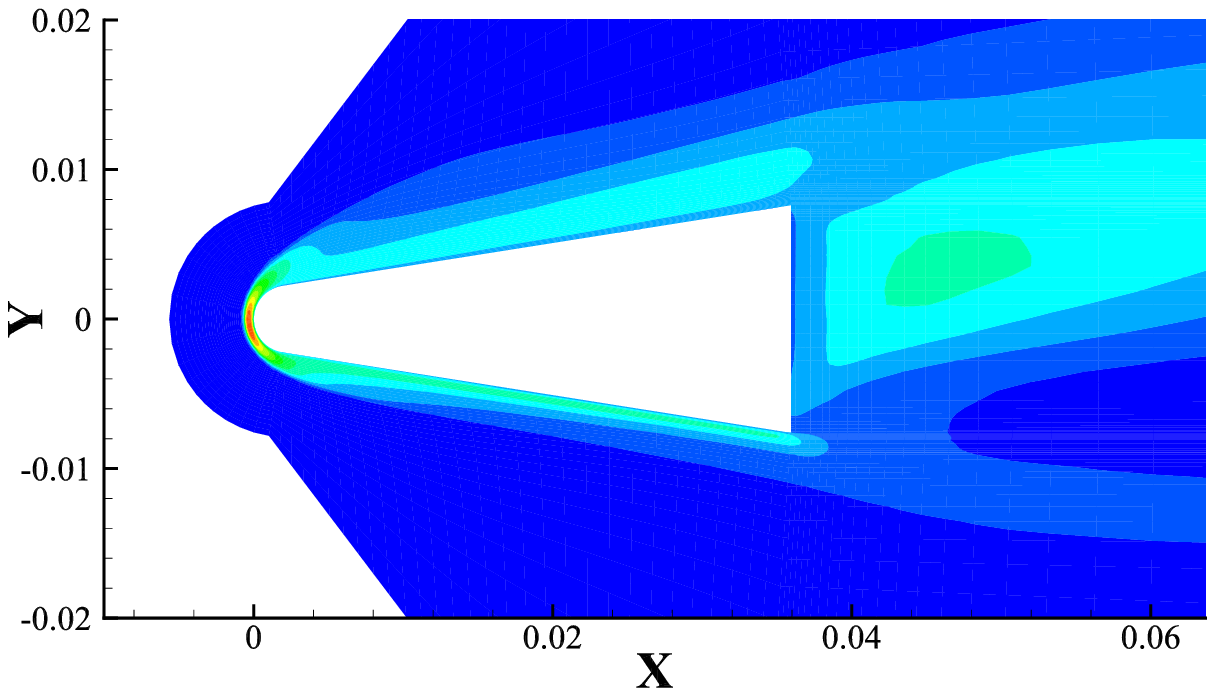}
    	}
    \subfigure[]{
			\includegraphics[width=0.22 \textwidth]{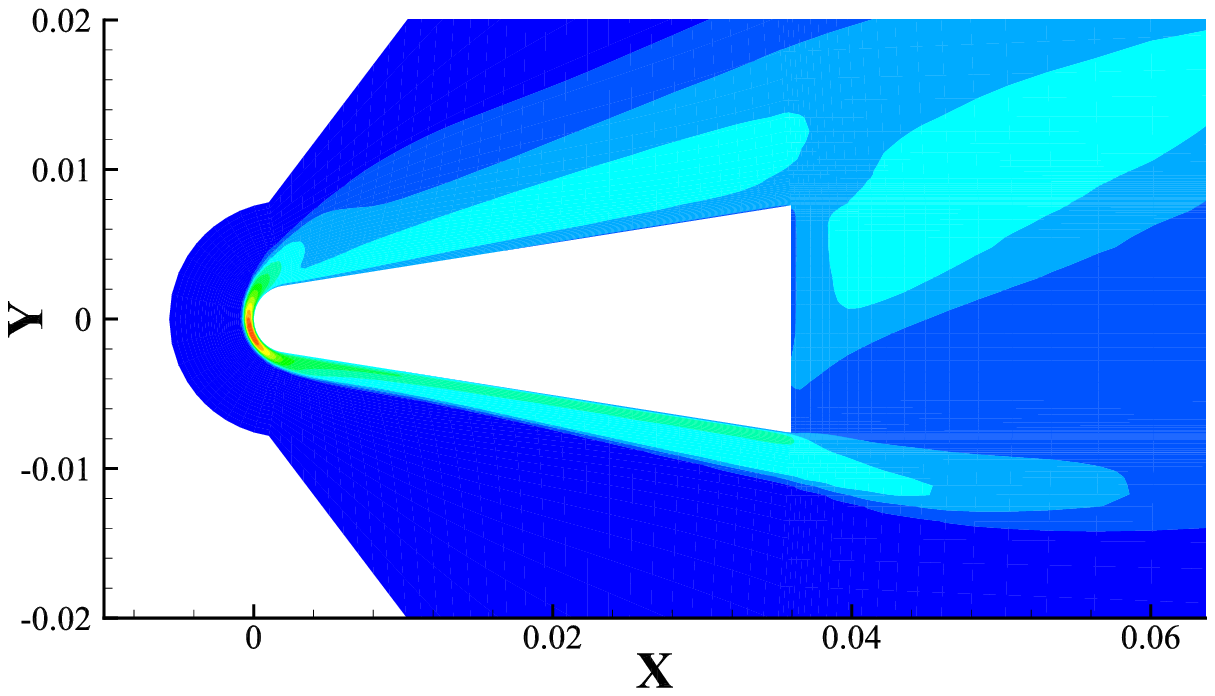}
		}
    \subfigure[]{
    		\includegraphics[width=0.22 \textwidth]{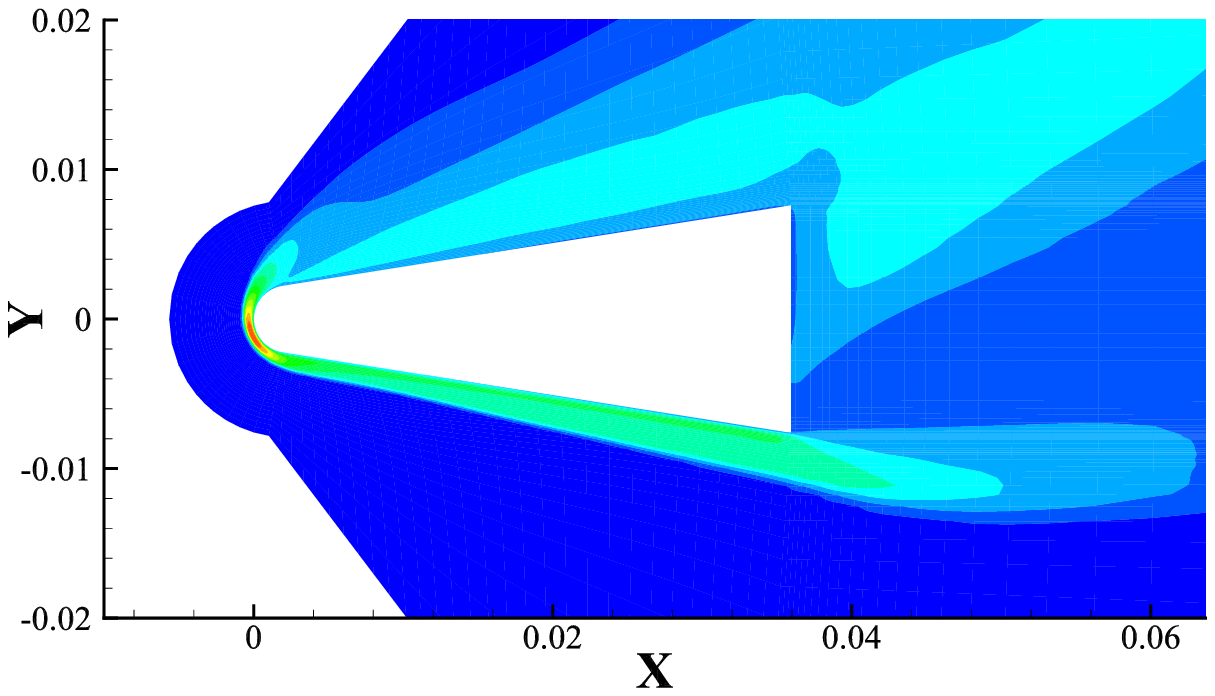}
    	}
	\caption{\label{cone-ar-t} Temperature contours at the symmetry plane of hypersonic flow over a blunt cone at four angles of attack ($Ar$, ${\rm{Ma}}_{\infty}=10.15$, ${\rm{Kn}}_{{\rm{HS}},\infty}=0.002$, $H\approx70{\rm{km}}$, dimensions in meters): (a) $0^{\circ}$, (b) $10^{\circ}$, (c) $20^{\circ}$, (d) $25^{\circ}$.}
\end{figure}

\begin{figure}[H]
	\centering
	\subfigure[]{
			\includegraphics[width=0.30 \textwidth]{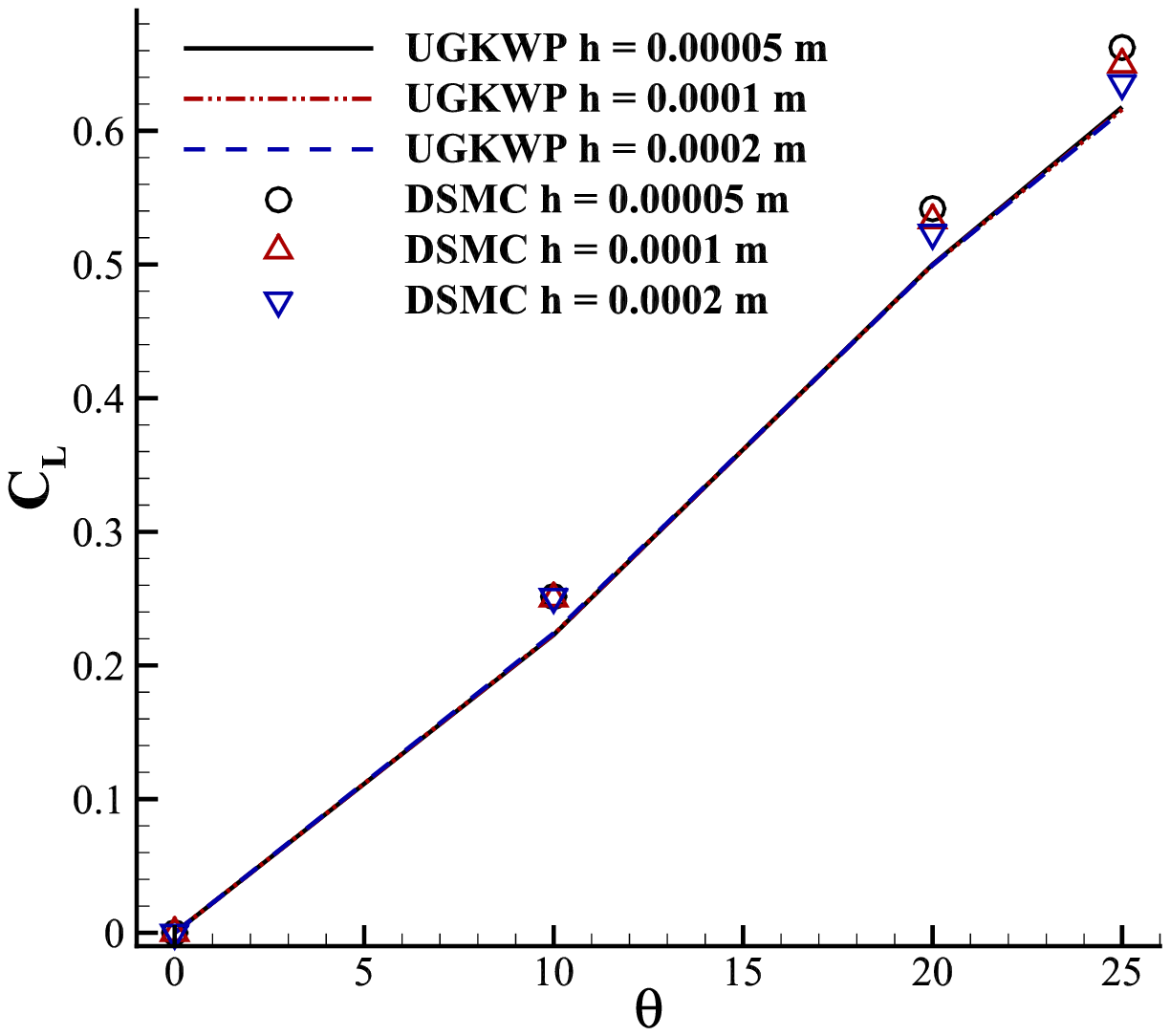}
		}
    \subfigure[]{
    		\includegraphics[width=0.30 \textwidth]{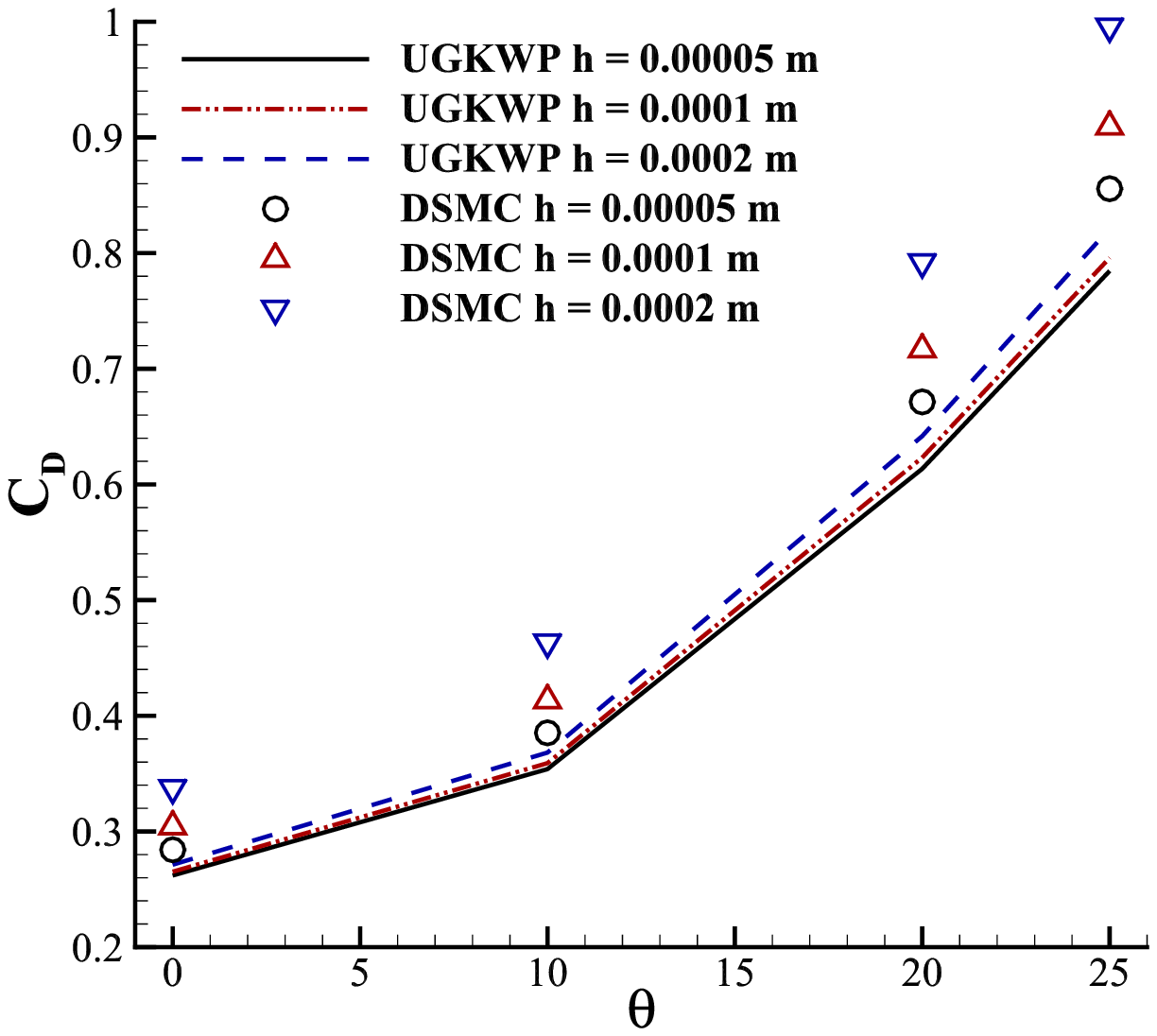}
    	}
    \subfigure[]{
			\includegraphics[width=0.30 \textwidth]{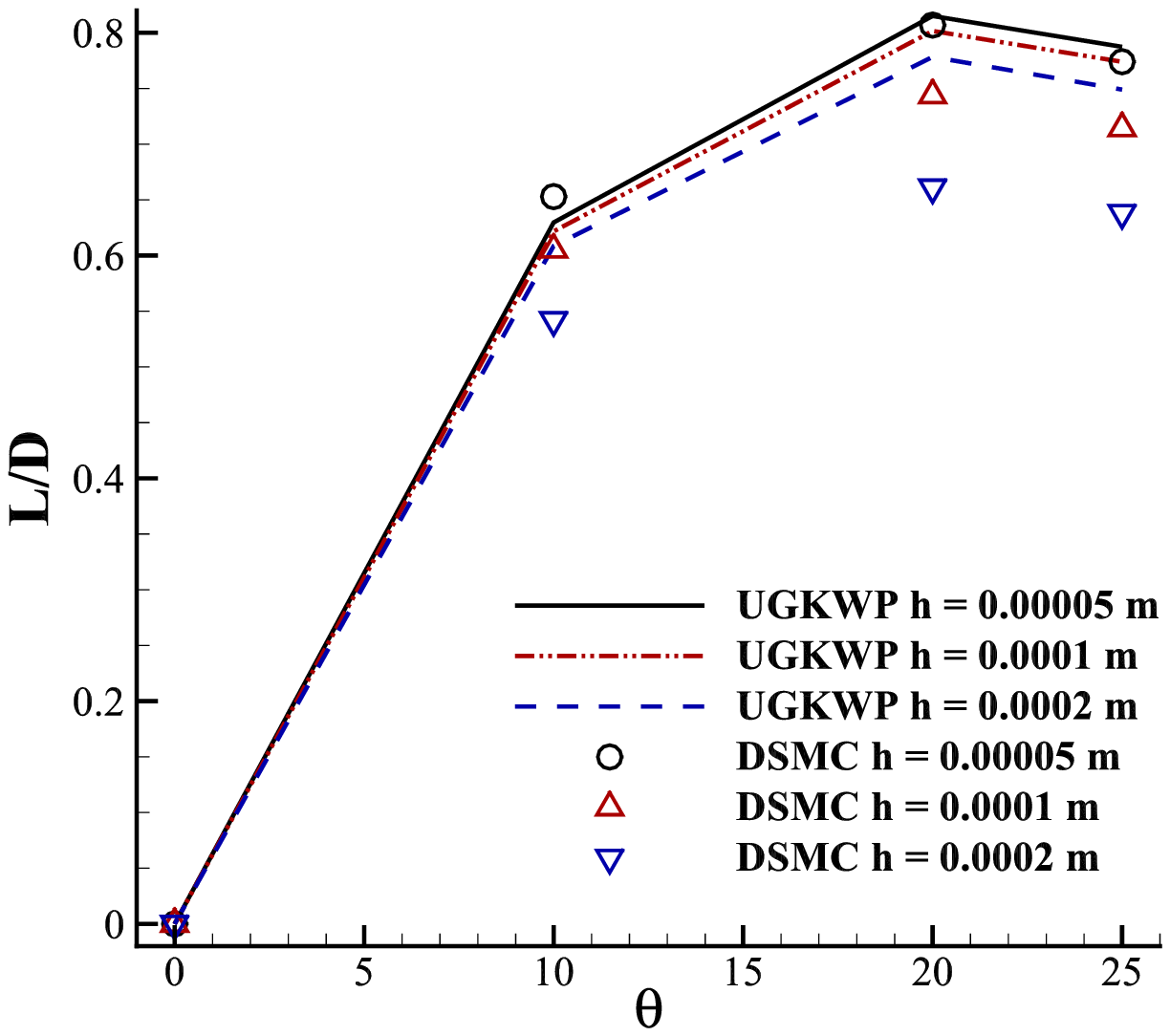}
		}
	\caption{\label{cone-ar-aero} Aerodynamic force coefficients of hypersonic flow over a blunt cone ($Ar$, ${\rm{Ma}}_{\infty}=10.15$, ${\rm{Kn}}_{{\rm{HS}},\infty}=0.002$, $H\approx70{\rm{km}}$): (a) $C_L$, (b) $C_D$, (c) $L/D$.}
\end{figure}

After validation against literature data, a more continuum-like multiscale case is conducted to examine mesh-independent performance. At an altitude of approximately $70{\rm{km}}$, ${\rm{Kn}}_{{\rm{HS}},\infty}$ is decreased to $0.002$, and the gas is argon ($Ar$) with $D=0$, ${\rm{Pr}}=0.667$, and $\omega=0.81$. Other parameters remain unchanged. The surface mesh is also identical to the previous case. As shown in Fig.~\ref{cone-ar-scale}, the nonequilibrium regions mainly lie in the shock region, the leeward-side expansion region at approximately $x=0.002{\rm{m}}$, and the wake flow region at $x>0.035941{\rm{m}}$. The windward side is more continuum-like, so the particle fraction is significantly reduced. The $C_P$ contours at the surface and temperature contours at the symmetry plane are shown in Figs.~\ref{cone-ar-cp} and \ref{cone-ar-t}, respectively. For the aerodynamic force coefficients, UGKWP and DSMC simulations are performed on identical meshes for each value of $h$. Because $C_D$ is more influenced by friction drag, it is more sensitive than $C_L$, which in turn makes $L/D$ highly sensitive. For $C_D$ and $L/D$, the UGKWP method exhibits substantially better mesh-independence behavior than DSMC, as shown in Fig.~\ref{cone-ar-aero}. Taking $L/D$ at an angle of attack of $25^\circ$ as an example, relative to the mesh-independent solution, the UGKWP deviation is within $5\%$ at $h=0.0002{\rm{m}}$, whereas the DSMC deviation is close to $20\%$.

\section{Conclusions}\label{sec:conclusion}
In this work, the mesh-independence behavior of the UGKWP method is systematically investigated for hypersonic multiscale flows. Several algorithmic improvements are introduced to preserve second-order spatial and temporal accuracy in the presence of particle noise. These include second-order particle sampling based on local macroscopic gradients, a revised Venkatakrishnan limiter for highly stretched boundary-layer cells, and conservation corrections after particle sampling. Moreover, the first-order Chapman--Enskog term is incorporated into the flux function of the wave part, enabling better recovery of the GKS in the near-continuum regime. A detailed validation of these modifications is conducted for hypersonic flow over a cylinder at ${\rm{Ma}}_{\infty}=5$ and ${\rm{Kn}}_{\infty}=0.01$. The mesh-independent performance is notably enhanced by the second-order particle sampling and the inclusion of the first-order Chapman--Enskog term in the wave flux function. Conversely, incorporating wall values in the weighted least-squares gradient reconstruction improves accuracy in coarse mesh cases. The mesh-independent behavior of the improved UGKWP method is highly consistent with the UGKS, although the heat flux coefficient remains more sensitive due to the effect of particle noise on gradient reconstruction. The influence of the reference particle number is also examined.

For a comprehensive comparison with the single-scale DSMC method, both two-dimensional hypersonic flow over a cylinder and three-dimensional flow over a blunt cone are analyzed. In the cylinder case at ${\rm{Ma}}_{\infty}=5$ and ${\rm{Kn}}_{\infty}=0.01$, wall pressure, shear stress, and heat flux coefficients ($C_P$, $C_F$, and $C_Q$) are examined. The results show that the multiscale UGKWP method exhibits significantly better mesh-independence behavior than DSMC for the mesh-sensitive quantities $C_F$ and $C_Q$. At $h=0.032$, the peak $C_F$ value predicted by UGKWP deviates by about $35\%$ from the finest-mesh solution, whereas the corresponding DSMC deviation is about $130\%$. The peak $C_Q$ value deviates by less than $6\%$ for UGKWP, but by more than $85\%$ for DSMC. The deviation of DSMC increases rapidly when the first-layer mesh height and time step exceed the local microscopic scales $\lambda$ and $\tau$. In addition, apart from the first-layer height $h$, UGKWP requires fewer circumferential cells than DSMC to achieve a similar level of accuracy. In the three-dimensional blunt-cone case at ${\rm{Ma}}_{\infty}=10.15$ and ${\rm{Kn}}_{{\rm{HS}},\infty}=0.002$ (at an altitude of approximately $70{\rm{km}}$), UGKWP provides substantially better mesh-independence performance than DSMC for $C_D$ and $L/D$. For example, at an angle of attack of $25^\circ$ and $h=0.0002\,{\rm m}$, the deviation of $L/D$ from the mesh-independent solution is within $5\%$ for UGKWP, whereas the DSMC deviation is close to $20\%$.

Overall, this study demonstrates that the proposed improvements significantly enhance the mesh-independence performance of the UGKWP method for hypersonic multiscale flows. Compared with DSMC, the improved UGKWP method allows coarser near-wall meshes and fewer circumferential cells without sacrificing accuracy in mesh-sensitive quantities such as $C_F$, $C_Q$, $C_D$, and $L/D$. This makes UGKWP a highly efficient and robust tool for hypersonic flow simulations around near-space vehicles, providing practical guidance for mesh design in aerodynamic and thermal protection applications.

\section*{Acknowledgements}
The current research is supported by the National Key R$\&$D Program of China (Grant No. 2022YFA1004500), the National Natural Science Foundation of China (92371107), and the Hong Kong Research Grants Council (16208324).

\section*{Declaration of competing interest}
The authors declare that they have no known competing financial interests or personal relationships that could have appeared to influence the work reported in this paper.

\section*{Data availability}
The data that support the findings of this study are available from the corresponding author upon reasonable request.

\bibliography{ugkwp2ndref}

\end{document}